\documentclass[12pt]{article}

\usepackage[totalheight = 23cm, totalwidth = 17cm]{geometry}
\usepackage{amssymb,amsmath,amsfonts,amsbsy,graphicx}
\usepackage{bm,color}

\def\mathbi#1{\textbf{\em #1}}

\newcommand{\mpl}{m_{\rm Pl}}
\newcommand{\fnl}{f_{\rm NL}}

\newcommand{\calB}{{\cal B}}
\newcommand{\calD}{{\cal D}}
\newcommand{\calE}{{\cal E}}
\newcommand{\calF}{{\cal F}}

\newcommand{\calH}{{\cal H}}
\newcommand{\calJ}{{\cal J}}
\newcommand{\calL}{{\cal L}}

\newcommand{\calN}{{\cal N}}
\newcommand{\calO}{{\cal O}}
\newcommand{\calP}{{\cal P}}
\newcommand{\calR}{{\cal R}}
\newcommand{\calS}{{\cal S}}

\renewcommand{\theequation}{\arabic{section}.\arabic{equation}}

\begin{document}

\begin{titlepage}

\rightline{\footnotesize{APCTP-Pre2016-012}} 
\vspace{-0.2cm}

\begin{center}

\vskip 1.0 cm

{\Large \bf Multi-field inflation and cosmological perturbations}

\vskip 1.0cm

{\large
Jinn-Ouk Gong
}

\vskip 0.5cm

\small{\it
Asia Pacific Center for Theoretical Physics, Pohang 37673, Korea and
\\
Department of Physics, Postech, Pohang 37673, Korea
}

\vskip 1.2cm

\end{center}

\begin{abstract}

We provide a concise review on multi-field inflation and cosmological perturbations. We discuss convenient and physically meaningful bases in terms of which perturbations can be systematically studied. We give formal accounts on the gauge fixing conditions and present the perturbation action in two gauges. We also briefly review non-linear perturbations.

\end{abstract}

\end{titlepage}

\newpage
\setcounter{page}{1}

\tableofcontents

\newpage

\section{Introduction}

Cosmology, a comprehensive study on the origin and evolution of the universe, has become a branch of physical sciences only in twentieth century. With its theoretical foundation being provided by general relativity~\cite{Einstein:1916vd}, the hot big bang cosmology emerged in 1920s in which the universe has ever been expanding from an extremely hot and dense initial state in very far past~\cite{lemaitre}. Supported by a series of observational discoveries including the expansion of the universe from the relation between distance and redshift of galaxies~\cite{Hubble:1929ig}, and especially the cosmic microwave background (CMB) in 1965~\cite{Penzias:1965wn}, now the hot big bang cosmology is regarded as the standard model of cosmology.

CMB was generated when the universe has become cooled down due to the expansion so that the coupling between electrons and photons by Thompson scattering could not be maintained any longer~\cite{Dicke:1965zz}. This happened about 380,000 years after the big bang, thus the CMB may be regarded as a snapshot of the universe at this very early moment. On the whole observable range, the CMB exhibits an almost perfect black body spectrum corresponding to a homogeneous temperature of $2.725$ K with an accuracy of $\calO(10^{-5})$. This observation suggests that when the CMB was generated, the observable patch of the universe was in thermal equilibrium in a single causally connected patch. However, as we will see, naively the universe at that time was composed of a huge number of causally disconnected regions and thus there is no reason for them to have the same temperature with $\calO(10^{-5})$ accuracy. That is, the hot big bang cosmology is plagued by an extremely finely tuned initial condition to reproduce the universe as we observe now~\cite{Misner:1967uu}, and this is the so-called horizon problem. There are similar problems of initial conditions with fine tuning such as flatness problem.

In early 1980s\footnote{The idea of exponential expansion was known even in 1960s~\cite{gliner}, and just before the term ``inflation'' appeared it was discussed in different contexts, e.g. phase transition~\cite{Sato:1980yn} or singularity problem~\cite{Starobinsky:1980te}.}, it was realized that all the necessary initial conditions for successful hot big bang cosmology can be naturally provided by a period of rapid expansion of the universe at very early times, called ``inflation''~\cite{inflation}. Furthermore, inflation not only solves the otherwise finely tuned initial conditions like the horizon problem, but also provides the seed of subsequent structure formation in the universe~\cite{Mukhanov:1981xt}: during inflation, quantum mechanical fluctuations are stretched due to the rapid expansion to become classical perturbations. After inflation, these perturbations become the seed for the temperature anisotropies in the CMB and the inhomogeneous distribution of galaxies on large scales. Thus, by observing them closely we can study the primordial cosmic inflation and the underlying physics. Indeed, an important prediction of inflation is that the primordial perturbations produced during inflation have more or less the same amplitude on different length scales, i.e. scale-invariant, since the expansion is so fast that no appreciable change in inflationary dynamics happened on the whole observable scales relevant for, say, the CMB observations. And this prediction has been verified with very high confidence by most recent CMB observations including the Planck missions~\cite{Ade:2015oja}.

An immediate question that follows would be how to realize inflation concretely. Since the energy scale of the very early universe when cosmic inflation occurred is likely to be extremely high, as large as $10^{15}$ GeV. This is far exceeding the energy scale we can probe with terrestrial particle accelerators such as the Large Hadron Collider at CERN, with which we have confirmed the standard model of particle physics up to TeV scale. We regard, however, the standard model of particle physics as a low-energy effective field theory of a parent theory relevant for higher energy scales due to a number of reasons: small but non-zero neutrino masses are not explained, gravity is not included, and so on. Thus, inflation is likely to be described in the context of theories beyond the standard model of particle physics~\cite{bsm-inflation} such as supersymmetric theories and string theory. One common feature of those theories is that there are a multiple number of degrees of freedom~\cite{Nilles:1983ge} that could be relevant for inflation. Usually inflation is supposed to be driven by a single scalar field dominating the energy density of the universe during inflation, called inflaton. There are many scalar, possibly light, fields as the so-called superpartners in supersymmetric theories and as moduli fields in string theory, which all can in principle participate in inflationary dynamics and thus can be identified as the inflatons. Therefore we do have good theoretical motivations to consider more than single degree of freedom during inflation. Furthermore, the existence of other degrees of freedom can give rise to a phenomenology full of rich and interesting observational consequences that can be further constrained or even detected in near future.

The aim of this article is to provide a concise and easily accessible review for inflation driven by a multiple number of fields, complementing many excellent review articles~\cite{bsm-inflation,reviews} and textbooks~\cite{books} on related topics. The outline of this article is as follows. In Section~\ref{sec:inflation}, we briefly recall the basic of inflationary cosmology. In Section~\ref{sec:BG} we move to our main topic of multi-field inflation by starting with the background dynamics. Section~\ref{sec:pert} is devoted to the conventional approach to the cosmological perturbations produced during multi-field inflation. In Section~\ref{sec:pert-general} we present more careful considerations on the formulation of perturbations with alternative choices of gauge. We devote Section~\ref{sec:non-linear} to discuss concisely non-linear perturbations. Then we conclude in Section~\ref{sec:conc}.

\section{Inflation}
\label{sec:inflation}
\setcounter{equation}{0}

Before we begin our main topic of multi-field inflation, we quickly recall inflation: what it is, why it is attractive and how it occurs.

\subsection{Background equations}

We begin with the so-called Friedmann-Robertson-Walker (FRW) metric of a flat universe
\begin{equation}
\label{eq:ds}
ds^2 = -dt^2 + a^2(t)\delta_{ij}dx^idx^j \, .
\end{equation}
This metric describes a flat, homogeneous, isotropic and expanding universe parametrized by the scale factor $a(t)$. The spatial distance with the scale factor being singled out is described by $\delta_{ij}dx^idx^j$, which is called {\em comoving} distance. On the contrary, the {\em physical} distance is multiplied by the scale factor.

We first want the key equations. These are given by the Einstein equation 
\begin{equation}
\label{eq:Einstein-eq}
G_{\mu\nu} = \frac{T_{\mu\nu}}{\mpl^2} \, .
\end{equation}
We first consider the left hand side of \eqref{eq:Einstein-eq}, namely, the Einstein tensor
\begin{equation}
\label{eq:Einstein_tensor}
G_{\mu\nu} = R_{\mu\nu} - \frac{1}{2}g_{\mu\nu}R \, .
\end{equation}
We can immediately write each component of  the metric tensor $g_{\mu\nu}$ and its inverse $g^{\mu\nu}$ as
\begin{equation}
\begin{split}
& g_{00} = -1 \, , \hspace{0.5cm} g_{ij} = a^2\delta_{ij} \, ,
\\
& g^{00} = -1 \, , \hspace{0.5cm} g^{ij} = a^{-2}\delta^{ij} \, .
\end{split}
\end{equation}
To compute the Einstein tensor $G_{\mu\nu}$, we need to calculate the Christoffel symbol, the Ricci tensor and the Ricci scalar:
\begin{equation}
\begin{split}
\Gamma^\rho_{\mu\nu} & = \frac{1}{2}g^{\rho\sigma} \left( g_{\mu\sigma,\nu} +
g_{\sigma\nu,\mu} - g_{\mu\nu,\sigma} \right) \, ,
\\
R_{\mu\nu} & = \Gamma^\alpha_{\mu\nu,\alpha} - \Gamma^\alpha_{\mu\alpha,\nu} +
\Gamma^\alpha_{\sigma\alpha}\Gamma^\sigma_{\mu\nu} -
\Gamma^\alpha_{\sigma\nu}\Gamma^\sigma_{\mu\alpha} \, ,
\\
R & = g^{\mu\nu}R_{\mu\nu} \, .
\end{split}
\end{equation}
The non-zero components of the Christoffel symbols are, after some calculations,
\begin{align}
\Gamma^0_{ij} & = a^2H\delta_{ij} \, ,
\\
\Gamma^i_{0j} = \Gamma^i_{j0} & = H\delta^i{}_j \, ,
\end{align}
with $H = \dot{a}/a$ being the Hubble parameter, otherwise zero. Then, easily we have
\begin{align}
R_{00} & = -3 \left( H^2 + \dot{H} \right) \, ,
\\
R_{ij} & = a^2 \left( 3H^2 + \dot{H} \right)\delta_{ij} \, ,
\\
R & = 6 \left( \dot{H} + 2H^2 \right) \, .
\end{align}
Thus, the non-zero components of the Einstein tensor \eqref{eq:Einstein_tensor}, or more frequently $G^\mu{}_\nu = g^{\mu\rho}G_{\rho\nu}$, are
\begin{align}
G_{00} & = 3H^2 \, ,
\\
G_{ij} & = -a^2 \left( 2\dot{H} + 3H^2 \right)\delta_{ij} \, ,
\\
G^0{}_0 & = -3H^2 \, ,
\\
G^i{}_j & = - \left( 2\dot{H} + 3H^2 \right)\delta^i{}_j \, .
\end{align}

As can be read from \eqref{eq:Einstein-eq}, the Einstein tensor which describes the structure of the space-time should be matched with the energy-momentum tensor which describes the matter residing in the space-time. On the assumption of the homogeneous and isotropic background, we may regard at the background level that the energy-momentum tensor is that of perfect fluid\footnote{In terms of the general hydrodynamical matter fluid, the energy-momentum tensor is written as
\begin{equation*}
\label{energy-momentum_tensor_HDmatter}
T_{\mu\nu} = (\rho+p)u_\mu u_\nu + pg_{\mu\nu} \, ,
\end{equation*}
where $u^\mu$ is the fluid 4-velocity which satisfies
\begin{equation*}
\label{4-velocity_constraint}
u^\mu u_\mu = g_{\mu\nu}u^\mu u^\nu = -1 \, ,
\end{equation*}
so that $u^\mu$ is a time-like, unit 4-vector. Thus we can set $u^\mu = (1,0,0,0)$. Using these we can trivially find \eqref{eq:energy-momentum_tensor_BG}.}, i.e.
\begin{equation}
\label{eq:energy-momentum_tensor_BG}
T^\mu{}_\nu = \mathrm{diag}(-\rho,p,p,p) \, .
\end{equation}

Now we can write each component of \eqref{eq:Einstein-eq}:
\begin{align}
\label{eq:Friedmann_eq}
00 \mbox{ component: } & H^2 = \frac{\rho}{3\mpl^2} \, ,
\\
\label{eq:BGij}
ij \mbox{ component: } & -3H^2 - 2\dot{H} = \frac{p}{\mpl^2} \, .
\end{align}
\eqref{eq:Friedmann_eq} is called the Friedmann equation, which relates the Hubble parameter to the energy density. Using \eqref{eq:Friedmann_eq} for \eqref{eq:BGij} to replace $H^2$ with $\rho$, we can find the time variation of $H$ as
\begin{equation}
\label{eq:acc_eq2}
\dot{H} = -\frac{\rho+p}{2\mpl^2} \, .
\end{equation}
Or, explicitly in terms of the time derivatives of the scale factor, 
\begin{equation}
\label{eq:acc_eq}
\frac{\ddot{a}}{a} = -\frac{\rho+3p}{6\mpl^2} \, .
\end{equation}
We will refer to this equation soon. Note that by taking a time derivative of \eqref{eq:Friedmann_eq} and using \eqref{eq:BGij} to eliminate $\dot{H}$, we can derive energy conservation equation
\begin{equation}
\label{eq:BGconservation}
\dot\rho + 3H(\rho+p) = 0 \, .
\end{equation}
This is what we can find from the conservation of energy-momentum tensor: from
\begin{equation}
T^\mu{}_{\nu;\mu} = T^\mu{}_{\nu,\mu} - \Gamma^\rho_{\mu\nu}T^\mu{}_\rho + \Gamma^\mu_{\rho\mu}T^\rho{}_\nu = 0 \, ,
\end{equation}
we can trivially check that $\nu=0$ component gives \eqref{eq:BGconservation}. $\nu=i$ component vanishes identically.

\subsection{Cosmic microwave background}

\subsubsection{Generation of the CMB}

With the necessary background equations, now let us see what happened in the past when the temperature was high. First, we note that from the conservation equation \eqref{eq:BGconservation} that different species scale differently: ordinary particles (electron, proton, neutron...) have very large rest energy compared to the kinetic energy, so they are called pressureless matter and $p=0$. Meanwhile, photons, or more generally relativistic particles, have $p=\rho/3$ and are called radiation. Plugging these relations into \eqref{eq:BGconservation}, we find
\begin{align}
\label{eq:rhom_scale}
\rho_\text{matter} & \propto a^{-3} \, ,
\\
\rho_\text{radiation} & \propto a^{-4} \, .
\end{align}
We may understand that the energy density of pressureless matter is inversely proportional to the volume $\sim a^3$ which contains the matter particles, and for radiation the energy density is also proportional to the frequency, or the inverse of the wavelength, so we have one more power of the scale factor. What this tells us is that, in the past, the universe was dominated by radiation.

\begin{figure}[h]
 \centering
  \includegraphics[width=12cm]{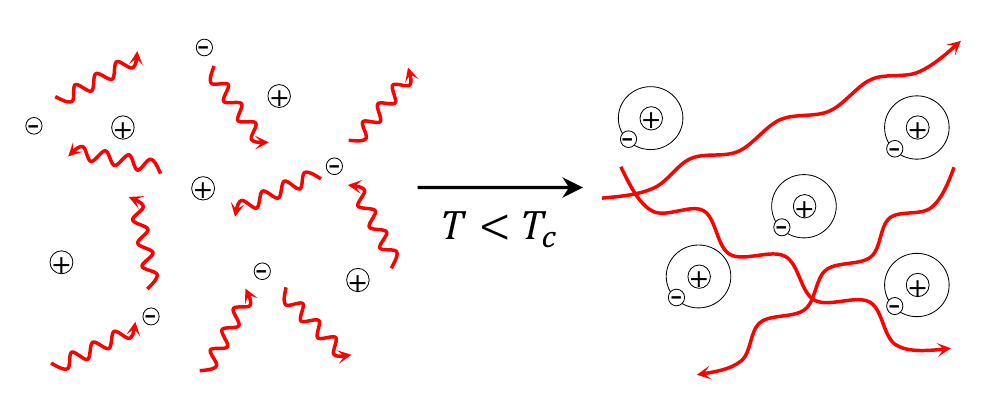}
  \caption{When $T>T_c$, electrons were free and constantly scattered off photons, so that the universe was ``foggy''. After the temperature drops below $T_c$, electrons are all captured by protons, and photons can propagate without scattering.}
 \label{fig:decoupling}
\end{figure}

More radiation in the past means, of course, the universe was hotter. It was too hot to maintain neutral molecules, like hydrogen: because of the very hot temperature, electrons were energetic enough to overcome the binding energy to protons, so that the universe was filled by radiation (mostly photons), free electrons and nuclei (and dark matter). During this stage, the mean free path of photons was very short because of the Thomson scattering between free electrons and photons, maintaining thermal equilibrium. Thus, the universe was very ``foggy'' for photons: exactly like we cannot see very far away when the weather is very foggy. This stage continued until the universe was cooled to a critical temperature $T_c \sim 3000$ K. Below this temperature, the binding energy between electrons and protons could overcome thermal background and there remained no free electron. Thus, from this time on, the universe has become transparent to photons and they could reach us after propagating for a long long time. This situation is depicted in Figure~\ref{fig:decoupling}. These very old photons, which have traveled since the moment of this ``last scattering'', are the cosmic microwave background (CMB). It was observed in 1965 by Penzias and Wilson by chance.

\begin{figure}[h]
 \centering
  \includegraphics[width=6cm]{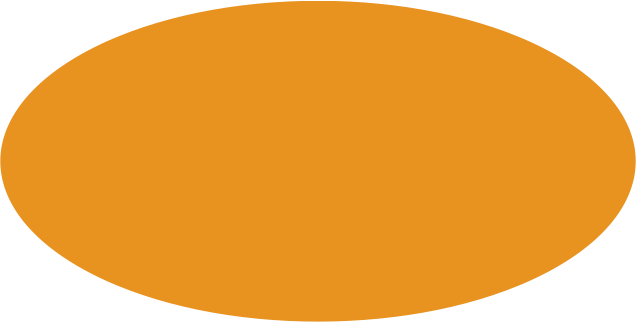}
  \hspace{1cm}
  \includegraphics[width=6.3cm]{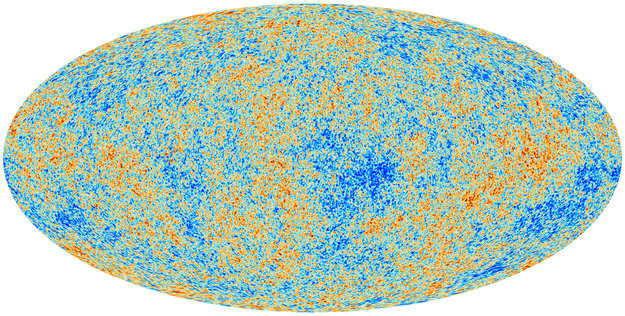}
  \caption{(Left) the cosmic microwave background is observed to be extremely homogeneous and isotropic with the average temperature $T_0 \sim 2.7$ K. (Right) however, it contains genuine temperature fluctuations with respect to $T_0$ of the magnitude $\delta{T}/T_0\sim10^{-5}$. The temperature fluctuation map is taken by the Planck satellite~\cite{Adam:2015rua}.}
 \label{fig:cmb}
\end{figure}

The observations tell us that the CMB is extremely homogeneous and isotropic, i.e. we observe the same average temperature $T_0 \sim 2.7$ K no matter which part or direction of the sky we observe. Since photons were constantly scattering off free electrons and thus in thermal equilibrium, the temperature spectrum of the CMB exhibits that of almost perfect blackbody radiation. Moreover, the CMB could be generated only when the universe was hotter in the past. Thus the discovery of the CMB was the knockdown blow for the steady state cosmology which was competing against the hot big bang model in 60's. Note that, after removing all the contaminations and foreground effects, we have genuine temperature fluctuations of the magnitude $\delta{T}/T_0 \sim 10^{-5}$. We will return to this point later. In Figure~\ref{fig:cmb} we show the background and fluctuation temperature maps of the CMB.

\subsubsection{Horizon problem}

The CMB has brought, with the triumph of the hot big bang cosmology, big mysteries at the same time. Let us consider 1 of them, namely, why the CMB is so much homogeneous. For this, it is very convenient to introduce the conformal time $\tau$, defined by 
\begin{equation}
d\tau \equiv \frac{dt}{a} \, .
\end{equation}
With $\tau$, the line element \eqref{eq:ds} is written as
\begin{equation}
ds^2 = a^2(\tau) \left( -d\tau^2 + \delta_{ij}dx^idx^j \right) \, ,
\end{equation}
so that the metric is written as a product of the static Minkowski metric times the scale factor. What does the conformal time mean? Let us consider the radial propagation of light, which is the null geodesic $ds^2=0$. Then, using the spherical coordinate we can write the radial distance $r$ a photon has traveled from some initial moment in terms of the conformal time as
\begin{equation}
r = \tau \, ,
\end{equation}
i.e. the conformal time measures the (comoving) distance a photon has traveled.

\begin{figure}[h]
 \centering
  \includegraphics[width=15cm]{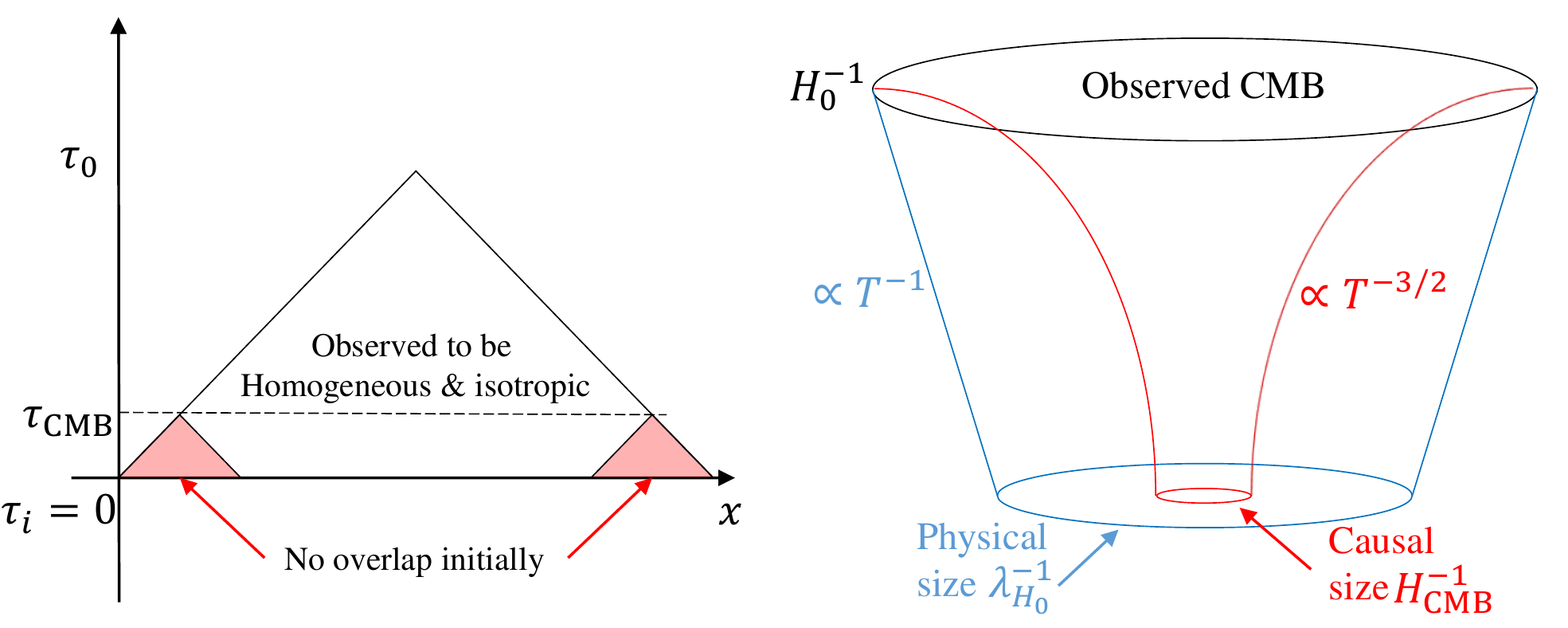}
  \caption{(Left) conformal diagram of the universe. From the cosmic singularity ($\tau_i=0$) until the moment of the CMB generation ($\tau_\text{CMB}$) there was no time for the CMB to achieve causal communication to have the same temperature $T_0$. (Right) as a sample calculation, we can see that at that time the universe was filled with $10^4-10^5$ causally disconnected patches.}
 \label{fig:horizon}
\end{figure}

Then what's the trouble with the CMB? We can straightforwardly find that from an initial moment $i$ till present $0$, the conformal time (i.e. the distance photons have traveled)
\begin{equation}
\tau = \int_i^0 \frac{dt}{a} = \int_i^0 \frac{1}{a}\frac{dt}{da}da = \int_i^0 \frac{1}{aH} d\log{a} \propto a^{1/2}|_i^0 \, ,
\end{equation}
where for each equality we have used 1) the scale factor is a function of time solely, $a=a(t)$, 2) the definition of the Hubble parameter, $\dot{a}=aH$, and 3) assumption of a matter dominated universe, $H \sim \rho_\text{matter}^{1/2} \sim a^{-3/2}$. Now, without loss of generality, we can take initial moment as the initial singularity $a(t_i)=0$, where also $\tau=0$, so that simply $\tau \propto a^{1/2}$. Further, using the relation between the scale factor which is normalized to $a_0=1$ at present and the redshift $z$
\begin{equation}
a = \frac{1}{1+z} \, ,
\end{equation}
we can find $\tau \propto (1+z)^{-1/2}$. Using $z_0=0$ and $z_\text{CMB} \sim 1100$, we can easily find
\begin{equation}
\frac{\tau_\text{CMB}}{\tau_0} \sim \frac{1}{\sqrt{1100^3}} \sim 0.03 \, .
\end{equation}
Thus, at the moment when the CMB was generated, the past light cones stemming from the two end points do not have any overlapping region initially, i.e. those two points were never in causal communication and thus there is no reason they should have the same temperature with the accuracy of $10^{-5}$: we must impose a heavy fine tuning over $10^4 - 10^5$ causally disconnected patches at the moment of the last scattering unless we provide a natural way for them to have the same temperature. This is the so-called horizon problem. It is depicted in the left panel of Figure~\ref{fig:horizon}. Note that the spatial distance shown in the figure is the comoving one, thus the physical distance is obtained by multiplying the scale factor $a(t)$ which vanishes as we approach the cosmic singularity, currently at $\tau=0$.

To have a more concrete idea, let us assume that the observed CMB size coincides with the current Hubble patch $1/H_0$, within which causal communications are possible. Then let us ask whether they were the same when the CMB was generated, or if different how much they were different. First, what is $\lambda_{H_0^{-1}}$, the {\em physical} size that corresponds to $1/H_0$? Physical sizes simply scale with the scale factor $a(t)$, which is inversely proportional to the temperature $T$. Thus, we can easily find
\begin{equation}
\lambda_{H_0^{-1}} = H_0^{-1}\frac{a_\text{CMB}}{a_0} = H_0^{-1}\frac{T_0}{T_\text{CMB}} \, .
\end{equation}
Meanwhile, $H$ evolves according to the Friedmann equation \eqref{eq:Friedmann_eq}. It is important to notice at this moment that $H$ depends on the energy density, i.e. which types of matter contents are there. For simplicity we assume the universe is dominated by matter that is inversely proportional to the physical volume as can be read from \eqref{eq:rhom_scale}, and thus $H$ is proportional to $T^3$:
\begin{equation}
H^2 \propto \rho_\text{matter} \propto a^{-3} \propto T^3 \, ,
\end{equation}
so that we can find $H_\text{CMB}^{-1}$, the Hubble horizon radius when the CMB was generated, as
\begin{equation}
H_\text{CMB}^{-1} = H_0^{-1} \left( \frac{T_0}{T_\text{CMB}} \right)^{3/2} \, .
\end{equation}
Thus, if we compare the ratio of these volumes,
\begin{equation}
\frac{\lambda_{H_0^{-1}}^3}{\left( H_\text{CMB}^{-1} \right)^3} = \left( \frac{T_\text{CMB}}{T_0} \right)^{3/2} \sim 4 \times 10^4 \, .
\end{equation}
That is, assuming that at present the Hubble horizon size and the CMB scale are the same, when the CMB was generated, the corresponding physical volume was filled with $10^4$ - $10^5$ causally disconnected patches: see the right panel of Figure~\ref{fig:horizon}. Thus, it is a tremendous fine tuning that these disconnected patches all turn out to have the same temperature with the accuracy of $10^{-5}$ as the current observations on the CMB demand.

\subsection{Inflation}

\subsubsection{Inflation: what and how}

Thus, we see that at the heart of the horizon problem lies the fact that the Hubble horizon $1/H = 1/\left(\dot{a}/a\right)$ always expands faster than the physical length scale $\lambda \sim a$,
\begin{equation}
\frac{d}{dt} \left( \frac{\lambda}{H^{-1}} \right) \sim \frac{d}{dt} \left[ \frac{a}{\left(\dot{a}/a\right)^{-1}} \right] = \ddot{a} < 0 \, ,
\end{equation}
irrespective of whether the universe is dominated by matter or radiation. Thus, we can just turn upside down and make the physical size expands faster than the Hubble horizon: then physical scales expand faster than the horizon so causal communication could be possible during this stage. This tells us
\begin{equation}
\label{eq:inf-def}
\frac{d}{dt} \left( \frac{\lambda}{H^{-1}} \right) > 0 \quad \longleftrightarrow \quad  \ddot{a}>0 \, .
\end{equation}
That is, the universe experiences an {\em accelerated} expansion. This period of accelerated expansion is called ``{\em inflation}''.

How can we more quantitatively say if it's inflation or not? We can rewrite \eqref{eq:acc_eq} as
\begin{equation}
\frac{\ddot{a}}{a} = \frac{2\rho}{6\mpl^2} - \frac{3\rho+3p}{6\mpl^2} = H^2 + \dot{H} > 0 \, ,
\end{equation}
where the 2nd equality follows by applying \eqref{eq:Friedmann_eq} and \eqref{eq:acc_eq2}, and the last inequality is the definition of inflation \eqref{eq:inf-def}. Thus, inflation occurs when the following condition is satisfied:
\begin{equation}
\label{eq:SRepsilon}
\epsilon \equiv -\frac{\dot{H}}{H^2} < 1 \, .
\end{equation}
This parameter, which tells whether it's inflation or not, is called ``slow-roll'' parameter, in the context of slow-roll inflation: see the next section.

So with what kind of matter can we have inflation? From \eqref{eq:acc_eq}, we see that to have $\ddot{a}>0$ we need a special form of matter which has a negative pressure, $p < -\rho/3$, or in terms of the equation of state $w$,
\begin{equation}
w \equiv \frac{p}{\rho} < -\frac{1}{3} \, .
\end{equation}
Clearly usual pressureless matter ($w=0$) or radiation ($w=1/3$) cannot support inflation. The simplest candidate is the so-called cosmological constant $\Lambda$, which has
\begin{equation}
p_\Lambda = -\rho_\Lambda \quad \left( w_\Lambda = -1 \right) \, .
\end{equation}
Then the Friedmann equation \eqref{eq:Friedmann_eq} is trivially solved: since $\Lambda$ is, as the name suggests, a constant and so is $H$:
\begin{equation}
H^2 = \frac{\Lambda}{3\mpl^2} = \text{constant} \, .
\end{equation}
Thus we can see that the scale factor increases exponentially during inflation as
\begin{equation}
a \propto \exp \left( \sqrt{\frac{\Lambda}{3\mpl^2}}t \right) \, .
\end{equation}

\subsubsection{Horizon problem revisited}

So the question is: how does inflation solve the horizon problem? Now we can move to the conformal time to see a clear visualization how inflation solves the horizon problem. During inflation, for convenience driven by a cosmological constant $\Lambda$ so that $H$ is constant, the conformal time is given by
\begin{equation}
\tau = \int \frac{dt}{a} = \int \frac{e^{-Ht}}{a_0} dt = -\frac{1}{aH} < 0 \, .
\end{equation}
That is, the conformal time is {\em negative} during inflation. Further, now the cosmic singularity $a=0$ can be pushed to $\tau = -\infty$. Thus, even the two end points at $\tau=\tau_\text{CMB}$ have no overlap at $\tau=0$, now $\tau$ can be extended to negative infinity during inflation so that there could be ample overlapping region enough to explain the homogeneity of the CMB.

\begin{figure}[h]
 \centering
  \includegraphics[width=15cm]{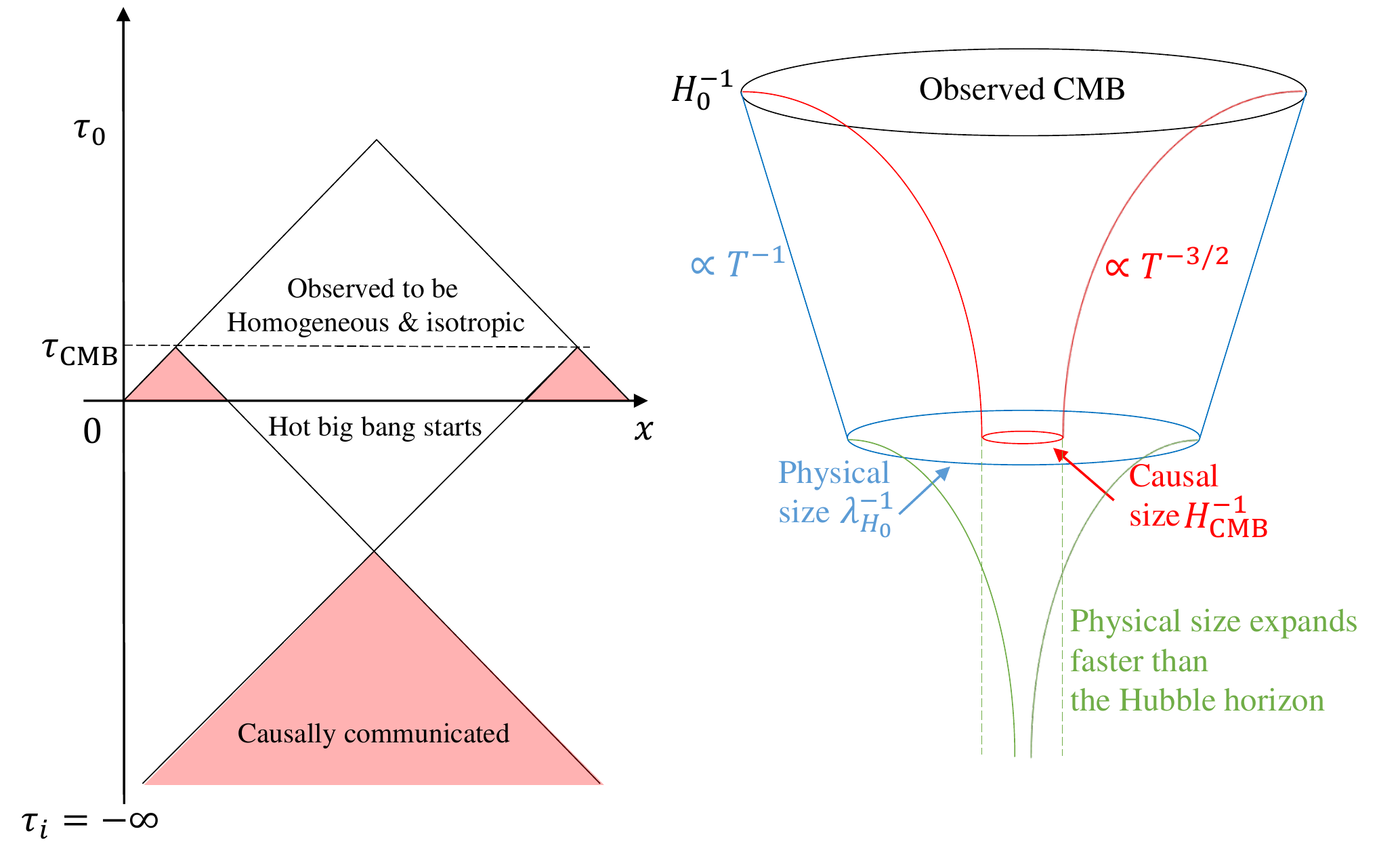}
  \caption{(Left) conformal diagram of the universe, this time including inflation. Inflation extends $\tau$ to $-\infty$, giving ample room for causal communication well before the onset of hot big bang evolution at $\tau=0$. (Right) inflation corresponds to the period when the physical size $\lambda \sim a$ expands faster than the Hubble horizon $1/H$.}
 \label{fig:horizon-sol}
\end{figure}

As is clear from Figure~\ref{fig:horizon-sol}, the longer inflation lasts, the larger the overlapping region becomes. Thus we need a certain duration of inflation to explain the homogeneous CMB. The amount of inflation is quantified by the number of $e$-folds $N$ between some initial ($i$) and final ($f$) moments, which is given by
\begin{equation}
\label{eq:e-folds}
N \equiv \int_i^f H dt = \int_i^f \frac{da}{a} = \log \left( \frac{a_f}{a_i} \right) \, .
\end{equation}
Thus, with a given $N$, the final scale factor is related to the initial scale factor by $a_f = a_i e^N$, i.e. the universe has expanded by $e^N$ times. Now we can compute how large $N$ should be for the CMB. The most natural way is that at the beginning of inflation (or the part of inflation relevant for our observable universe) the physical length scale $\lambda_{H_0^{-1}}$ is smaller than the Hubble horizon during inflation $H_I$ so that causal communication has been established within $\lambda_{H_0^{-1}}$ to have the same temperature. This gives
\begin{equation}
\lambda_{H_0^{-1}} = H_0^{-1} \frac{a_i}{a_0} = H_0^{-1} \frac{a_f}{a_0} \frac{a_i}{a_f} = H_0^{-1} \frac{T_0}{T_f} e^{-N} < H_I^{-1} \, .
\end{equation}
Thus, solving for $N$ from the last inequality, we obtain
\begin{equation}
N > \log \left( \frac{T_0}{H_0} \right) - \log \left( \frac{T_f}{H_I} \right) \sim 67 - \log \left( \frac{T_f}{H_I} \right) \, ,
\end{equation}
where we have used $H_0 \sim 10^{-42}$ GeV and $T_0 \sim 10^{-13}$ GeV. Thus, assuming that the logarithmic term which includes two unknown factors give a number of $\calO(1)$, we require that
\begin{equation}
N \gtrsim 60 \, .
\end{equation}
That is, to explain the homogeneity of the CMB, i.e. to solve the horizon problem, we need 60 $e$-folds of expansion: during inflation the universe should have expanded by $e^{60} \sim 10^{26}$ times.

\subsubsection{Single field inflation and slow-roll approximation}

The cosmological constant is obviously the simplest candidate that drives inflation, but the problem is that if this is the case, inflation never ends and we cannot recover the universe in which we live with stars, galaxies, clusters of galaxies and so on. Thus, we need some different material which can mimic the cosmological constant and at the same time provide a ``graceful exit'' from inflation. This is usually achieved by a scalar field $\phi$. For simplicity here we assume that this scalar field, named ``inflaton'' in the sense that it drives inflation, is minimally coupled to gravity and has canonical kinetic term. Then the action is the sum of the gravitational sector, which we take the Einstein-Hilbert action, and the matter sector:
\begin{equation}
\label{eq:action}
S = \int d^4x \sqrt{-g} \frac{\mpl^2}{2}R + \int d^4x \sqrt{-g} \bigg[ \underbrace{ -\frac{1}{2}g^{\mu\nu}\partial_\mu\phi\partial_\nu\phi - V(\phi) }_{\equiv \calL_m} \bigg] \, .
\end{equation}
The corresponding energy-momentum tensor $T_{\mu\nu}$ of $\phi$ can be obtained by perturbing the matter Lagrangian with respect to $g^{\mu\nu}$,
\begin{equation}
\label{eq:phiEMtensor}
T_{\mu\nu} = -\frac{2}{\sqrt{-g}} \frac{\delta \left( \sqrt{-g}\calL_m \right)}{\delta{g}^{\mu\nu}} = \partial_\mu\phi\partial_\nu\phi - g_{\mu\nu} \left[ \frac{1}{2}g^{\rho\sigma}\partial_\rho\phi\partial_\sigma\phi + V(\phi) \right] \, .
\end{equation}
Then we can easily compute $00$ and $ii$ components which can then be matched to the energy density and pressure respectively\footnote{The relations between fluid quantities in terms of one or more scalar fields are highly non-trivial. See~\cite{field-fluid} for example.} [see \eqref{eq:energy-momentum_tensor_BG}]:
\begin{align}
\label{eq:BGphi_rho}
\rho & = -T^0{}_0 = \frac{1}{2}\dot\phi^2 + \frac{1}{2}\frac{(\nabla\phi)^2}{a^2} + V \, ,
\\
\label{eq:BGphi_p}
p & = \frac{1}{3}T^i{}_i = \frac{1}{2}\dot\phi^2 - \frac{1}{6}\frac{(\nabla\phi)^2}{a^2} - V \, .
\end{align}
Thus, if potential dominates over the kinetic energy ($\partial_\mu\phi\partial^\mu\phi \ll V$) these simplify to $\rho \approx V \approx -p$, thus the inflaton provides a nearly cosmological constant, leading to an exponential expansion of the universe, viz. inflation.

Let us write the background equation of motion for $\phi$. From this we can find a number of useful formulae which do not resort to the dynamics of $\phi$ but to $V$ and its derivatives only. The equation of motion for $\phi$ can be found from the Euler-Lagrange equation,
\begin{equation}
\partial_\mu \left[ \frac{\partial\calL}{\partial(\partial_\mu\phi)} \right] = \frac{\partial\calL}{\partial\phi} \, .
\end{equation}
This gives
\begin{equation}
-\Box\phi + \frac{\partial{V}}{\partial\phi} = 0 \, ,
\end{equation}
where
\begin{equation}
\Box \equiv \frac{1}{\sqrt{-g}} \partial_\mu \left( \sqrt{-g}g^{\mu\nu}\partial_\nu \right) = -\frac{\partial^2}{\partial t^2} - 3H\frac{\partial}{\partial t} + \frac{\Delta}{a^2} \, ,
\end{equation}
with $\Delta \equiv \delta^{ij}\partial_i\partial_j$ being the spatial Laplacian operator. Here for the last equality we have taken the background metric. Thus, the background field $\phi = \phi(t)$ follows the equation of motion
\begin{equation}
\label{eq:BGphi_eom}
\ddot\phi + 3H\dot\phi + \frac{\partial{V}}{\partial\phi} = 0 \, .
\end{equation}

So how this equation for $\phi$ simplifies? From \eqref{eq:acc_eq}, using \eqref{eq:BGphi_rho} and \eqref{eq:BGphi_p} we require
\begin{equation}
\frac{\ddot{a}}{a} = -\frac{\dot\phi^2-V}{3\mpl^2} > 0 \, .
\end{equation}
Note that we can again precisely find \eqref{eq:SRepsilon} by using  \eqref{eq:BGphi_rho} and \eqref{eq:BGphi_p} for \eqref{eq:Friedmann_eq} and \eqref{eq:acc_eq2}. Then this means
\begin{equation}
\dot\phi^2 < V \, .
\end{equation}
Taking a time derivative on both sides, this says $\ddot\phi < \partial{V}/\partial\phi$. Thus, \eqref{eq:BGphi_eom} is simplified to
\begin{equation}
\label{eq:SRBGphi_eq}
3H\dot\phi + \frac{\partial{V}}{\partial\phi} = 0 \, .
\end{equation}
Thus we can replace $\dot\phi$, or more generally the dynamics of $\phi$, with the derivatives of the potential $V$.

Then now let us consider the slow-roll parameter $\epsilon$, \eqref{eq:SRepsilon}. Applying \eqref{eq:SRBGphi_eq}, we find
\begin{equation}
\label{eq:epsilon-SR}
\epsilon = -\frac{\dot{H}}{H^2} \approx \frac{\dot\phi^2/(2\mpl^2)}{V/(3\mpl^2)} \approx \frac{3}{2} \frac{1}{V} \frac{{V'}^2}{9H^2} \approx \frac{\mpl^2}{2} \left( \frac{V'}{V} \right)^2 \, ,
\end{equation}
where $V' \equiv \partial{V}/\partial\phi$. Thus, $\epsilon$ in the slow-roll approximation tells us how steep the potential slope is. We can introduce another important slow-roll parameter $\eta$, which describes how quickly $\epsilon$ evolves:
\begin{equation}
\label{eq:eta-SR}
\eta \equiv \frac{\dot\epsilon}{H\epsilon} \approx \left[ H \frac{\mpl^2}{2} \left( \frac{V'}{V} \right)^2 \right]^{-1} \mpl^2 \frac{V'}{V} \left[ \frac{V''}{V} - \left( \frac{V'}{V} \right)^2 \right] \dot\phi \approx 2\mpl^2\frac{V''}{V} + 4\epsilon \, .
\end{equation}
Also note that in the slow-roll approximation the $e$-fold $N$ can be written in terms of the potential solely:
\begin{equation}
\label{eq:N_SR}
N = \int_i^f H dt = \int_i^f H \frac{dt}{d\phi}d\phi = \int_i^f \frac{H}{\dot\phi}d\phi \approx \frac{1}{\mpl^2} \int_f^i \frac{V}{V'}d\phi \, .
\end{equation}

\section{Background dynamics}
\label{sec:BG}
\setcounter{equation}{0}

In the previous section, we have briefly recalled the basic of inflation driven by a single inflaton field. Being well acquainted with the prerequisite knowledge, now we move to our main topic of multi-field inflation. Explicitly, we consider a $n$-dimensional multi-field system with a generic field space metric $G_{ab}$ coupled to the Einstein gravity:
\begin{equation}
\label{eq:action}
S = \int d^4x \sqrt{-g} \left[ \frac{\mpl^2}{2}R - \frac{1}{2}G_{ab}g^{\mu\nu}\partial_\mu\phi^a\partial_\nu\phi^b - V(\phi) \right] \, .
\end{equation}
We may well consider more general possibilities, such as $f(R)$ gravity~\cite{f(R)} or $P(G_{ab},X^{ab},\phi^a)$ with $X^{ab} \equiv -g^{\mu\nu}\partial_\mu\phi^a\partial_\nu\phi^b/2$~\cite{k-inflation}. But the discussions and considerations presented in this article can be straightly applied and extended, so we restrict our discussions to \eqref{eq:action}.

We begin with the background dynamics. The energy-momentum tensor derived from \eqref{eq:action} is of the form
\begin{equation}
\label{eq:EMtensor}
T_{\mu\nu} = G_{ab}\partial_\mu\phi^a\partial_\nu\phi^b - g_{\mu\nu} \left[ \frac{1}{2}G_{ab}g^{\rho\sigma}\partial_\rho\phi^a\partial_\sigma\phi^b + V(\phi) \right] \, ,
\end{equation}
from which we can combine the $00$ and $ij$ components of the Einstein equation to find
\begin{align}
\label{eq:Friedmann}
H^2 & = \frac{1}{3\mpl^2} \left[ \frac{1}{2}\dot\phi_0^2 + V(\phi_0) \right] \, ,
\\
\label{eq:Hdot}
\dot{H} & = -\frac{\dot\phi_0^2}{2\mpl^2} \, ,
\end{align}
where $\dot\phi_0^2 \equiv G_{ab}\dot\phi_0^a\dot\phi_0^b$ with $\phi_0^a$ being the background value. The background equation of motion for the scalar fields can be obtained either by the variation of the action \eqref{eq:action} with respect to $\phi^a$ or $\nu=0$ component of the conservation of the energy-momentum tensor $T^\mu{}_{\nu;\mu}=0$ as
\begin{equation}
\label{eq:BGphi}
D_t\dot\phi_0^a + 3H\dot\phi_0^a + G^{ab}V_b = 0 \, ,
\end{equation}
where 
\begin{equation}
D_t\dot\phi_0^a = \frac{D\dot\phi_0^a}{dt} \equiv \frac{d\dot\phi_0^a}{dt} + \Gamma^a_{bc}\dot\phi_0^b\dot\phi_0^c
\end{equation}
is a covariant time derivative with $\Gamma^a_{bc}$ being the Christoffel symbol constructed by the field space metric $G_{ab}$. For more geometric discussions, see e.g.~\cite{geometry}.

Before we proceed further to discuss perturbations, it is very useful to consider the change of basis in the field space. As in space-time, $\phi^a$ plays the role of coordinates in the field space\footnote{This is of particular caution when we discuss non-linear perturbations, for which we will further return to this point in Section~\ref{sec:non-linear}.}. Thus we are free to choose and/or transform to a convenient basis. One possible choice is a local orthogonal frame: we can introduce a complete set of vielbeins $e_a^I = e_a^I(t)$ which maps the general, arbitrary basis denoted by the index $a$ into a local orthogonal frame denoted by the superscript $I$ as
\begin{equation}
e^I_ae^J_bG^{ab} = \delta^{IJ} \quad \text{and} \quad e^I_ae^J_b\delta_{IJ} = G_{ab} \, .
\end{equation}
Note that here we have not specified the new, orthogonal $IJ$ basis: we may choose whatever frame we like as long as it is orthogonal. A physically important one is the so-called ``kinematic basis'', which is set along and perpendicular to the field trajectory~\cite{geometry,basis,Achucarro:2010da}. The unit tangent vector $T^a$ is defined by 
\begin{equation}
\label{eq:Ta}
T^a \equiv \frac{\dot\phi_0^a}{\dot\phi_0} \, .
\end{equation}
The normal vector $N^a$ which satisfies $G_{ab}T^aN^b = 0$ is naturally proportional to the derivative of $T^a$, i.e. $D_tT^a \propto N^a$. Taking a time derivative to $T^a$ and using \eqref{eq:BGphi}, we find
\begin{equation}
D_tT^a = -\frac{\ddot\phi_0}{\dot\phi_0}T^a - \frac{1}{\dot\phi_0} \left( 3H\dot\phi_0^a + V^a \right) \, .
\end{equation}
Projecting this equation along $T^a$ and $N^a$, we obtain respectively
\begin{align}
\label{eq:BGtangent}
& \ddot\phi_0 + 3H\dot\phi_0 + V_T = 0 \, ,
\\
& D_tT^a = -\frac{V_N}{\dot\phi_0}N^a \equiv \dot\theta N^a \, ,
\end{align}
where $V_T \equiv V_aT^a$ and $V_N \equiv V_aN^a$ are the projection of the potential derivative onto the tangential and perpendicular direction to the trajectory respectively, and we have {\em defined} the proportionality parameter as the angular velocity $\dot\theta$ of the trajectory. Note that the background equation along the tangent direction \eqref{eq:BGtangent} is precisely the same as that in single field inflation \eqref{eq:BGphi_eom}. Thus we are naturally led to identify the tangential component of the field fluctuations as what is associated with the curvature perturbation, as we will see in later sections. Also note that the kinematic basis is not the only sensible choice for an orthogonal basis. We will discuss another physically illuminating choice in the next section.

\section{Dynamics of perturbations}
\label{sec:pert}
\setcounter{equation}{0}

Up to now, we have considered background dynamics only. But as mentioned before, the observables in the universe such as the temperature anisotropies of the CMB and the inhomogeneous distribution of galaxies on large scales are originated from the primordial perturbations. During inflation, the whole observable patch of the universe was once deep inside the horizon, subject to quantum fluctuations around the homogeneous background. These fluctuations exit the horizon and become classical perturbations before they disappear due to rapid expansion. Since the perturbation in the matter sector is equivalent to that in space-time, there are small perturbations in otherwise smooth three-hypersurfaces. Once inflation is over and Hubble horizon expands faster than the physical scales, these perturbations enter the horizon and matter piles up, leading eventually to gravitational collapse when the amplitude of density perturbation exceeds certain critical value. Thus cosmological perturbations produced during inflation lies at the heart of the observational tests. In this section, we first take a conventional approach to the cosmological perturbations in multi-field inflation to gain solid idea. More careful considerations on cosmological perturbations will be given in the following section.

\subsection{Solutions of constraints}

We begin with the Arnowitt-Deser-Misner (ADM) form of the metric~\cite{Arnowitt:1962hi},
\begin{equation}
ds^2=-\calN^2 dt^2+\gamma_{ij}(\beta^i dt + dx^i)(\beta^j dt + dx^j) \, ,
\end{equation}
where $\calN$ is the lapse function and $\beta^i$ is the shift vector and $\gamma_{ij}$ is the spatial metric. With the extrinsic curvature
\begin{equation}
\label{eq:extrinsiccurvature}
K_{ij} = \frac{1}{2\calN} \left( \partial_t\gamma_{ij} - \beta_{i;j} - \beta_{j;i} \right) \, ,
\end{equation}
with the covariant derivative being with respect to $\gamma_{ij}$, the action \eqref{eq:action} is rewritten as 
\begin{equation}
\label{eq:ADMaction}
S = \int d^4x \sqrt{\gamma} \calN \left[ \frac{\mpl^2}{2}  \left( R^{(3)}+K_{ij}K^{ij}-K^2 \right) - \frac{1}{2}G_{ab}g^{\mu\nu}\partial_\mu\phi^a\partial_\nu\phi^b - V(\phi) \right] \, ,
\end{equation}
where $R^{(3)}$ is the curvature of three-dimensional hypersurface of constant $t$, constructed from $\gamma_{ij}$, and $K \equiv K^i{}_i$.

Varying the action \eqref{eq:ADMaction} with respect to $\calN$ and $\beta^i$ respectively yield the constraints
\begin{align}
R{}^{(3)} - \left( K^{ij}K_{ij}-K^2 \right) & = \frac{2}{\mpl^2} \left[ \frac{1}{2}G_{ab}g^{\mu\nu}\partial_\mu\phi^a\partial_\nu\phi^b + V(\phi) \right] \, ,
\\
\left( K^{ij}-\gamma^{ij}K \right)_{;j} & = \frac{G_{ab}}{\mpl^2\calN} \left( \dot\phi^b - \beta^j\partial_j\phi^b \right) \partial^i\phi^a \, .
\end{align}
Solving these constraints give the solutions of unphysical perturbation variables. As advertised, in this section we work in the flat gauge in which the scalar sector of the spatial metric $\gamma_{ij}$ is unperturbed:
\begin{equation}
\label{eq:flatgauge}
\gamma_{ij} = a^2(t) \left( \delta_{ij} + h_{ij}^{TT} \right) \, ,
\end{equation}
where the pure tensor $h_{ij}^{TT}$ is transverse and traceless:
\begin{equation}
{h^{TT}}^i{}_i = {h^{TT}}^{ij}{}_{,j} = 0 \, .
\end{equation}
For our discussion in this section at linear order, it is sufficient to obtain the linear solutions of the constraints. Writing the perturbation parts of the variables as
\begin{align}
\calN & = 1 + \alpha \, ,
\\
\beta_i & = \chi_{,i} + \beta_i^T \, ,
\\
\phi^a & = \phi_0^a + Q^a \, ,
\end{align}
where the transverse vector $\beta_i^T$ satisfies $\partial^i\beta_i^T = 0$, we find the solutions of the constraints as
\begin{align}
\label{eq:lapse-sol}
\alpha & = \frac{G_{ab}}{2\mpl^2H}\dot\phi_0^bQ^a \equiv \calN_aQ^a \, ,
\\
\label{eq:shift-sol}
-2\mpl^2H\frac{\Delta}{a^2}\chi & = 6\mpl^2H^2\alpha - \alpha\dot\phi_0^2 + G_{ab}\dot\phi_0^aD_tQ^b + V_aQ^a \, ,
\end{align}
and $\beta_i^T = 0$. Plugging these solutions back into the action, we obtain the quadratic action for the field fluctuations $Q^a$ relevant for linear perturbation theory.

We can also arrive at the same solutions from the conventional perturbed Einstein equation. We can write the perturbed metric as
\begin{equation}
\label{eq:metric-pert}
ds^2 = -(1+2A)dt^2 + 2a\calB_idtdx^i + a^2 \left[ (1+2\varphi)\delta_{ij} + 2\calE_{ij} \right]dx^idx^j \, ,
\end{equation}
where the $0i$ and $ij$ components can be decomposed into 
\begin{align}
\label{eq:0ipert}
\calB_i & = B_{,i} + S_i \, , 
\\
\label{eq:spatialmetricpert}
\calE_{ij} & = H_{T,ij} + F_{(i,j)} + \frac{1}{2}h_{ij}^{TT} \, ,
\end{align}
with $A = \alpha$ and $B = \chi/a$. Here the pure vectors $S_i$ and $F_i$ are transverse: 
\begin{equation}
S^i{}_{,i} = F^i{}_{,i} = 0 \, .
\end{equation}
Then at linear order the scalar, vector and tensor equations are all decoupled and we can consider them independently from each other. The flat gauge condition \eqref{eq:flatgauge} corresponds to setting $\varphi = H_T = 0$, the $00$ and $0i$ components of the Einstein equation are respectively, with the energy-momentum tensor \eqref{eq:EMtensor},
\begin{align}
6H^2A + 2H\frac{\Delta}{a}B & = \frac{1}{\mpl^2} \left( -G_{ab}\dot\phi_0^aD_tQ^b + \dot\phi_0^2A - V_aQ^a \right) \, ,
\\
2HA_{,i} & = \frac{1}{\mpl^2}G_{ab}\dot\phi_0^a\partial_iQ^b \, ,
\end{align}
from which we can find the same solutions for $A$ and $B$ as \eqref{eq:lapse-sol} and \eqref{eq:shift-sol}, respectively.

\subsection{Quadratic action for perturbations}

Having found the linear solutions of the constraints, now we can use them to write the quadratic action of the field fluctuations $Q^a$ and the tensor perturbation $h_{ij}^{TT}$. After straightforward manipulations, we can obtain the quadratic action as~\cite{Achucarro:2010da,Seery:2005gb,Langlois:2008mn,Gong:2011uw,Elliston:2012ab}
\begin{equation}
\label{eq:S2-1}
S_2 = \int \! d^4x \frac{a^3}{2} \left\{ G_{ab}D_tQ^aD_tQ^b - \frac{\delta^{ij}}{a^2}G_{ab}\partial_iQ^a\partial_jQ^b - M^2_{ab}Q^aQ^b + \frac{\mpl^2}{4} \left[ \left( \dot{h}_{ij}^{TT} \right)^2 - \frac{1}{a^2}\partial^kh_{ij}^{TT}\partial_kh_{ij}^{TT} \right] \right\} \, ,
\end{equation}
where 
\begin{equation}
\label{eq:Mab}
M^2_{ab} \equiv V_{ab} - \mathbb{R}_{acdb}\dot\phi_0^c\dot\phi_0^d + (3-\epsilon)\frac{\dot\phi_{0a}}{\mpl}\frac{\dot\phi_{0b}}{\mpl} + \frac{1}{\mpl^2H} \left( \dot\phi_{0a}V_{b} + \dot\phi_{0b}V_{a} \right) \, ,
\end{equation}
with $V_{ab} \equiv V_{;ab}$, $\mathbb{R}^a_{bcd} \equiv \Gamma^a_{bd,c} - \Gamma^a_{bc,d} + \Gamma^a_{ce}\Gamma^e_{bd} - \Gamma^a_{de}\Gamma^e_{bc}$ being the Riemann tensor associated with the field space and $\epsilon$ being the slow-roll parameter defined in \eqref{eq:SRepsilon}. The field indices are raised and lowered by the field space metric $G_{ab}$. The equations of motion we can find by perturbing \eqref{eq:S2-1} with respect to $Q^a$ and $h_{ij}^{TT}$ are, respectively,
\begin{align}
\label{eq:Qeom1}
D_t^2Q^a + 3HD_tQ^a - \frac{\Delta}{a^2}Q^a + (M^2)^a{}_bQ^b & = 0 \, ,
\\
\label{eq:tensoreom}
\ddot{h}_{ij}^{TT} + 3H\dot{h}_{ij}^{TT} - \frac{\Delta}{a^2}h_{ij}^{TT} & = 0 \, .
\end{align}
We see that in general the field fluctuations are coupled to each other through the effective mass matrix $M^2_{ab}$. Also note that the tensor sector is exactly the same as that in single field inflation.

While we will solve \eqref{eq:Qeom1} and \eqref{eq:tensoreom} later in Section~\ref{sec:P}, at the moment for \eqref{eq:Qeom1} it is convenient to make use of the local orthogonal vielbeins $e^I_a$ introduced in Section~\ref{sec:BG}. Using the orthogonal basis is particularly useful to connect the field fluctuations to the curvature perturbation. The field fluctuation $Q^a$ in an arbitrary basis can be transformed into the one in the orthogonal frame by incorporating $e^I_a$ as
\begin{equation}
Q^I = e^I_aQ^a \, .
\end{equation}
Further, introducing $u^I \equiv aQ^I$ and moving to the conformal time $d\tau = dt/a$, we obtain
\begin{equation}
\label{eq:S2}
S_2 = \int \! d\tau d^3x \frac{1}{2} \left[ \delta_{IJ} \left( {u^I}'{u^J}' + 2{u^I}'Z^J{}_Ku^K + Z^I{}_KZ^J{}_Lu^Ku^L - \delta^{ij}\partial_iu^I\partial_ju^J \right) - a^2 M^2_{IJ} u^Iu^J \right] \, ,
\end{equation}
where a prime denotes a derivative with respect to the conformal time, $M^2_{IJ} \equiv e^a_Ie^b_JM^2_{ab} - H^2(2-\epsilon)\delta_{IJ}$, and $Z^I{}_J \equiv e^I_aD_\tau e^a_J$. The resulting equation of motion is
\begin{equation}
\label{eq:uIeom}
{u^I}'' + 2Z^I{}_J{u^J}' + {Z^I{}_J}'u^J - Z^I{}_JZ^J{}_Ku^K - \Delta u^I + a^2(M^2)^I{}_Ju^J = 0 \, .
\end{equation}
Thus we see that while the basis is orthogonal, the form of the equation of motion at first looks more complicated. This is because the vielbeins that map from an arbitrary basis to the orthogonal one are time-dependent, $e^I_a = e^I_a(t)$, so that the change of the vielbeins itself is reflected in the antisymmetric matrix $Z_{IJ} = -Z_{JI}$.  Note, however, that we have used in \eqref{eq:uIeom} usual partial derivatives, not covariant ones. Indeed, using $Z_{IJ}$ we can define a new covariant derivative $\calD_\tau$ acting on quantities such as $v^I$ labelled with the $I$-index as
\begin{equation}
\label{eq:newcovderiv}
\calD_\tau u^I \equiv \frac{du^I}{d\tau} + Z^I{}_Ju^J \, .
\end{equation}
Then the quadratic action \eqref{eq:S2} becomes very simple~\cite{Achucarro:2010da}:
\begin{equation}
S_2 = \int \! d\tau d^3x \frac{1}{2} \left[ \delta_{IJ} \left( \calD_\tau u^I \calD_\tau u^J - \delta^{ij}\partial_iu^I\partial_ju^J \right) - a^2M^2_{IJ}u^Iu^J \right] \, ,
\end{equation}
so is the equation of motion:
\begin{equation}
\calD_\tau^2u^I - \Delta u^I + a^2(M^2)^I{}_Ju^J = 0 \, .
\end{equation}

\subsection{Explicit calculations: two-field case}

Having discussed the general aspects, in this section we consider explicitly two-field inflation case. The benefit is three-fold: first, we will perform explicit calculations in detail so that we see how the previous discussions become materialized. Second, two-field inflation is the simplest and thus most intuitive case with multiple number of fields, yet captures many important aspects of multi-field inflation. Lastly, we can visualize the geometric implications very easily.

\subsubsection{Kinematic basis}

First we consider the kinematic basis. For two-field case, one of the vielbeins corresponds to the tangent vector $T^a$ and the other to the normal vector $N^a$,
\begin{equation}
e^a_T = T^a \quad \text{and} \quad e^a_N = N^a \, .
\end{equation}
Then the parallel and normal perturbations with respect to the inflationary trajectory are given respectively by
\begin{align}
u^T & = aQ^T = aT_aQ^a \, ,
\\
u^N & = aQ^N = aN_aQ^a \, .
\end{align}
By choosing this frame, we find that $Z_{TN} = -Z_{NT} = -a\dot\theta = -\theta'$. Then, the quadratic action \eqref{eq:S2} becomes
\begin{align}
\label{S2:kinematic2field}
S_2 = \int d\tau d^3x \frac{1}{2} & \left[ {{u^T}'}^2 - (\nabla u^T)^2 - \left( a^2M^2_{TT} - {\theta'}^2 \right){u^T}^2 \right.
\nonumber\\
& \quad + {{u^N}'}^2 - (\nabla u^N)^2 - \left( a^2M^2_{NN} - {\theta'}^2 \right) {u^N}^2 
\nonumber\\
& \left. \quad - 4\theta'{u^T}'u^N  - 2 \left( a^2M^2_{TN} + \theta'' \right) u^Tu^N \right] \, ,
\end{align}
where the symmetric matrix $M^2_{IJ}$ consists of the following elements:
\begin{align}
M^2_{TT} & = V_{TT} + 2H^2\epsilon(3-\epsilon) + \frac{2}{\mpl^2H}\dot\phi_0V_T - (2-\epsilon)H^2 \, ,
\\
M^2_{NN} & = M^2 - (2-\epsilon)H^2 \, ,
\\
M^2_{TN} & = V_{TN} - 2\epsilon\dot\theta H \, ,
\end{align}
where $M^2 \equiv V_{NN} + \epsilon\mpl^2H^2\mathbb{R}$ is the (effective) mass squared of the orthogonal mode $u^N$ with $\mathbb{R}$ being the Ricci scalar parametrizing the geometry of the field space, and the projections of the potential derivatives can be written as
\begin{align}
V_{TT} & = H^2 \left( 3\epsilon - 3\delta_1 - \delta_2 + \frac{\dot\theta^2}{H^2} \right) \, ,
\\
\label{eq:VTN}
V_{TN} & = H\dot\theta \left( -3 + 2\epsilon - 2\delta_1 + \frac{\ddot\theta}{H\dot\theta} \right) \, ,
\end{align}
where we have defined 
\begin{equation}
\label{eq:SR-delta}
\delta_n \equiv \frac{1}{H^n\dot\phi_0}\frac{d^{n+1}\phi_0}{dt^{n+1}} \, .
\end{equation}
The action \eqref{S2:kinematic2field} in the kinematic basis leads to the coupled equations of motion describing the evolution of both modes $u^T$ and $u^N$:
\begin{align}
\label{eq:uT}
{u^T}'' - 2\theta'{u^N}' + \left( -\Delta + a^2M^2_{TT} - {\theta'}^2 \right)u^T + \left( a^2M^2_{TN} - \theta'' \right) u^N & = 0 \, ,
\\
\label{eq:uN}
{u^N}'' + 2\theta'{u^T}' + \left( -\Delta + a^2M^2_{NN} - {\theta'}^2 \right)u^N + \left( a^2M^2_{NT} + \theta'' \right) u^T & = 0 \, .
\end{align}

\subsubsection{Mass basis}

\begin{figure}[h]
 \centering
  \includegraphics[width=10cm]{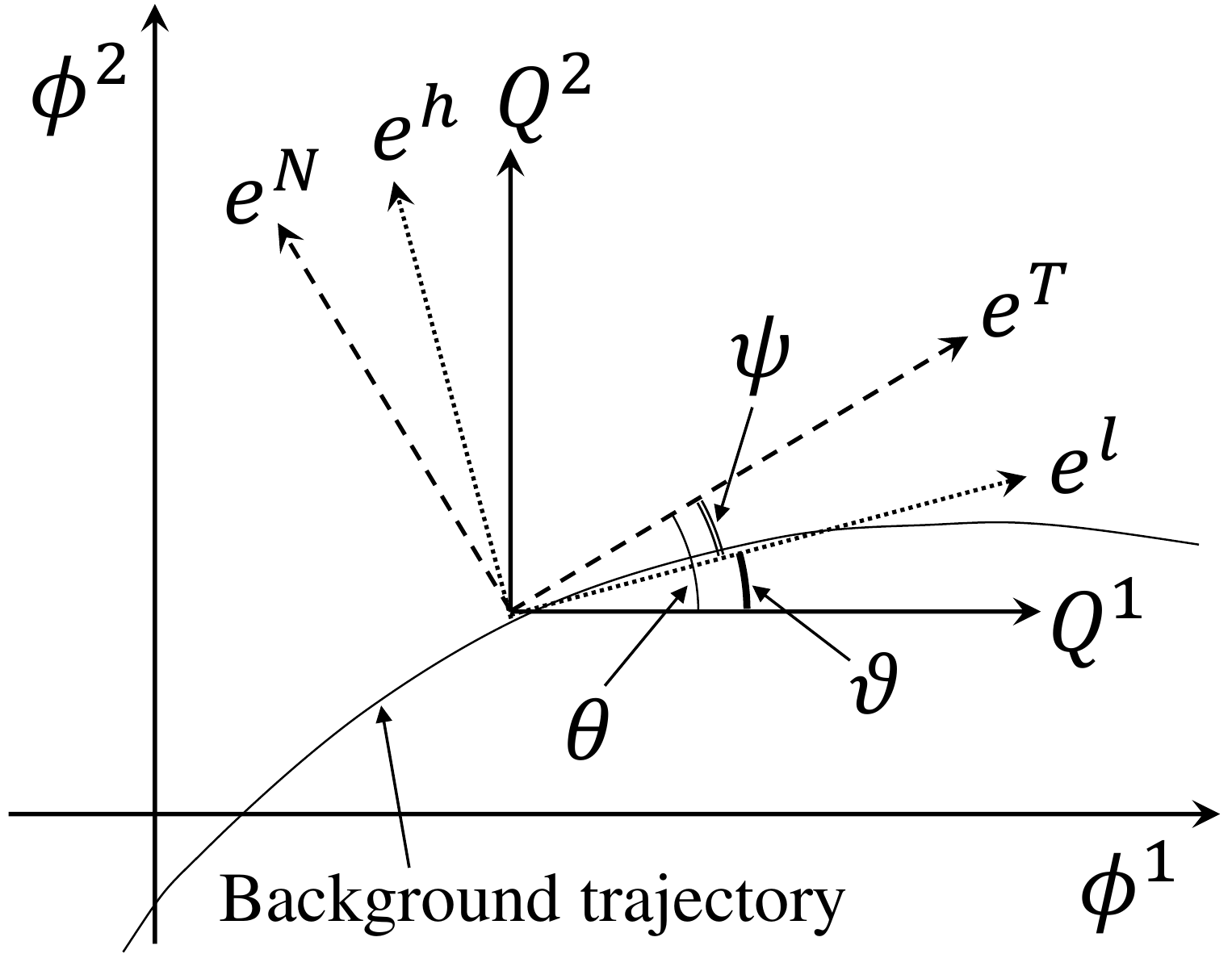}
  \caption{A schematic plot showing the relation between the arbitrary field basis ${\phi^1,\phi^2}$, kinematic basis $\{e^T,e^N\}$ and mass basis $\{e^l,e^h\}$. The field fluctuations in the field basis are decomposed into $\{Q^1,Q^2\}$. The angular velocity of the kinematic basis is $\dot\theta$ while that of the mass basis is $\dot\vartheta$, so that $\theta$, $\vartheta$ and $\psi$ are related by $\theta = \vartheta + \psi$, although for $\theta$ and $\vartheta$ only their rates of change are important since the field basis is arbitrary, with respect to which $\theta$ and $\vartheta$ are set.}
 \label{fig:basis}
\end{figure}

In the very previous section we have considered the perturbations in the kinematic basis, set along and perpendicular to the background inflationary trajectory. In general for a curved trajectory, which naturally incorporates ``heavy'' and ``light'' degrees of freedom, in the mass matrix in the kinematic basis which can be written as
\begin{equation}
\label{eq:VIJ}
V_{IJ} = 
\begin{pmatrix}
V_{TT} & V_{TN} 
\\
V_{TN} & V_{NN}
\end{pmatrix}
\, ,
\end{equation}
the off-diagonal component of the mass matrix $V_{TN}$ is non-zero, as can be read from \eqref{eq:VTN}. Thus, regarding heaviness (or lightness) of the relevant degrees of freedom, it is most convenient to adopt another set of basis in which the mass matrix $V_{IJ}$ becomes diagonal. We may call this as the ``mass basis''~\cite{massbasis}. Then the two eigenvalues of the mass basis correspond to the light and heavy masses along the trajectory respectively. Notice that if one performs the change-of-basis around the bottom of the potential~\cite{Burgess:2012dz}, no kinematic information is required because geometry determines everything. Explicit diagonalization of \eqref{eq:VIJ} gives two eigenvalues,
\begin{equation}
\label{eq:VllVhh}
\lambda_\pm = \frac{1}{2} \left[ V_{NN}+V_{TT} \pm \left( V_{NN}-V_{TT} \right) \sqrt{1+\beta^2} \right] \quad \text{where} \quad \beta \equiv \frac{2V_{TN}}{V_{NN}-V_{TT}} \, ,
\end{equation}
so that $\lambda_-$ ($\lambda_+$) corresponds to $V_{ll}$ ($V_{hh}$). The corresponding eigenvectors transformed from the kinematic basis are then 
\begin{equation}
\begin{split}
e_l^a & = T^a\cos\psi - N^a\sin\psi = 
\begin{pmatrix}
\cos\psi
\\
-\sin\psi
\end{pmatrix}
\, ,
\\
e_h^a & = T^a\sin\psi + N^a\cos\psi =
\begin{pmatrix}
\sin\psi
\\
\cos\psi
\end{pmatrix}
\, ,
\end{split}
\end{equation}
with
\begin{equation}
\cos\psi \equiv \frac{1+\sqrt{1+\beta^2}}{\sqrt{2\left(1+\beta^2+\sqrt{1+\beta^2}\right)}}
\quad \text{and} \quad
\sin\psi \equiv \frac{\beta}{\sqrt{2\left(1+\beta^2+\sqrt{1+\beta^2}\right)}} \, .
\end{equation}
So what does the angle $\psi$ mean? This becomes transparent if we write the change-of-basis matrix $P$ is, with $e_l^a$ and $e_h^a$ on the first and second column and row respectively is
\begin{equation}
\label{eq:k-to-m}
P = \begin{pmatrix}
\cos\psi & \sin\psi
\\
-\sin\psi & \cos\psi
\end{pmatrix}
\, .
\end{equation}
Thus, given the kinematic basis $\{T^a,N^a\}$, we can rotate it by $\psi$ to obtain the mass basis. This situation is depicted in Figure~\ref{fig:basis}. Notice that when we discuss about the kinematic basis, we take care of not $\theta$, which is the rotation angle from the general basis $\{\phi^1,\phi^2\}$ to the kinematic one, but only $\dot\theta$. This is because the field space coordinates $\phi^a$ can be set totally arbitrary, so important is only the rate of change of the kinematic basis with respect to the general basis, i.e. the ``angular velocity'' of the trajectory. But $\psi$ denotes the misalignment between the kinematic and mass bases, thus not only its rate of change but also its value itself are important to describe the dynamics along the trajectory: if $\psi \neq 0$ so that the two bases are misaligned, oscillations of the trajectory may be caused until the misalignment disappears.

Finally, $M^2_{IJ}$ in \eqref{eq:S2} in the mass basis is 
\begin{align}
\label{eq:coupling-Mll}
M_{ll}^2 & = V_{ll} - 2\epsilon H^2 \left[ \left( 3-\epsilon+\frac{\dot\epsilon}{H\epsilon} \right)\cos^2\psi - \frac{\dot\theta}{H}\sin(2\psi) + \mpl^2\mathbb{R}\sin^2\psi \right] - H^2(2-\epsilon) \, ,
\\
\label{eq:coupling-Mlh}
M_{lh}^2 & = 2\epsilon H^2 \left[ -\frac{1}{2}\left( 3-\epsilon+\frac{\dot\epsilon}{H\epsilon} \right)\sin(2\psi) - \frac{\dot\theta}{H}\cos(2\psi) + \frac{1}{2}\mpl^2\mathbb{R}\sin(2\psi) \right] \, ,
\\
\label{eq:coupling-Mhh}
M_{hh}^2 & = V_{hh} - 2\epsilon H^2 \left[ \left( 3-\epsilon+\frac{\dot\epsilon}{H\epsilon} \right)\sin^2\psi + \frac{\dot\theta}{H}\sin(2\psi) + \mpl^2\mathbb{R}\cos^2\psi \right] - H^2(2-\epsilon) \, .
\end{align}
Then the quadratic action \eqref{eq:S2} becomes explicitly~\cite{Gao:2015aba}
\begin{align}
\label{eq:S2-2}
S_2 & = \int \! d\tau d^3x \frac{1}{2} \left[ {u_l'}^2 - (\nabla u_l)^2 - \left( a^2M_{ll}^2-{\vartheta'}^2 \right)u_l^2 \right.
\nonumber\\
& \qquad\qquad\qquad + {u_h'}^2 - (\nabla u_h)^2 - \left( a^2M_{hh}^2-{\vartheta'}^2 \right)u_h^2
\nonumber\\
& \qquad\qquad\qquad - 4\vartheta'u_l'u_h - 2 \left( a^2M_{lh}^2 + \vartheta'' \right) u_lu_h \bigg] \, ,
\end{align}
where $\vartheta' = (\theta-\psi)'$ denotes the angular velocity of the mass basis. The first and second lines denote the free terms of the light and heavy modes respectively, while the third line is the interaction between them. Note that the form of the action \eqref{eq:S2-2} is precisely the same as that in the kinematic basis \eqref{S2:kinematic2field}. This should be obviously the case, since both the kinematic and mass bases are orthogonal ones.

\section{Formulation of perturbations}
\label{sec:pert-general}
\setcounter{equation}{0}

In the previous section, we have considered the perturbations in the flat gauge to extract physical scalar degrees of freedom. Let us recall the general arguments: on top of the matter sector which contains $n$ scalar fields, the gravitational sector brings additional 4 scalar, 4 vector and 2 tensor degrees of freedom. However, not all of them are physical, as we have the freedom to choose an arbitrary coordinate system with the same physics. That is, gravity allows gauge degrees of freedom. These fictitious gauge degrees of freedom can be removed by imposing appropriate gauge conditions and by solving the constraints -- naively we have $n+4$ scalar variables: $n$ from $n$ scalar field components and 4 from the metric. Since there are 1 temporal and 1 spatial gauge transformations in the scalar sector, we can eliminate 2 of them. In the flat gauge discussed in the previous section, we impose the conditions that the perturbations of three-dimensional spatial metric on each time slice vanish. The remaining metric degrees of freedom are perturbations of the lapse function and the shift vector. They further can be removed by solving 2 constraint equations, so that after all $n$ degrees of freedom are left. Namely, we can write all the physical degrees of freedom solely in terms of the field fluctuations $Q^a$.

This however is not necessarily the only sensible choice. Especially, we do not directly observe the inflaton fields -- the temperature fluctuations in the CMB and the distribution of galaxies on large scales reflect the initial perturbation in the curvature of the constant-time spatial hypersurfaces. This curvature perturbation becomes manifest in other gauge choices, not in the flat gauge we adopted in the previous section in which by definition there is no perturbation on spatial hypersurfaces. But the curvature perturbation is associated with the metric, so in this case we have to give 1 degree of freedom to the gravitational sector while the remaining $n-1$ to the matter sector that contains $n$ fields. But how can we do this conveniently, since the gravitational sector receives contributions from the total matter contents? Thus, we need to reconsider the formulation of perturbations more formally to establish possible gauge choices on concrete ground. We can discuss formally the issue of gauge fixing in the Hamiltonian formulation, which is also useful for path integral quantization of cosmological perturbations~\cite{pathintegral}. In this section, we consider the Hamiltonian analysis of cosmological perturbations to study the gauge conditions systematically.

\subsection{First-order form}

For the Hamiltonian formulation, we begin with the canonical momentum of $\gamma_{ij}$ as
\begin{equation}
\Pi^{ij} = \frac{\delta S_G}{\delta \partial_t \gamma_{ij}}=\frac{\mpl^2}{2}\sqrt{\gamma}(K^{ij}-\gamma^{ij}K) \, .
\end{equation}
The pure gravity part of the action then is written in the first-order form,
\begin{equation}
S_G = \int d^4x \left( \Pi^{ij}\partial_t{\gamma}_{ij} - \calN\calH_G - \beta_i \calH_G^i \right) \, ,
\end{equation}
where 
\begin{align}
\calH_G & = \frac{2}{\mpl^2\sqrt{\gamma}} \left( \Pi^{ij} \Pi_{ij} - \frac12 \Pi^2 \right) - \frac{\mpl^2}{2}\sqrt{\gamma} R^{(3)},
\\
\calH_G^i & = -2 \left( \partial_j\Pi^{ij} + \Gamma^i_{jk}\Pi^{jk} \right) \, ,
\end{align}
with $\Pi \equiv \Pi^i{}_i$. Here, the lapse $\calN$ and the shifts $\beta_i$ play the role of Lagrangian multipliers for the constraints $\calH$ and $\calH^i$ respectively, which are interpreted as the generators of diffeomorphism. Likewise, we can write the matter part of the action in the first-order form as
\begin{equation}
S_M = \int d^4x \left( \Pi_a\partial_t{\phi}^a - \calN\calH_M - \beta_i\calH_M^i \right) \, ,
\end{equation}
where the canonical momentum $\Pi_a$ and the constraints $\calH_M$ and $\calH_M^i$ are
\begin{align}
\Pi_a & = \frac{\sqrt{\gamma}}{\calN} G_{ab} \left( \partial_t\phi^b - \beta^i\partial_i\phi^b \right) \, ,
\\
{\cal H}_M & = \frac{G^{ab}}{2\sqrt{\gamma}}\Pi_a\Pi_b+\frac{1}{2}\sqrt{\gamma}\gamma^{ij}G_{ab}\partial_i\phi^a \partial_j \phi^b +\sqrt{\gamma} V(\phi^a) \, ,
 \\
{\cal H}_M^i & = \Pi_a \partial^i \phi^a \, .
\end{align}
Thus, the action in the first-order form is given by
\begin{equation}
S = \int d^4x \left[ \Pi^{ij}\partial_t{\gamma}_{ij} + \Pi_a\partial_t{\phi}^a - \calN\left(\calH_G+\calH_M \right) - \beta_i \left( \calH_G^i + \calH_M^i \right) \right] \, .
\end{equation}
It is clearly seen that unphysical $\calN$ and $\beta_i$ become Lagrange multipliers accompanying constraints $\calH_G+\calH_M$ and $\calH_G^i+\calH_M^i$, respectively.

We are interested in the cosmological perturbations around the time-dependent classical background. For gravity sector, we separate the FRW background quantities and their perturbations as:
\begin{align}
\Pi^{ij} & = \frac{P(t)}{6a(t)} \left[ \delta^{ij}+\pi^{ij}(t,\mathbi{x}) \right] \, ,
\\
\gamma_{ij} & = a^2(t) \left[ \delta_{ij}+h_{ij}(t,\mathbi{x}) \right] \, ,
\\
\calN & = \calN_0(t)+\alpha(t,\mathbi{x}) \, ,
\\
\beta^i & = \beta^i(t,\mathbi{x}) \, .
\end{align}
Note that at the moment we have kept the background lapse function denoted by $\calN_0(t)$, not simply setting it to 1. This is because we can obtain a background equation by varying the zeroth order action with respect to $\calN_0$ [see \eqref{eq:S0} and \eqref{eq:BGNbar}]. For the matter sector, the perturbations around the classical backgrounds $\phi^a_0(t)$ are written as
\begin{align}
\phi^a & = \phi_0^a(t) + Q^a(t,\mathbi{x}) \, ,
\\
\label{eq:conjmom-phi}
\Pi_a & = P_a(t) + \pi_a(t,\mathbi{x}) \, .
\end{align}
We can obtain the classical solutions from the zeroth order action,
\begin{equation}
\label{eq:S0}
S_0 = \int d^4x \left[ P\partial_t a + P_a\partial_t\phi_0^a - \calN_0 \left( -\frac{P^2}{12\mpl^2 a}+\frac{G^{ab}P_aP_b}{2a^3}+a^3V \right) \right] \, .
\end{equation}
Varying this with respect to the classical backgrounds, we obtain the background equations of motion as:
\begin{alignat}{4}
\frac{\delta S_0}{\delta P} = 0 & : & \qquad \dot{a} & = -\frac{P}{6\mpl^2a} \, ,
\\
\frac{\delta S_0}{\delta a} = 0 & : & \qquad \dot{P} & = -\frac{P^2}{12\mpl^2a^2} + \frac{3G^{ab}P_aP_b}{2a^4} - 3a^2V \, ,
\\
\label{eq:BGmomphi-sol}
\frac{\delta S_0}{\delta P_a} = 0 & : & \qquad \dot\phi_0^a & = \frac{G^{ab}P_b}{a^3} \, ,
\\
\frac{\delta S_0}{\delta\phi_0^a} = 0 & : & \qquad \dot{P}_a & = -a^3V_a - \frac{1}{2a^3} G^{cd}{}_{,a} P_cP_d \, ,
\\
\label{eq:BGNbar}
\frac{\delta S_0}{\delta\calN_0} = 0 & : & \qquad \frac{P^2}{12\mpl^2a} & = \frac{G^{ab}P_aP_b}{2a^3} + a^3V \, ,
\end{alignat}  
Combining these equations, we obtain the familiar background equations \eqref{eq:Friedmann}, \eqref{eq:Hdot} and \eqref{eq:BGphi}. Note that from the linear order action 
\begin{align}
S_1 = \int d^4x \calN_0 \bigg[ & \frac{a}{6} h \left( -\dot{P}+HP+\frac{P^2}{12 \mpl^2 a^2}+\frac{3G^{ab}P_aP_b}{2a^4}-3a^2 V \right)
+ \frac{P}{3}\pi \left( \dot{a}+\frac{P^2}{6\mpl^2 a} \right)
\nonumber\\
& + Q^a \left( \dot{P}_a+\frac{1}{2a^3}G^{cd}{}_{,a}P_cP_d+a^3V_a \right) + \pi_a \left( \dot{\phi}_0^a-\frac{G^{ab}P_b}{a^3} \right)
\nonumber\\
& + \frac{\alpha}{\calN_0} \left( \frac{P^2}{12\mpl^2a} - \frac{G^{ab}P_aP_b}{2a^3} - a^3V \right) \bigg] \, ,
\end{align}
where $h \equiv h^i{}_i$ and $\pi \equiv \pi^i{}_i$, we can immediately read the same background equations, which we obtained by perturbing the zeroth order action with respect to the background variables, as the constraints for the perturbation variables.

\subsection{Gauge fixing conditions}

Having found the first-order form of the action, now we can proceed formal discussion on the possible choice of gauge fixing. Before we begin explicitly with the quadratic action for cosmological perturbations, let us recall how to remove properly the unphysical degrees of freedom for a constrained system, viz. when constraints are present~\cite{const-H}. Let us consider a Hamiltonian with $r$ constraints in a $2f$-dimensional phase space $(q_1,\cdots,q_f; p_1,\cdots,p_f)$,
\begin{equation}
H_\text{total}(q_i,p_i) = H(q_1,\cdots,q_f; p_1,\cdots,p_f) + \sum_{m=1}^r \lambda_m\chi_m(q_1,\cdots,q_f; p_1,\cdots,p_f) \, ,
\end{equation}
where the Lagrangian multipliers $\lambda_m$ are combinations of unphysical degrees of freedom. Since there are $r$ constraints $\chi_m$, the number of degrees of freedom in the phase space is reduced to $2f-r$ by solving $r$ constraints. But the phase space should be even-dimensional, so total remaining degrees of freedom should be not $2f-r$ but $2f-2r$. That is, still $r$ unphysical degrees of freedom remain. This is because different values of $\lambda_m$ correspond to different copies of physical degrees of freedom in the phase space on $\chi_m = 0$. Thus we need $r$ more, {\em gauge-fixing} conditions $\psi_m(q_1,\cdots,q_f; p_1,\cdots,p_f) = 0$. In quantum field theory, these gauge fixing conditions are subject to $\text{det}([\chi_m,\psi_n])\neq0$ so that just one set of physical degrees of freedom is chosen. By choosing gauge fixing conditions satisfying $[\psi_m,\psi_n]=0$, we can identify $\psi_m$ as unphysical canonical momenta $p_m$ then the remaining $f-r$ momenta $(p_1^*,\cdots,p_{f-r}^*)$ are physical. Then $\text{det}([\chi_m,\psi_n]) = \text{det}([\chi_m,p_n]) = \text{det}(i\partial\chi_m/\partial{q}_n) \neq 0$ so that it is always possible to identify unphysical variables $q_m$ by inverting $\chi_m(q_1,\cdots,q_f;p_1^*,\cdots,p_{f-r}^*;p_1=\cdots=p_r=0)=0$. Thus, we are finally left with $2f-2r$ physical, constrained variables $(q_1^*,\cdots,q_{f-r}^*;p_1^*,\cdots,p_{f-r}^*)$.

Now we return to our discussion on the cosmological perturbations around the classical solutions. At quadratic order, the first-order form of the action is written as
\begin{equation}
\label{eq:Sfree}
S_2 = \int d^4x  \left( \pi_a\partial_tQ^a - 2a^3 H \pi^{ij}\partial_t h_{ij} - \calH_2 + \alpha C^0_1 + \beta_i C^i_1 \right) \, ,
\end{equation}
where $\calH_2$ is the quadratic Hamiltonian, and $C_1^\mu$ denote the constraints linear in perturbations. They are given by
\begin{align}
\calH_2 & = 4a^3H^2\frac{\mpl^2}{2} \left[ \frac12 \pi^{ij}A_{ijkl}\pi^{kl} + \pi^{ij} \left( 2h_{ij} - \frac12\delta_{ij}h \right) \right] + \frac{G^{ab}}{2a^3} \left( \pi_a\pi_b - h P_a\pi_b \right)
\nonumber\\
& \quad + \frac{a}{2}G_{ab}\partial_iQ^a\partial_iQ^b + \frac{a^3}{2} \left( V_{ab}Q^aQ^b + h V_aQ^a \right)
\nonumber\\
& \quad + \left( \frac{5\mpl^2}{4} a^3H^2 + \frac{G^{ab}}{8a^3}P_aP_b \right) h^{ij}h_{ij} - \left( \frac{3\mpl^2}{8} a^3 H^2 - \frac{G^{ab}}{16 a^3} P_aP_b \right) h^2
\nonumber\\
& \quad + a\frac{\mpl^2}{2} \left( \frac14 h\Delta h - \frac12h h^{ij}{}_{,ij} + \frac12 h^{ij}\partial^k\partial_ih_{jk} - \frac14 h^{ij}\Delta h_{ij} \right) + a^3\frac{V}{4} \left( \frac{h^2}{2} - h^{ij}h_{ij} \right) \, ,
\\
C^{0}_1 & = a^3h \left( \frac{G^{ab}P_aP_b}{4a^6} + \frac{\mpl^2}{2}H^2 \right) + a\frac{\mpl^2}{2} \left( h^{ij}{}_{,ij} - \Delta h \right) - a^3 V_aQ^a - \frac{G^{ab}}{a^3}P_a\pi_b + 2\mpl^2a^3H^2\pi - \frac{a^3}{2}hV \, ,
\label{Eq:C0(1)}
\\
C^i_1 & = -\frac{1}{a^2}P_a \partial^i Q^a - 2aH\mpl^2 \left( \partial_j\pi^{ij} + \partial_j h^{ij} - \frac12 \partial^i h \right) \, ,
\end{align}
with $A_{ijkl} \equiv \delta_{ik}\delta_{jl} + \delta_{il}\delta_{jk} - \delta_{ij}\delta_{kl}$. Notice that going beyond cubic order, which is necessary for non-linear perturbation theory, we can schematically write
\begin{equation}
S_{\geq3} = \int d^4x \left( -\calH_{\geq3} + \alpha C^0_{\geq 2} + \beta_i C^i_{\geq 2} \right) \, ,
\end{equation}
where the subscript $\geq n$ denotes $n$-th order and beyond in perturbations.

Since gravity has four constraints, we need four gauge fixing conditions $\psi_\mu$ ($\mu=0,1,2,3$) satisfying
\begin{equation}
{\rm det}\left( \left\{C^\mu,\psi_\nu\right\} \right)\ne 0 \, ,
\label{Eq:FPcondition}
\end{equation}
where curly brackets denote the Poisson brackets. To extract the possible gauge fixing conditions, we need the Poisson brackets of the constraints $C^\mu$ with the fluctuations $h_{ij}$ and $Q^a$. We can explicitly find
\begin{align}
\left\{ h_{ij}(t, \mathbi{x}), C^0(t, \mathbi{y}) \right\} & = -2H{\tilde\gamma}^{-1/2} \Big\{ \delta_{ij} + \left( -h_{ij}+\delta_{ij}h-2\pi^{ij}+\delta_{ij}\pi \right)
\nonumber\\
& \qquad\qquad\qquad + \left[ (h+\pi)h_{ij}-2h_{ik}h_{jk}-2(h_{il}\pi^{jl}+h_{jl}\pi^{il})+\pi^{kl}h_{kl}\delta_{ij} \right]
\nonumber\\
& \qquad\qquad\qquad + (-2 h_{il}\pi^{lk}h_{jk}+h_{kl}\pi^{kl}h_{ij}) \Big\} \delta^{(3)}(\mathbi{x}-\mathbi{y}) \, ,
\\
\left\{ h_{ij}(t, \mathbi{x}), C^k(t, \mathbi{y}) \right\} & = -\frac{2}{a^2} \left( \delta_{k(i}\partial^x_{j)}-\Gamma^k_{ij} \right) \delta^{(3)}(\mathbi{x}-\mathbi{y}) \, ,
\\
\left\{ Q^a(t, \mathbi{x}), C^0(t, \mathbi{y}) \right\} & = -\frac{\tilde{\gamma}^{-1/2}}{a^3}(P_a+\pi_a) \delta^{(3)}(\mathbi{x}-\mathbi{y}) \, ,
\\
\left\{ Q^a(t, \mathbi{x}), C^i(t, \mathbi{y}) \right\} & = -\frac{1}{a^3}\tilde{\gamma}^{ik}\partial_kQ^a \delta^{(3)}(\mathbi{x}-\mathbi{y}) \, ,
\end{align}
where $\tilde\gamma_{ij} = \gamma_{ij}/a^2$. Since the participating degrees of freedom are the perturbed metric $h_{ij}$ and the field fluctuations $Q^a$, we can think of four possibilities for the gauge fixing conditions:
\begin{itemize}
 \item[1.] Both the $\psi^0$ and the $\psi^i$ conditions come from $h_{ij}$.
 \item[2.] Both the $\psi^0$ and the $\psi^i$ conditions come from $Q^a$.
 \item[3.] The $\psi^0$ condition comes from $h_{ij}$, whereas the $\psi^i$ condition comes from $Q^a$.
 \item[4.] The $\psi^0$ condition comes from $Q^a$, whereas the $\psi^i$ condition comes from $h_{ij}$.  
\end{itemize}
However, the 3-vector condition made up of $Q^a$, say, $\psi_i=\psi_i(Q^a,\partial_k Q^a)$ satisfies
\begin{align}
\left\{ \psi_j(Q)(t, \mathbi{x}), C^i(t, \mathbi{y}) \right\} & = -\frac{\tilde{\gamma}^{ik}}{a^2}  \partial_k Q^a (t, \mathbi{y}) \left[ \frac{\partial\psi_j}{\partial Q^a}(t, \mathbi{x}) + \frac{\partial \psi_j}{\partial (\partial_lQ^a)}(t, \mathbi{x})\partial^x_l \right] \delta^{(3)}(\mathbi{x}-\mathbi{y}) 
\nonumber\\
&= -\frac{\tilde\gamma^{ik}}{a^2} D_k\psi_j \delta^{(3)}(\mathbi{x}-\mathbi{y}) \, .
\end{align}
Thus it vanishes under $\psi_i=0$, and \eqref{Eq:FPcondition} is violated. That is, the spatial vector constraints $C^i$ cannot be satisfied with the field fluctuations $Q^a$. Hence, plausible gauge fixing conditions should be chosen between the possibilities 1 and 4.

For the first possibility, the gauge fixing is imposed entirely from the metric fluctuation. One simple example is
\begin{equation}
\label{eq:flatgauge}
\begin{split}
\psi_0 & = 0 \, ,
\\
\psi_i & = \partial^j \left( h_{ij}-\frac{\delta_{ij}}{3}h \right) \, .
\end{split}
\end{equation}
In terms of the metric decomposition in \eqref{eq:metric-pert} and \eqref{eq:spatialmetricpert} the gauge fixing by $\psi_\mu=0$ in this choice is to set the scalar components of $h_{ij}$ zero, so that there is no scalar perturbation in the spatial metric as in \eqref{eq:flatgauge}. Thus in this gauge condition, the curvature of the three-hypersurface is uniform so it is called uniform-curvature gauge. Since we are considering spatially flat universe, we may call it ``flat'' gauge condition. In the flat gauge, the scalar degrees of freedom are entirely given to $Q^a$. Meanwhile, as an example for the fourth possibility, we can make use of the conditions
\begin{equation}
\label{eq:comgauge}
\begin{split}
\psi_0 & = G_{ab}\dot{\phi}_0^a Q^b \, ,
\\
\psi_i & = \partial^j \left( h_{ij}-\frac{\delta_{ij}}{3}h \right) \, .
\end{split}
\end{equation}
We can note from \eqref{eq:EMtensor} that the condition $\psi_0=0$ is equivalent to $T^0{}_i=0$. In terms of hydrodynamic fluid, the 3-velocity of the fluid vanishes in this gauge, $u_i = 0$. Thus in this gauge condition we do not observe momentum flux, and we move together with the cosmic fluid. So this gauge condition is known and the ``comoving'' gauge condition, and one scalar degree of freedom is associated with the metric -- the curvature perturbation. But notice that all field contents contribute to $\psi_0$. A more convenient way of implementing $\psi_0=0$ will be discussed in the following sections.

\subsection{Quadratic action in the comoving gauge}

Having discussed the possible gauge conditions, now we consider the quadratic action \eqref{eq:Sfree} for which we have not yet applied any gauge yet. For this, we decompose the metric perturbation $h_{ij}$ using the scalar, vector and tensor components as we already did in \eqref{eq:metric-pert} and \eqref{eq:spatialmetricpert}\footnote{We may write the scalar components as $2H_L\delta_{ij} + 2 ( \partial_i\partial_j - \Delta\delta_{ij}/3 )H_T$ so that $H_L$ is the only longitudinal contribution~\cite{Bardeen:1980kt}. Note that in this decomposition $\varphi = H_T - \Delta{H_T}/3$.}:
\begin{equation}
\label{Eq:metricdec}
h_{ij} = 2\varphi\delta_{ij} + 2H_{T,ij} + 2\partial_{(i}F_{j)} + h_{ij}^{TT} \, .
\end{equation}
The quadratic action \eqref{eq:Sfree} looks horribly complex, so we first need to arrange terms for physical clarity. For this, we introduce the Mukhanov-Sasaki variable~\cite{MSvariable} \`a la single field inflation as
\begin{equation}
\label{Eq:multiredef}
\tilde{Q}^a \equiv Q^a - \frac{\dot{\phi}_0^a}{H} \varphi \, .
\end{equation}
Then, up to irrelevant auxiliary field terms which can be eliminated from dynamics after appropriate redefinitions, we obtain a surprisingly simple form:
\begin{equation}
\label{Eq:quadratic}
S_2 = \int d^4x \frac{a^3}{2} \left\{ G_{ab}D_t\tilde{Q}^aD_t\tilde{Q}^b - \frac{G_{ab}}{a^2}\partial^i \tilde{Q}^a\partial_i \tilde{Q}^b - M_{ab}^2 \tilde{Q}^a\tilde{Q}^b + \frac{\mpl^2}{4} \left[ \left( \dot{h}_{ij}^{TT} \right)^2 - \frac{1}{a^2} \partial^k{h}_{ij}^{TT} \partial_k{h}_{ij}^{TT} \right] \right\} \, ,
\end{equation}
where $M_{ab}^2$ is given by \eqref{eq:Mab}.

Since the pure tensor action in \eqref{Eq:quadratic} is the same as that in single field inflation, we from now on concentrate on the scalar sector. As mentioned at the beginning of this section, it is very conventional to adopt the flat gauge for multi-field inflation in which $\varphi = H_T = 0$ so that simply $\tilde{Q}^a = Q^a$. This is because the number of the physical degrees of freedom, after eliminating the unphysical ones, is the same as the number of field contents. But in the comoving gauge, where one physical degree of freedom is the curvature perturbation $\varphi$, we need to fix a non-trivial temporal gauge condition $C^0$ for which all the field fluctuations contribute as can be read from $\psi_0$ in \eqref{eq:comgauge}. Thus, \eqref{Eq:multiredef} is, while it gives a very simple form of the quadratic action \eqref{Eq:quadratic}, not most convenient to implement the comoving gauge condition. A more convenient alternative is to decompose $Q^a$ into the directions along and orthogonal to time evolution~\cite{Achucarro:2012sm} as 
\begin{equation}
\label{eq:decomposition2}
Q^a(t,\mathbi{x}) = Q^a_\bot(t,\mathbi{x}) + \dot\phi^a_0(t)\tilde\pi(t,\mathbi{x}) \, ,
\end{equation}
with the orthogonality condition 
\begin{equation}
\label{eq:orthogonality}
G_{ab}\dot\phi^a_0Q_\bot^b = 0 \, .
\end{equation}
We can then rewrite the temporal gauge condition of \eqref{eq:comgauge} very simply as as
\begin{equation}
\psi_0 = \dot\phi_0^2\tilde\pi \, .
\end{equation}
Thus, we can impose the the comoving gauge conveniently by $\tilde\pi=0$. Note that the linear gauge transformation $Q^a \to Q^a - \dot\phi^a_0\xi^0$ tells us
\begin{equation}
\tilde\pi \to \tilde\pi - \xi^0 \quad \text{and} \quad Q^a_\bot \to Q^a_\bot \, .
\end{equation}
This means that $\tilde\pi$ is the fluctuation in the direction of the time translation itself, and is thus interpreted as the Goldstone mode resulting from the spontaneous breaking of the time translation invariance~\cite{Cheung:2007st}. Meanwhile, the orthogonal fluctuations $Q^a_\bot$, which are usually called ``isocurvature'' modes, are gauge invariant\footnote{Note that in models with non-minimal couplings to gravity~\cite{nm-multi-models}, for single field case the curvature perturbation remains invariant~\cite{single-conformalinv} under the conformal transformation by which the gravitational sector becomes the minimal Einstein-Hilbert one~\cite{conformal}, but for multi-field this is not the case~\cite{multi-conformalinv}.}. Then, we can write the multi-field version of the Mukhanov-Sasaki variable \eqref{Eq:multiredef} as
\begin{equation}
\label{eq:MS-comoving}
\tilde{Q}^a(t,\mathbi{x}) = Q^a_\bot(t,\mathbi{x}) - \frac{\dot\phi^a_0}{H} \left( \varphi - H\tilde\pi \right)(t,\mathbi{x}) \equiv Q^a_\bot(t,\mathbi{x}) - \frac{\dot\phi^a_0}{H} \pi(t,\mathbi{x}) \, .
\end{equation}
Note that due to the orthogonality condition, $n-1$ out of total $n$ $Q_\bot^a$'s are independent in the comoving gauge: the remaining single degree of freedom is $\pi$, which is also gauge invariant. With this decomposition, the scalar quadratic action is rewritten in terms of gauge invariant variables $\pi$ and $Q^a_\bot$ as, using the time derivative of \eqref{eq:orthogonality} to eliminate $D_tQ^a$ in the mixing term with $\pi$, 
\begin{align}
\label{eq:S2scalar}
S_2 = \int d^4x \frac{a^3}{2} \bigg[ & G_{ab}D_t{Q}_\bot^aD_t{Q}_\bot^b - \frac{G_{ab}}{a^2}\partial^iQ_\bot^a\partial_iQ_\bot^b - M^2_{ab}Q_\bot^aQ_\bot^b 
\nonumber\\
& \left. + 2\epsilon\mpl^2 \left( \dot\pi^2 - \frac{1}{a^2}\partial^i\pi\partial_i\pi \right) - \frac{4}{H}V_aQ^a_\bot\dot\pi \right] \, .
\end{align}

We close this section by writing the curvature perturbation in a form convenient for computing higher order correlation functions that we will discuss later. We can non-linearly generalize the metric perturbation \eqref{Eq:metricdec} by exponentiating it as~\cite{Lyth:2004gb}
\begin{equation}
\label{eq:R}
\gamma_{ij} = a^2(t)e^{2\varphi(t,\mathbi{x})} \left[ e^{h^{TT}} \right]_{ij} \, ,
\end{equation}
where $\text{det}\left[e^{h^{TT}}\right] = 1$, i.e. the matrix $h^{TT}$ is traceless and represents tensor perturbations, so that the scalar perturbation is isolated as $\varphi$. Notice that this form also means the scalar perturbation can be interpreted as the ``local expansion rate'' which can be written as the determinant of $\gamma_{ij}$: see \eqref{eq:localH2}. In the comoving gauge in which $\varphi = \calR$, the transformation from $\pi$ to $\calR$ is given by~\cite{Maldacena:2002vr,Noh:2004bc}
\begin{align}
\label{eq:pi->R}
\calR & = \pi + \left( \epsilon-\frac{\delta}{2} \right)\pi^2 + \frac{1}{H}\pi \dot{\pi} - \frac{1}{4a^2H^2} \left[ \partial^i\pi\partial_i\pi - \partial^i\partial^j \Delta^{-1} \left( \partial_i\pi\partial_j\pi \right) \right]
\nonumber\\
& \quad + \frac{\epsilon}{H} \left[ \partial^i\pi \partial_i\Delta^{-1}\dot{\pi} - \partial^i\partial^j\Delta^{-1} \left( \partial_i\pi \partial_j\Delta^{-1} \dot{\pi} \right) \right] - \frac{1}{4H}\dot{h}_{ij}^{TT}\partial^i\partial^j \pi + \cdots \, .
\end{align}
In terms of $\calR$, the quadratic action remains the same with $\pi$ replaced by $\calR$. But the transformation \eqref{eq:pi->R} does give rise to additional contributions to the higher order action in terms of $\pi$ as we will see in Section~\ref{sec:non-linear}.

\subsection{Two-field case}

Now we consider two-field case explicitly. Due to the orthogonality condition, we know that $Q_\bot^a$ is proportional to the normal vector $N^a$: $Q_\bot^a \propto N^a$. We denote the amplitude of $Q_\bot$ as $\calF$, and use the (linear) relation \eqref{eq:pi->R} to replace $\pi$ simply with $\calR$. That is, the gauge-invariant variable $\tilde{Q}^a$ is written in terms of $\calR$ and $\calF$ as 
\begin{equation}
\tilde{Q}^a = N^a\calF - \frac{\dot\phi_0^a}{H}\calR \, .
\end{equation}
Then the quadratic action becomes
\begin{equation}
\label{eq:S2-R&F}
S_2 = \int d^4x \frac{a^3}{2} \left\{ 2\epsilon\mpl^2 \left[ \dot\calR^2 - \frac{(\nabla\calR)^2}{a^2} \right] + \dot\calF^2 - \frac{(\nabla\calF)^2}{a^2} - M_\text{eff}^2 \calF^2 + 4\dot\theta\frac{\dot\phi_0}{H}\dot\calR\calF \right\} \, .
\end{equation}
where $M_\text{eff}^2 = V_{NN} + \epsilon\mpl^2H^2\mathbb{R} - \dot\theta^2$. Identifying $u^T = -\dot\phi_0\calR/H$ and $u^N = \calF$, after some manipulations we can recover \eqref{S2:kinematic2field}.

\subsubsection{Effective single field theory}

Note that if a hierarchy of scales is present in the mass matrix, then we can compute a fairly reliable effective single field theory: see~\cite{Chluba:2015bqa} for a concrete review on this subject. If we do not have any light mode other than $\calR$, we can integrate out the heavy isocurvature modes $\calF$ by performing formally the path integral over $\calF$. Writing \eqref{eq:S2-R&F} schematically as
\begin{equation}
\label{eq:S2schematic}
S_2 = \frac{1}{2} \int \calR\Delta_{\calR\calR}\calR + \frac{1}{2} \int \calF\Delta_{\calF\calF}\calF + \int \calF\Delta_{\calR\calF}\calR \, ,
\end{equation}
we can evaluate the Gaussian integral over the heavy field $\calF$ as
\begin{align}
e^{-S_\text{eff}[\calR]} & = \exp \left( \frac{1}{2} \int \calR\Delta_{\calR\calR}\calR \right) \int \calD\calF \exp \left( \frac{1}{2} \int \calF\Delta_{\calF\calF}\calF + \int \calF\Delta_{\calR\calF}\calR \right)
\nonumber\\
& = \exp \left\{ \frac{1}{2} \int \calR\Delta_{\calR\calR}\calR - \frac{1}{2} \int \Big[ \left( \Delta_{\calR\calF}\calR \right) \Delta_{\calF\calF}^{-1} \left( \Delta_{\calR\calF}\calR \right) \Big] \right\} \text{det}\left[\Delta_{\calF\calF}\right]^{-1/2} \, .
\end{align}
Then the effective action for $\calR$ reads
\begin{equation}
S_\text{eff}[\calR] = \frac{1}{2} \int \calR\Delta_{\calR\calR}\calR - \frac{1}{2} \int d^4x d^4y \Delta_{\calR\calF}\calR(x) \Delta_{\calF\calF}^{-1}(x,y) \Delta_{\calR\calF}\calR(y) + S_\text{counter} \, ,
\end{equation}
where $S_\text{counter}$ denotes the contributions from the functional determinant and does not depend on $\calR$. Note that instead of formally performing the Gaussian integral, we could solve for $\calF$ via its equations of motion
\begin{equation}
\label{eq:solF}
\Delta_{\calF\calF}\calF = -\Delta_{\calR\calF}\calR \, ,
\end{equation}
which has the formal solution $\calF = -\Delta_{\calF\calF}^{-1}\Delta_{\calR\calF}\calR$. Substituting this solution into \eqref{eq:S2schematic} gives
\begin{equation}
S_\text{eff}[\calR] = \frac{1}{2} \int \calR\Delta_{\calR\calR}\calR - \frac{1}{2} \int \Delta_{\calR\calF}\calR \Delta_{\calF\calF}^{-1} \Delta_{\calR\calF}\calR \, ,
\end{equation}
with the only difference from the formal result being the absence of $S_\text{counter}$. Thus at quadratic order, we can simply substitute the solution for $\calF$ back into the original action \eqref{eq:S2-R&F} to obtain the desired effective action for $\calR$ solely~\cite{Achucarro:2012sm}.

The effective action obtained in this manner is intrinsically non-local, as the operator $\Delta_{\calF\calF} = -\Box + M_\text{eff}^2$ contains derivative operators. However, given that the adiabaticity condition~\cite{adia-condition}
\begin{equation}
\left| \frac{\ddot\theta}{\dot\theta} \right| \ll M_\text{eff}
\end{equation}
is satisfied, we can expand the solution of \eqref{eq:solF} as a power series of $M_\text{eff}^2$ as
\begin{equation}
\label{eq:F-expansion}
\calF = \left( -\Box + M_\text{eff}^2 \right)^{-1} 2\dot\theta\frac{\dot\phi_0}{H}\dot\calR = \frac{1}{M_\text{eff}^2} \left( 1 + \frac{\Box}{M_\text{eff}^2} + \cdots \right) 2\dot\theta\frac{\dot\phi_0}{H}\dot\calR \, ,
\end{equation}
making the non-local theory to the one with higher derivatives. Especially, the leading term gives a simple modified kinetic term:
\begin{equation}
S_2 = \int d^4x a^3\epsilon\mpl^2 \left[ \left( 1 + \frac{4\dot\theta^2}{M_\text{eff}^2} \right)\dot\calR^2 - \frac{(\nabla\calR)^2}{a^2} \right] \, ,
\end{equation}
which is interpreted as the speed of sound $c_s^{-2}$~\cite{Achucarro:2010da,Achucarro:2012sm,singleeft-cs}. That is, the propagation speed of the adiabatic mode $\calR$ is reduced by the interaction with the heavy isocurvature mode $\calF$, as the kinetic energy of $\calR$ is extracted to excite $\calF$. Including higher order derivative operators in the expansion \eqref{eq:F-expansion} leads to a modified dispersion relation in such a way that the validity of the effective theory is improved~\cite{single-eft-next}.

\section{Power spectrum}
\label{sec:P}
\setcounter{equation}{0}

\subsection{Free solutions of mode functions and power spectrum}

Having discussed various forms of the quadratic action for perturbations, now we solve the derived equations of motion for those perturbations. What we have seen previously is that in general there are always interactions between different components. These mixing terms lead to the coupled set of differential equations, for example \eqref{eq:uT} and \eqref{eq:uN}. Thus finding exact solutions of these coupled equations is rather non-trivial. But given that the interactions are small enough, we can perturbatively find the solutions and corrections to the power spectrum of the perturbation of our interest. Presuming that the interactions are sufficiently small, we conveniently split the quadratic part of the action as
\begin{equation}
S_2 = S_\text{2,free} + S_\text{2,int} \, ,
\end{equation}
where $S_\text{2,free}$ contains only free, decoupled terms for a certain component $\Psi$ without interaction with other degrees of freedom, while $S_\text{2,int}$ includes quadratic interactions among them. The leading solution of a certain perturbation component $\Psi$ is coming from the decoupled free quadratic action for $\Psi$, which has the following schematic form as a harmonic oscillator:
\begin{equation}
\label{eq:Sfree}
S_\text{free} = \int d\tau d^3x \frac{1}{2} \left[ {\Psi'}^2 - (\nabla\Psi)^2 - m^2\Psi^2 \right] \, ,
\end{equation}
with $m^2$ being time-dependent in general. To obtain this form we need to perform further manipulations. For example, for the decoupled free part of the quadratic action for the curvature perturbation $\calR$ [or equivalently $\pi$ at quadratic order: see \eqref{eq:pi->R}] we rescale $\calR$ as
\begin{equation}
\Psi = \frac{a\dot\phi_0}{H} \calR \equiv z\calR \, ,
\end{equation}
which gives $m^2 = z''/z$. The resulting equation of motion we can derive from \eqref{eq:Sfree} is
\begin{equation}
\label{eq:eom-Psi}
\Psi'' - \Delta\Psi - m^2\Psi = 0 \, .
\end{equation}
We can write the Fourier mode $\Psi(\tau,\mathbi{k})$, which will be more convenient for the subsequent study, as
\begin{equation}
\label{eq:Psi-Fourier}
\Psi(\tau,\mathbi{x}) = \int \frac{d^3k}{(2\pi)^3} e^{i\mathbi{k}\cdot\mathbi{x}} \Psi(\tau,\mathbi{k}) \, ,
\end{equation}
then the equation of motion \eqref{eq:eom-Psi} becomes, for the Fourier mode,
\begin{equation}
\label{eq:freeeq}
\Psi'' + (k^2 + m^2)\Psi = 0 \, .
\end{equation}
Identifying $k^2 + m^2 \equiv \omega_k^2(\tau)$, we can see that \eqref{eq:freeeq} describes a harmonic oscillator with time dependent frequency $\omega_k^2(\tau)$. Being a canonically normalized harmonic oscillator, we can follow the standard quantization procedure for $\Psi$. That is, we promote $\Psi$ and the conjugate momentum $\Pi_\Psi = \Psi'$ to operators $\widehat\Psi$ and $\widehat\Pi_\Psi$ and imposes the canonical commutation relations between them.

Since $\Psi$ is a free field, we can expand the operator $\widehat\Psi$ in terms of the creation and annihilation operators in the Fourier space. The Fourier mode given by \eqref{eq:Psi-Fourier} is promoted to the operator $\widehat\Psi(\tau,\mathbi{k})$, which we can expand in terms of the creation and annihilation operators:
\begin{equation}
\label{eq:Psi-opexp}
\widehat\Psi(\tau,\mathbi{k}) = a(\mathbi{k})\Psi_k(\tau) + a^\dag(-\mathbi{k})\Psi_k^*(\tau) \, ,
\end{equation}
where the creation and annihilation operators satisfy the standard commutation relations
\begin{equation}
\label{eq:commutation}
\left[ a(\mathbi{k}), a^\dag(\mathbi{q}) \right] = (2\pi)^3 \delta^{(3)}(\mathbi{k}-\mathbi{q}) \, ,
\end{equation}
otherwise zero. Now we require that the canonical conjugate variables $\widehat\Psi$ and $\widehat\Pi_\Psi$ satisfy the equal time canonical commutation relation
\begin{equation}
\left[ \widehat\Psi(\tau,\mathbi{x}), \widehat\Pi_\Psi(\tau,\mathbi{y}) \right] = i\delta^{(3)}(\mathbi{x}-\mathbi{y}) \, .
\end{equation}
Using the Fourier mode \eqref{eq:Psi-Fourier} and the expansion \eqref{eq:Psi-opexp} with the commutation relation \eqref{eq:commutation}, we can see that the mode function $\Psi_k$ satisfies the normalization condition
\begin{equation}
\label{eq:normalization}
\Psi_k\frac{d\Psi_k^*}{d\tau} - \frac{d\Psi_k}{d\tau}\Psi_k^* = i \, .
\end{equation}

To determine the mode function $\Psi_k(\tau)$, which amounts to fix the vacuum state $|0\rangle$ defined by
\begin{equation}
a(\mathbi{k})|0\rangle = 0 \quad \text{for all } k \, ,
\end{equation}
we impose the vacuum boundary condition when the mode is deep inside the horizon $\tau \to -\infty$, i.e. $k \gg aH$ where the mode function solution is the positive frequency mode with $\omega_k = k$:
\begin{equation}
\label{eq:boundary}
\Psi_k = \frac{1}{\sqrt{2k}} e^{-ik\tau} \, .
\end{equation}
With this boundary condition and the normalization condition \eqref{eq:normalization}, the general solution for \eqref{eq:freeeq} can be written in terms of the Bessel functions, and for later convenience we use the Hankel function:
\begin{equation}
\label{eq:massivesol}
\Psi_k(\tau) = \sqrt{-\tau} \left[ c_1(k)H^{(1)}_\nu(-k\tau) + c_2(k)H^{(2)}_\nu(-k\tau) \right] \, ,
\end{equation}
where $c_1(k)$ and $c_2(k)$ are coefficients to be determined, and 
\begin{equation}
\nu^2 \equiv \frac{9}{4} - \frac{m^2}{H^2}
\end{equation}
is assumed to be constant. To fix the coefficients, we require that when the mode is deep inside the horizon $\tau \to -\infty$ we recover the vacuum boundary solution \eqref{eq:boundary}. This can be found by taking the argument of the Hankel function $-k\tau$ very large,
\begin{equation}
H^{(1)}_\nu(z) \underset{z\gg1}{\longrightarrow} \sqrt{\frac{2}{\pi z}} \exp \left[ i \left( z - \frac{\pi}{2}\nu - \frac{\pi}{4} \right) \right] \, ,
\end{equation}
with $H^{(2)}_\nu$ being the complex conjugate of $H^{(1)}_\nu$. Thus, to match \eqref{eq:boundary}, $c_1(k)$ and $c_2(k)$ should be
\begin{equation}
\label{eq:Hankel-coeff}
c_1(k) = \frac{\sqrt{\pi}}{2} e^{i(\nu + 1/2)\pi/2} \quad \text{and} \quad c_2(k) = 0 \, .
\end{equation}
A particularly important and simple case is when $\nu = 3/2$ exactly, which corresponds to the massless limit:
\begin{equation}
\label{eq:massless-sol}
\Psi_k = \frac{1}{\sqrt{2k}} \left( 1-\frac{i}{k\tau} \right) e^{-ik\tau} \, .
\end{equation}

Given the free solution of $\Psi_k$, now we can compute the power spectrum of $\Psi_k$, defined by
\begin{equation}
\label{eq:Pdef}
\langle0| \Psi(\mathbi{k})\Psi(\mathbi{q}) |0\rangle \equiv (2\pi)^3 \delta^{(3)}(\mathbi{k}+\mathbi{q}) \frac{2\pi^2}{k^3} \calP_\Psi(k)\, ,
\end{equation}
so that in terms of the mode function $\Psi_k$ 
\begin{equation}
\calP_\Psi(k) = \frac{k^3}{2\pi^2} \left| \Psi_k \right|^2 \, ,
\end{equation}
which can be evaluated at a convenient time, e.g. the moment of horizon crossing. For example, the power spectrum of the curvature perturbation can be directly found from its free quadratic action as
\begin{equation}
\calP_\calR =\frac{k^3}{2\pi^2} \left| \frac{\Psi_k}{z} \right|^2 = \left( \frac{H}{2\pi} \right)^2 \frac{1}{2\epsilon\mpl^2} \, .
\end{equation}
Alternatively, using the kinematic basis discussed in Sections~\ref{sec:BG} and \ref{sec:pert} in which $\calR$ is related to the tangential component $u^T$ in the standard manner as
\begin{equation}
\calR = -\frac{H}{\dot\phi_0} Q^T = -\frac{H}{\dot\phi_0} \frac{u^T}{a} \, ,
\end{equation}
we can write
\begin{equation}
\calP_\calR(k,\tau) = \frac{H^2}{\dot\phi_0^2} \calP_{TT}(k,\tau) = \frac{k^3}{4\pi^2a^2\mpl^2\epsilon} \left| {u}^T_k \right|^2 \, .
\end{equation}
where $m^2$ for $u^T$ is given by, as can be read from \eqref{eq:uT}, $m^2 = a^2M_{TT}^2 - {\theta'}^2$. Using the mass basis instead gives similar results.

\subsection{Corrections to the power spectrum}

In the previous section, we have obtained the free power spectrum with the quadratic interactions between different components being ignored. Now we consider the effects of the interactions to the power spectrum perturbatively. Having perturbative interactions, we can compute readily the corrections due to the interaction terms between different fields using the in-in formalism~\cite{in-in}, which let us briefly recall here.

One crucial point when interactions exist, viz. the system is evolving, is that any (vacuum) expectation values should be taken with respect to the interaction vacuum state $|\Omega\rangle$, i.e. the {\em actual} vacuum state of the theory, {\em not} the free vacuum state $|0\rangle$ defined in the previous section. For this description, we resort to the interaction picture. In quantum mechanics, the interaction picture (or Dirac picture) is an intermediate between the Schr\"odinger picture and the Heisenberg picture. Whereas in the other two pictures either the state vector or the operators carry time dependence, in the interaction picture both carry part of the time dependence of observables. The purpose of the interaction picture is to shunt all the time dependence due to the free Hamiltonian $H_0$ onto the operators, leaving only the interaction Hamiltonian $H_\mathrm{int}$ affecting the time-dependence of the state vectors.

Now, we denote by $\left\langle \widehat{\mathcal{O}}(t) \right\rangle$ the expectation value evaluated at a time $t$ of a time-dependent operator
\begin{equation}\label{interaction_operator}
\widehat{\mathcal{O}}(t) = \left( e^{-i\int_{t_\mathrm{in}}^t H_0(t') dt'}
\right)^\dag \widehat{\mathcal{O}} \left( e^{-i\int_{t_\mathrm{in}}^t H_0(t'') dt''}
\right) \, ,
\end{equation}
where $t_\mathrm{in}$ is some early ``in'' time when the interaction is turned on. This expectation value is taken with respect to the vacuum state at that time $|\Omega(t)\rangle$ that has evolved from an ``in'' state, which we take as the vacuum $|0\rangle$, according to
\begin{equation}
\label{eq:interaction_vector}
|\Omega(t)\rangle = e^{-i\int_{t_\mathrm{in}}^t H_\mathrm{int}(t') dt'}|0\rangle \, .
\end{equation}
Given appropriate normalizations, now let us consider $H_\mathrm{int}$ as a small perturbation to the free Hamiltonian $H_0$. Expanding the exponential in \eqref{eq:interaction_vector}, in terms of $H_\mathrm{int}$, we obtain\footnote{Here, we omit for simplicity $\lim_{t_\mathrm{in}\to-\infty(1-i\varepsilon)}$ with $\varepsilon \ll 1$, but $t_\mathrm{in}$ is evaluated in this limit after all.}
\begin{align}
\label{eq:inin-expansion}
\left\langle \widehat{\mathcal{O}}(t) \right\rangle = & \left\langle0\left|\left[ 1 -
i \int_{t_\mathrm{in}}^t H_\mathrm{int}(t')dt' + \frac{1}{2} \left( -i
\int_{t_\mathrm{in}}^t H_\mathrm{int}(t')dt' \right)^2 + \cdots \right]^\dag
\widehat{\mathcal{O}}(t) \right.\right.
\nonumber\\
& \left.\left. \hspace{0.5cm} \times \left[ 1 - i \int_{t_\mathrm{in}}^t
H_\mathrm{int}(t')dt' + \frac{1}{2} \left( -i \int_{t_\mathrm{in}}^t
H_\mathrm{int}(t'')dt'' \right)^2 + \cdots \right] \right|0\right\rangle
\nonumber\\
= & \left\langle0\left| \widehat{\mathcal{O}}(t) \right|0\right\rangle + i
\int_{t_\mathrm{in}}^t dt_1 \left\langle0\left| \left[ H_\mathrm{int}(t'),
\widehat{\mathcal{O}}(t) \right] \right|0\right\rangle
\nonumber\\
& - \int_{t_\mathrm{in}}^t dt_2 \int_{t_\mathrm{in}}^{t_1} dt_2 \left\langle0\left|
\left[ H_\mathrm{int}(t_1), \left[ H_\mathrm{int}(t_2), \widehat{\mathcal{O}}(t)
\right] \right] \right|0\right\rangle + \cdots \, ,
\end{align}
where for the last equality we have used the commutator identity
\begin{equation}
AAB - 2ABA + BAA = [A,[A,B]] \, .
\end{equation}

Having briefly recalled the in-in formalism, now we consider the corrections to the free power spectrum of $\Psi$ due to the interaction terms between different fields, which we collectively denote by $\Phi$. Schematically we write the interaction terms as
\begin{equation}
\label{eq:quadratic-int}
S_\text{2,int} = \int d^4x a^3c(t) \calO^{(\Phi)} \Phi \calO^{(\Psi)} \Psi \, ,
\end{equation}
where $c(t)$ is the time-dependent coupling between $\Psi$ and $\Phi$, and $\calO^{(X)}$ is the possible derivative operator for the field $X$. Since the interaction Hamiltonian contains two free field contents, the first non-vanishing contribution to the two-point function comes from the interaction Hamiltonian squared. Correspondingly, the correction terms now involve two free propagators: one for $\Psi$ and the other for the coupled field $\Phi$. Thus, in terms of the free propagator 
\begin{equation}
i\Delta_{+-}^X(x,x') \equiv \left\langle 0 \left| X(x')X(x) \right| 0 \right\rangle \, ,
\end{equation}
where the subscript $+$ ($-$) denotes forward (backward) in time with $i\Delta_{+-} = -i\Delta_{-+}^*$, from \eqref{eq:inin-expansion} we can write as the two-point function of $\Psi$ including the leading corrections due to its interaction with $\Phi$ as
\begin{align}
\label{eq:general2point}
\left\langle \Omega \left| \Psi(t,\mathbi{x})\Psi(t,\mathbi{y}) \right| \Omega \right\rangle 
& = i\Delta^\Psi_{+-}
-\int^t_{t_i} dt_1d^3x_1 \int^t_{t_i} dt_2d^3x_2 (a^3c)(t_1)(a^3c)(t_2) 
\nonumber\\
& \hspace{5em} \times \calO_1^{(\Psi)} i\Delta^\Psi_{+-}(t, t_1) \calO_2^{(\Psi)} i\Delta^\Psi_{-+}(t,t_2) \calO_1^{(\Phi)}\calO_2^{(\Phi)} i\Delta^\Phi_{+-}(t_1, t_2)
\nonumber\\
& \hspace{1em} - \int^t_{t_i} dt_1 d^3x_1 \int^{t_1}_{t_i} dt_2 d^3x_2 (a^3c)(t_1)(a^3c)(t_2)
\nonumber\\
& \hspace{4em} \times \left[ \calO_1^{(\Psi)} i\Delta^\Psi_{+-}(t_1, t) \calO_2^{(\Psi)} i\Delta^\Psi_{+-}(t, t_2) \calO_1^{(\Phi)}\calO_2^{(\Phi)} i\Delta^\Phi_{+-}(t_1, t_2) \right.
\nonumber\\
& \hspace{5em} \left. + \calO_1^{(\Psi)} i\Delta^\Psi_{-+}(t, t_1) \calO_2^{(\Psi)} i\Delta^\Psi_{-+}(t_2, t) \calO_1^{(\Phi)}\calO_2^{(\Phi)} i\Delta^\Phi_{+-}(t_2, t_1) \right] \, .
\end{align}
This is most general expression for the two-point correlation function of $\Psi$ including the leading corrections due to the interaction with $\Phi$. The corresponding power spectrum can be found by taking the Fourier transformation of $\left\langle \Omega \left| \Psi(t,\mathbi{x})\Psi(t,\mathbi{y}) \right| \Omega \right\rangle$.

\subsubsection{Two-field case}

For definiteness, let us consider explicitly the corrections to the power spectrum of $\calR$ due to the interaction with $Q_\bot^a$ in the simplest two-field case. The total quadratic action including the interaction is \eqref{eq:S2-R&F}, from which we read the coefficient of the quadratic mixing term and the derivative operators as 
\begin{equation}
\label{Eq:quadmix}
c(t) = \sqrt{8\epsilon}\mpl\dot\theta \, , \quad  \calO^{(\calR)} = \partial_t \, , \quad \calO^{(\calF)} = 1 \, .
\end{equation}
Then the two-point correlation of $\calR$ is given by
\begin{align}
\left\langle \Omega \left| \calR(t,\mathbi{x})\calR(t,\mathbi{y}) \right| \Omega \right\rangle & = i\Delta^\calR_{+-}
-\int^t_{t_i} dt_1d^3x_1 \int^t_{t_i} dt_2d^3x_2 (a^3c)(t_1)(a^3c)(t_2) 
\nonumber\\
& \hspace{5em} \times \partial_{t_1} i\Delta^\calR_{+-}(t, t_1) \partial_{t_2} i\Delta^\calR_{-+}(t,t_2) i\Delta^\calF_{+-}(t_1, t_2)
\nonumber\\
& \hspace{1em} - \int^t_{t_i} dt_1 d^3x_1 \int^{t_1}_{t_i} dt_2 d^3x_2 (a^3c)(t_1)(a^3c)(t_2)
\nonumber\\
& \hspace{4em} \times \left[ \partial_{t_1} i\Delta^\calR_{+-}(t_1, t) \partial_{t_2} i\Delta^\calR_{+-}(t, t_2) i\Delta^\calF_{+-}(t_1, t_2) \right.
\nonumber\\
& \hspace{5em} \left. + \partial_{t_1} i\Delta^\calR_{-+}(t, t_1) \partial_{t_2} i\Delta^\calR_{-+}(t_2, t) i\Delta^\calF_{+-}(t_2, t_1) \right] \, .
\end{align}
The power spectrum is obtained by performing the Fourier transformation of the above result. In the very simple limit of constant $c(t)$, which corresponds to a constant turn, using the massive mode function solution \eqref{eq:massivesol} with \eqref{eq:Hankel-coeff} for $\calF$, we find
\begin{equation}
\calP_\calR = \left( \frac{H}{2\pi} \right)^2 \frac{1}{2\epsilon\mpl^2} \left( 1 + \frac{4c^2C}{\epsilon\mpl^2H^2} \right) \, ,
\end{equation}
where~\cite{qsf}
\begin{align}
\label{eq:coeff-C-int}
C \equiv \frac{\pi}{4} \Re & \left\{ \int^\infty_0 dx_1 \int^\infty_{x_1} dx_2 \left[ x_1^{-1/2}H_\nu^{(1)}(x_1)e^{ix_1}x_2^{-1/2}H_\nu^{(2)}(x_2)e^{-ix_2} \right.\right.
\nonumber\\
& \left. \hspace{8em} - x_1^{-1/2}H_\nu^{(1)}(x_1)e^{-ix_1}x_2^{-1/2}H_\nu^{(2)}(x_2)e^{-ix_2} \right] \bigg\} \, .
\end{align}
Especially, if $\calF$ is also very light so that $\nu \to 3/2$, we can perform the integral analytically using the massless mode function solution \eqref{eq:massless-sol} to find
\begin{equation}
\label{eq:coeff-C}
C = \frac{1}{2} (\alpha-\log{x})^2 + \frac{\pi^2}{8} - \frac{\gamma^2}{2} - \frac{3}{4} = \left. \frac{\alpha^2}{2} + \frac{\pi^2}{8} - \frac{\gamma^2}{2} - \frac{3}{4} \right|_{k=aH} \, ,
\end{equation}
where $\alpha \equiv 2-\log2-\gamma \approx 0.729637$ with $\gamma \approx 0.577216$ being the Euler-Mascheroni constant, and for the second equality we have evaluated at the moment of horizon crossing $k=aH$\footnote{In fact, if we integrate the outermost integral of \eqref{eq:coeff-C-int} from $x_e \equiv -k\tau_e$, we may interpret $-\log(-k\tau_e) = N_k $ as the number of $e$-folds elapsed between the moment of horizon crossing for the mode of our interest and the end of inflation $\tau_e$~\cite{Gong:2001he}. This seems diverging as $x_e\to0$, but see the discussion below \eqref{eq:zeta-expansion}.} [see also \eqref{eq:Q-2point}]. For the field fluctuations $Q^a$ in the flat gauge, we can find similar results.

\subsection{Alternative solutions}

In the previous sections, we have considered the free, decoupled and interaction terms separately, and have regarded the latter as perturbations to apply the in-in formalism. This is because fully solving the coupled differential equations is a non-trivial task. We can, however, take a different approach to solve the equations of motion directly. As we will see, we still treat the interaction terms perturbatively, so in this sense it is equivalent to what we have seen in the previous sections. One advantage of the alternative we present here is that all the corrections, including the deviations from perfect de Sitter limit $\tau = -1/(aH)$, are systematically taken into account. We will solve for the field fluctuations $Q^a$ in the flat gauge, but it is straightforwardly extended to the curvature perturbation $\calR$ in the comoving gauge.

Our starting point is the equation of motion for $Q^a$ \eqref{eq:Qeom1}. Defining $u^a \equiv aQ^a$ and 
\begin{equation}
\label{eq:variable-x}
x \equiv -k\tau = \frac{k}{aH} \left( 1 + \epsilon + \cdots \right) \, ,
\end{equation}
we can rewrite \eqref{eq:Qeom1} as~\cite{multi-eom}
\begin{equation}
\label{eq:Qeom2}
D_x^2u^a + \left( 1 - \frac{2}{x^2} \right) u^a = \frac{3}{x^2} \zeta^a{}_bu^b \, ,
\end{equation}
where $\zeta^{ab}$ is given by
\begin{equation}
\label{eq:zeta-ab}
\zeta^{ab} = G^{ab}\epsilon + \frac{\dot\phi_0^a}{H}\frac{\dot\phi_0^b}{H} + \frac{\mathbb{R}^a{}_{cd}{}^b}{3} \frac{\dot\phi_0^c}{H}\frac{\dot\phi_0^d}{H} - \frac{V^{ab}}{3H^2} + \cdots \, .
\end{equation}
Since $u^a$ canonically normalizes \eqref{eq:S2-1}, we can apply the canonical commutation relations between $u^a$ and its conjugate momentum $D_\tau u^a$:
\begin{equation}
\left[ u^a(\tau,\mathbi{x}), D_\tau u^b(\tau,\mathbi{y}) \right] = i\delta^{(3)}(\mathbi{x}-\mathbi{y})G^{ab} \, ,
\end{equation}
otherwise zero.

What we can see from the rescaled equation of motion \eqref{eq:Qeom2} is that the left hand side describes the evolution of a single component $u^a$ in perfect de Sitter background, while everything else -- including the deviations from perfect de Sitter background and mixing with other components -- is placed on the right hand side. Thus we can solve the homogeneous part of \eqref{eq:Qeom2},
\begin{equation}
D_x^2u_0^a + \left( 1 - \frac{2}{x^2} \right)u_0^a = 0 \, ,
\end{equation}
very easily with the boundary conditions, as in the previous sections, being imposed at $x\to\infty$ as
\begin{equation}
u_0^a(x,\mathbi{k}) = a^a(\mathbi{k})u_0(x) + {a^a}^\dag(-\mathbi{k})u_0^*(x) \, ,
\end{equation}
where the creation and annihilation operators satisfy the commutation relations
\begin{equation}
\left[ a^a(\mathbi{k}), {a^b}^\dag(\mathbi{q}) \right] = (2\pi)^3\delta^{(3)}(\mathbi{k}-\mathbi{q})G^{ab}
\end{equation}
and the mode function solution $u_0(x)$ is that of a massless field \eqref{eq:massless-sol}:
\begin{equation}
u_0(x) = \frac{1}{\sqrt{2k}} \left( 1 + \frac{i}{x} \right) e^{ix} \, .
\end{equation}
With these boundary conditions, along with the same normalization condition as \eqref{eq:normalization}, we can apply the standard Green's function method to solve the inhomogeneous equation as~\cite{Gong:2001he}
\begin{equation}
\label{eq:Green-sol}
u^a(x) = u_0^a(x) + \frac{3}{2}i \int_x^\infty \frac{du}{u^2} \zeta^a{}_b u^b(u) \left[ u_0^*(u)u_0(x) - u_0^*(x)u_0(u) \right] \, .
\end{equation}
The first correction terms are obtained by computing the integral with $u^b(u)$ being the homogeneous solution $u_0$, and the next corrections by plugging the solution with leading correction terms, and so on. We can iterate as many times as we like to find more and more accurate solutions.

To implement the slow-roll approximation evaluated at, say, the moment of horizon crossing $k = aH$ which is different from $x=1$ by $\calO(\epsilon)$ as can be read from \eqref{eq:variable-x}, we take the ansatz $\zeta^{ab} = \zeta^{ab}(\log{x})$ and expand $\zeta^{ab}$ as a power series in $\log{x}$:
\begin{equation}
\label{eq:zeta-expansion}
\zeta^{ab} = \sum_{n=0}^\infty \frac{\zeta_{n+1}^{ab}}{n!} (\log{x})^n \, ,
\end{equation}
with $\zeta_n^{ab} = \calO(\epsilon^n)$. This expansion is valid as long as the series converges, which is the case for a wide range of value for $x$ as $e^{-1/\calO(\epsilon)} \ll x \ll e^{1/\calO(\epsilon)}$. We can substitute the expansion ansatz \eqref{eq:zeta-expansion} into the Green's function solution \eqref{eq:Green-sol} and integrate iteratively to find the desired solution for $\varphi^a$ explicitly in terms of the slow-roll parameters. Especially, we can find the asymptotic solution at later time $x\to0$ as
\begin{align}
u^a(x) & \underset{x\to0}{\longrightarrow} \frac{i}{\sqrt{2k}x} \left\{ a^a(\mathbi{k}) - {a^a}^\dag(\mathbi{k}) + \left[ \left( \alpha + \frac{i\pi}{2} \right)\zeta_1^a{}_b - \zeta_1^a{}_b\log{x} + \cdots \right] a^b(\mathbi{k}) \right.
\nonumber\\
& \hspace{12em} \left. - \left[ \left( \alpha + \frac{i\pi}{2} \right)\zeta_1^a{}_b - \zeta_1^a{}_b\log{x} + \cdots \right]^* {a^b}^\dag(-\mathbi{k}) + \cdots \right\} \, ,
\end{align}
with $\alpha$ being the same numerical factor as encountered in \eqref{eq:coeff-C}. This gives
\begin{align}
\label{eq:Q-2point}
\left\langle Q^a(\mathbi{k}) {Q^b}(\mathbi{q}) \right\rangle & = \frac{1}{2ka^2x^2} \delta^{(3)}(\mathbi{k}+\mathbi{q}) \left( G^{ab} + 2\alpha\zeta_1^{ab} - 2\zeta_1^{ab}\log{x} + \cdots \right) 
\nonumber\\
& = \left. \frac{H^2}{2k^3} \delta^{(3)}(\mathbi{k}+\mathbi{q}) \Big[ (1-2\epsilon)G^{ab} + 2\alpha\zeta_1^{ab} + \cdots \Big] \right|_{k=aH} \, , 
\end{align}
where for the second equality we have evaluated the right hand side at the moment of horizon crossing $k=aH$. While we can proceed as much accurate as we like, what we have obtained is the power spectrum of the field fluctuations in the flat gauge, {\em not} that of the curvature perturbation $\calR$. Thus we need additional manipulation that relates $Q^a$ and $\calR$. Unlike single field case, in multi-field case it is non-trivial as we have seen from \eqref{Eq:multiredef} and \eqref{eq:MS-comoving}. Fortunately, there is a very simple geometric identity called the $\delta{N}$ formalism that relates $Q^a$ and $\calR$ in a very straightforward manner: see \eqref{eq:deltaN}. We postpone more detailed discussions on the $\delta{N}$ formalism to the following section.

\section{Towards non-linear perturbations}
\label{sec:non-linear}
\setcounter{equation}{0}

Up to now, we have considered linear cosmological perturbations. They are described by quadratic action presented in many different forms in the previous section like \eqref{Eq:quadratic}. The structure of the action is, with appropriate manipulations, essentially that of a harmonic oscillator, so we can directly apply our conventional wisdom in quantum field theory. Being described as quantum harmonic oscillators, linear cosmological perturbations are free, leading to the conclusion that the power spectra are the only non-vanishing correlation functions. The rapid observational advances during the last decade, especially the most recent Planck mission for the CMB temperature anisotropies, have constrained the properties of the power spectra as~\cite{Ade:2015oja}:
\begin{align}
\label{eq:P-obsbound}
\log\left( 10^{10}\calP_\calR \right) & = 3.094 \pm 0.034 \, ,
\\
\label{eq:nR-obsbound}
n_\calR & \equiv \frac{d\log\calP_\calR}{d\log{k}}+1 = 0.9645 \pm 0.0049 \, ,
\\
r & \equiv \frac{\calP_T}{\calP_\calR} \lesssim 0.12 \, .
\end{align}
Indeed, these observational constraints have been very powerful in discriminating different models of inflation and in ruling out the models inconsistent  with observations. For example, the simple quartic potential model is not favoured well beyond $2\sigma$ level, since this model predict too large tensor-to-scalar ratio of $r \gtrsim 0.2$.

With further developments in observations -- including the polarization of the CMB, gravitational waves and distribution of galaxies on large scales -- that allow us to probe a wide range of scales, we can hope to make use of higher-order correlation functions to constrain more tightly inflation models and to probe inflationary dynamics. The higher-order correlation functions, starting from three-point function or its Fourier transform, the bispectrum, incorporate non-linear perturbations described beyond quadratic action of the cosmological perturbations. Thus non-linear perturbations also tell us the ``interactions'' among the degrees of freedom participating in inflation, revealing the important physics underlying inflation that cannot be extracted from the power spectra\footnote{Of course higher-order correlation functions may well contribute to the power spectrum as loop corrections.}. The amplitude of the bispectrum, the first non-zero probe of non-linear perturbations, is conveniently parametrized by the so-called non-linear parameter $\fnl$~\cite{Komatsu:2001rj}, which is roughly speaking the ratio of the bispectrum to the power spectrum squared, and the current observational constraints from the CMB on different configurations of the bispectrum are~\cite{Ade:2015ava}
\begin{alignat}{3}
\label{eq:fNL-obsbound}
\fnl^\text{local} & = 2.5 \pm 5.7 \qquad & & (0.8 \pm 5.0) \, ,
\nonumber\\
\fnl^\text{equil} & = -16 \pm 70 \qquad & & (-4 \pm 43) \, ,
\\
\fnl^\text{ortho} & = -34 \pm 33 \qquad & & (-26 \pm 21) \, ,
\nonumber
\end{alignat}
where the numbers in the parentheses are obtained by combining the polarization data. So currently our universe is largely consistent with free, linear, Gaussian cosmological perturbations. The importance of non-linear perturbations however does not diminish and we can expect to further constrain or even detect the deviation from perfect Gaussian nature of the primordial perturbations, viz. the primordial non-Gaussianity in near future. In this section, we discuss how to describe non-linear perturbations in multi-field inflation.

\subsection{Issue of mapping}

\begin{figure}[h]
 \centering
 \includegraphics[width=10cm]{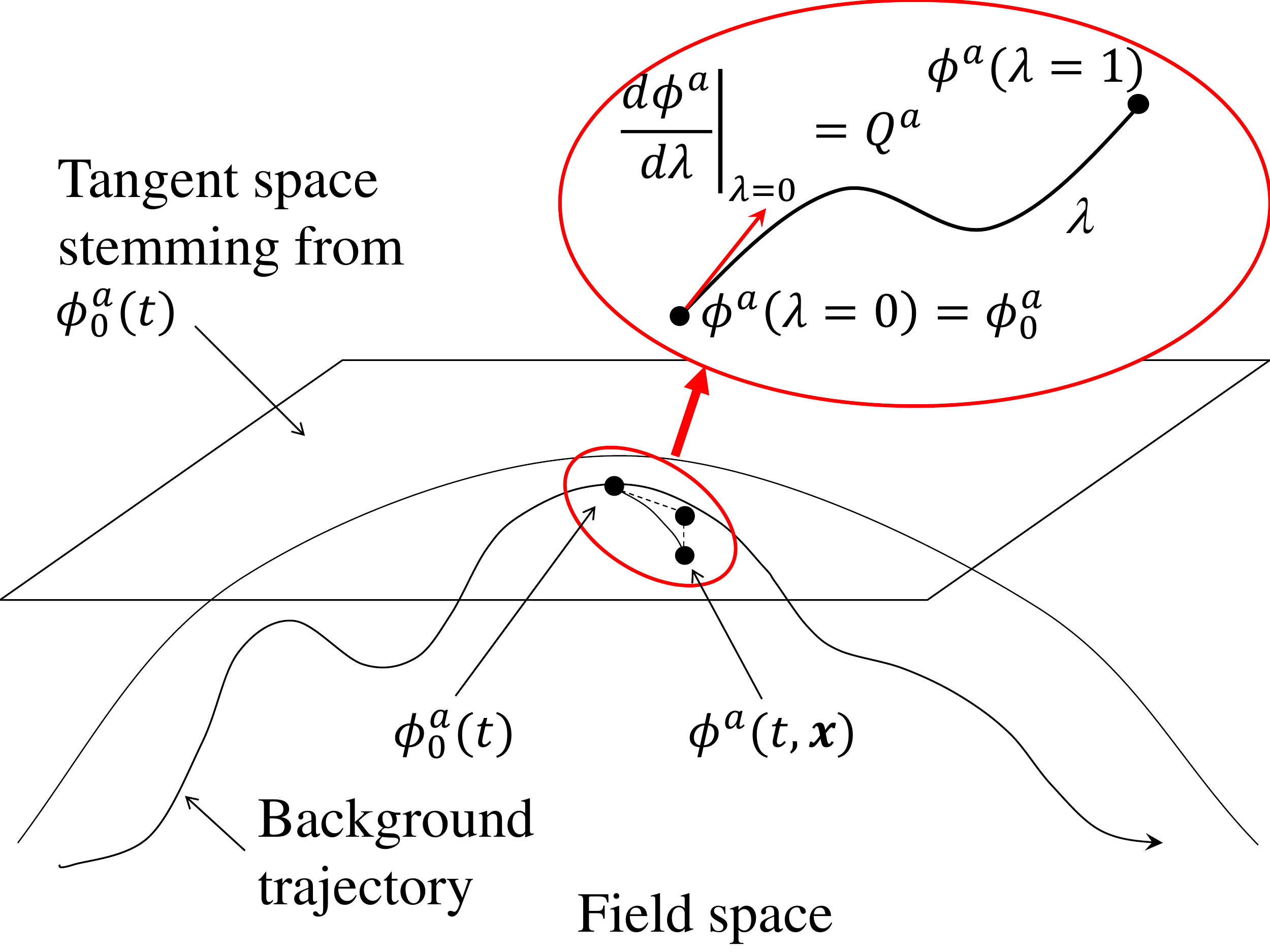}
 \caption{A schematic figure showing a physical field $\phi^a$ in the field space around the background trajectory $\phi_0^a(t)$. The geodesic connecting $\phi^a$ and $\phi_0^a$ is parametrized $\lambda$, which runs from 0 to $\varepsilon$.}
 \label{fig:field_space}
\end{figure}

An important point we have to take care of when we discuss non-linear field fluctuations in multi-field inflation in the flat gauge is that, as emphasized previously, in the field space $\phi^a$ plays the role of coordinates. Thus, as we do in general relativity, it is of crucial convenience to maintain the covariance of the formulation: the field fluctuations $\delta\phi^a$ around the background trajectory $\phi_0^a = \phi_0^a(t)$ are coordinate dependent -- for example, we are free to choose the time slicing condition as $\delta\phi^1 = 0$ -- and is thus not covariant. How then to formulate $\delta\phi^a$ covariantly? We begin by noting that the two points, $\phi_0^a(t)$ and $\phi^a = \phi_0^a + \delta\phi^a$, can be connected by a unique geodesic with respect to the field space metric $G_{ab}$~\cite{Gong:2011uw,Elliston:2012ab}. This geodesic is parametrized by $\lambda$ that runs from 0 to $\varepsilon$, corresponding to the two endpoints of the geodesic, $\phi_0^a$ and $\phi^a$ respectively. Here, $\varepsilon$ is a bookkeeping parameter to count the order of perturbation and we will set it to unity in the end. To specify the geodesic, we need the initial point $\phi_0^a$ and its velocity, which we denote by $Q^a$. This situation is depicted in Figure~\ref{fig:field_space}.

Denoting the covariant derivative with respect to $\lambda$ by $D_\lambda \equiv D/d\lambda$, we can write the geodesic equation for $\phi^a(\lambda)$ as
\begin{equation}
\label{eq:geodesic}
D_\lambda^2\phi^a = \frac{d^2\phi^a}{d\lambda^2} + \Gamma^a_{bc}\frac{d\phi^b}{d\lambda}\frac{d\phi^c}{d\lambda} = 0 \, ,
\end{equation}
with the following the initial conditions at $\lambda=0$:
\begin{align}
\left. \phi^a \right|_{\lambda=0} = & \phi_0^a \, ,
\\
\left. D_\lambda\phi^a \right|_{\lambda=0} = & \left. \frac{d\phi^a}{d\lambda} \right|_{\lambda=0} \equiv Q^a \, .
\end{align}
Expanding $\phi^a(\lambda=\varepsilon)$ as a power series with respect to $\varepsilon$ from $\lambda=0$, we find
\begin{equation}
\label{eq:mapping}
\phi^a(\lambda=\varepsilon) = \left. \phi^a \right|_{\lambda=0} + \left. \frac{d\phi^a}{d\lambda} \right|_{\lambda=0} \varepsilon + \left. \frac{1}{2!}\frac{d^2\phi^a}{d\lambda^2} \right|_{\lambda=0} \varepsilon^2 + \left. \frac{1}{3!}\frac{d^3\phi^a}{d\lambda^3} \right|_{\lambda=0} \varepsilon^3 + \cdots \, .
\end{equation}
Since the derivatives with respect to $\lambda$ here are {\em not} covariant ones, we can trade quadratic and higher derivatives with single ones by using the geodesic equation \eqref{eq:geodesic}. That is, we can replace the quadratic derivative with
\begin{equation}
\label{eq:geodesic3}
\frac{d^2\phi^a}{d\lambda^2} = -\Gamma^a_{bc}\frac{d\phi^b}{d\lambda}\frac{d\phi^c}{d\lambda} \, ,
\end{equation}
and the third order derivative with
\begin{equation}
\frac{d^3\phi^a}{d\lambda^3} = \left( \Gamma^a_{de}\Gamma^e_{bc} - \Gamma^a_{bc;d} \right) \frac{d\phi^b}{d\lambda}\frac{d\phi^c}{d\lambda}\frac{d\phi^d}{d\lambda} \, ,
\end{equation}
and so on. Thus, we can write \eqref{eq:mapping} as
\begin{equation}
\phi^a(\lambda=\varepsilon) = \phi^a_0 + Q^a\varepsilon - \frac{1}{2}\Gamma^a_{bc}Q^bQ^c \varepsilon^2 + \frac{1}{6} \left( \Gamma^b_{de}\Gamma^e_{bc} - \Gamma^a_{bc;d} \right) Q^bQ^cQ^d \varepsilon^3 + \cdots \, ,
\end{equation}
Setting $\varepsilon = 1$, we finally obtain
\begin{equation}
\label{eq:mapping2}
\phi^a-\phi_0^a \equiv \delta\phi^a = Q^a - \frac{1}{2}\Gamma^a_{bc}Q^bQ^c + \frac{1}{6} \left( \Gamma^a_{de}\Gamma^e_{bc} - \Gamma^a_{bc;d} \right) Q^bQ^cQ^d + \cdots \, .
\end{equation}
We can see that at linear order, the field fluctuations $\delta\phi^a$ and the vector $Q^a$ are identical. However, going beyond linear order they are manifestly different as we can see from the above equation. Only when we write the equations in terms of $Q^a$, they can be expressed in a covariant manner.

\subsection{Cubic order action}

Having found the covariant description of the field fluctuations, we can straightly calculate the third and higher-order action of the field fluctuations $Q^a$ in the flat gauge. Given the higher-order action of our interest, we can use the in-in formalism discussed in the previous section to compute the corresponding higher-order correlation functions. For example, to compute the bispectrum, we need the third order action for the first non-zero contributions. It is straightforward after some arrangement to find the third order action as
\begin{align}
\label{eq:S3-fieldfluc}
S_3 = \int d^4x a^3 & \left\{ 3\mpl^2H^2\alpha^3 + 2\mpl^2H\alpha^2\frac{\Delta}{a^2}\chi - \frac{\mpl^2}{2a^4} \left[ \chi^{,ij}\chi_{,ij} - (\Delta\chi)^2 \right] \alpha \right.
\nonumber\\
& + (g_1)_{abc}Q^aQ^bQ^c + (g_2)_{abc}D_tQ^aQ^bQ^c + (g_3)_{abc}D_tQ^aD_tQ^bQ^c
\nonumber\\
& \left. + \alpha G_{ab}\dot\phi_0^a\partial_iQ^b\partial^i\chi - G_{ab}D_tQ^a\partial_iQ^b\partial^i\chi - \frac{1}{2}\alpha G_{ab} \frac{\partial^iQ^a\partial_iQ^b}{a^2} \right\} \, ,
\end{align}
where the coefficients $(g_i)_{abc}$ are given by
\begin{align}
(g_1)_{abc} & = \frac{1}{6} \left( \mathbb{R}_{dabe;c}\dot\phi_0^d\dot\phi_0^e + V_{abc} \right) + \frac{1}{2}\calN_c \left( -\mathbb{R}_{dabe}\dot\phi_0^d\dot\phi_0^e + V_{ab} \right) - \frac{\dot\phi_0^2}{2} \calN_a\calN_b\calN_c \, ,
\\
(g_2)_{abc} & = \frac{1}{6} \left( \mathbb{R}_{dbca} + 3\mathbb{R}_{abcd} \right) \dot\phi_0^d + \dot\phi_{0a}\calN_b\calN_c \, ,
\\
(g_3)_{abc} & = -\frac{1}{2}G_{ab}\calN_c \, ,
\end{align}
and $\alpha$ (and $\calN_a$) and $\chi$ are the linear solutions of the lapse and shift given by \eqref{eq:lapse-sol} and \eqref{eq:shift-sol} respectively. Note that the first three terms of \eqref{eq:S3-fieldfluc} are coming from the gravity sector, while the rest from the matter sector. From \eqref{eq:S3-fieldfluc} we can compute the bispectrum of the field fluctuations $Q^a$, evaluated at the moment of the horizon crossing. We can calculate the bispectrum of the curvature perturbation by implementing the $\delta{N}$ formalism: see~\cite{Seery:2005gb,Elliston:2012ab}.

In the comoving gauge, we have already singled out the curvature perturbation as \eqref{eq:R} in a non-perturbative manner. So we can straightforwardly expand the action in terms of $\calR$ and the orthogonal modes $Q_\bot^a$. The only caution is that the Goldstone mode $\pi$, in terms of which the quadratic action \eqref{eq:S2scalar} is written, is related to $\calR$ non-linearly as \eqref{eq:pi->R} so that we have additional non-linear contributions from the quadratic action. Then we can find the leading action cubic in $\calR$ the same as the standard single field case~\cite{Maldacena:2002vr}:
\begin{align}
S_{\calR\calR\calR} = \int d^4x a^3\mpl^2 \Bigg[ & \left (-\epsilon^2+2\epsilon\delta_1 \right)\calR\dot\calR^2 - 2\epsilon^2 \dot\calR\partial^i\calR\Delta^{-1}\partial_i\dot\calR + \left( 3\epsilon^2-2\epsilon\delta_1 \right) \calR \frac{(\nabla\calR)^2}{a^2}
\nonumber\\
& \left. + \frac14 \left( \frac{\epsilon V_{ab}\dot{\phi}_0^a \dot{\phi}_0^b}{H^2} + \frac{V_{abc}\dot{\phi}_0^a \dot{\phi}_0^b \dot{\phi}_0^c}{3H^3} + \frac{G_{ab}\ddot{\phi}_0^a\ddot{\phi}_0^b}{H^2} \right) \calR^3 \right] + {\cal O}(\epsilon^3) \, .
\label{Eq:cubic}
\end{align}
Meanwhile, the $\pi$-$Q^a_\bot$ mixing term in the quadratic action leads to the following $\calR^2Q_\bot$ mixing in the cubic order action: 
\begin{align}
S_\text{quadratic mixing} = \int d^4x a^3 \frac{V_aQ_\bot^a}{H} \bigg\{ & 
(3\epsilon-\delta_1)\calR\dot{\calR} + \frac{1}{H} \left( \calR\ddot{\calR}+\dot{\calR}^2 \right)
- \frac{1-\epsilon}{4a^2H} \left[ (\nabla\calR)^2 - \partial^i\partial^j \Delta^{-1} \left( \partial_i\calR\partial_j\calR \right) \right]
\nonumber\\
& + \frac{1}{2a^2H^2} \left[ \partial^i\calR\partial_i\dot{\calR} - \partial^i\partial^j \Delta^{-1} \left( \partial_i\calR\partial_j\dot{\calR} \right) \right]
\nonumber\\
& \left. + \frac{\epsilon}{H} \left( 1 - \partial^i\partial^j\Delta^{-1} \right) \left( \partial_i\dot{\calR}\Delta^{-1}\partial_j\calR + \partial_i\calR\Delta^{-1}\partial_j\dot{\calR} \right) \right\} + {\cal O}(\epsilon^2) \, . 
\end{align} 
We can also find the other mixing terms -- $Q_\bot^3$, $\calR Q_\bot^2$ and $\calR^2Q_\bot$ -- straightly as
\begin{align}
S_\text{cubic mixing} = \int d^4x a^3 \bigg\{ & -\frac{V_{abc}}{6}Q_\bot^a Q_\bot^b Q_\bot^c -\frac{V_{abc}}{6 H^2} \left( Q_\bot^a \dot{\phi}_0^b \dot{\phi}_0^c+ \dot{\phi}_0^aQ_\bot^b  \dot{\phi}_0^c+ \dot{\phi}_0^a \dot{\phi}_0^bQ_\bot^c \right) \calR^2
\nonumber\\
& + \frac{V_{abc}}{6H} \left( Q_\bot^aQ_\bot^b\dot{\phi}_0^c + Q_\bot^a\dot{\phi}_0^bQ_\bot^c + \dot{\phi}_0^aQ_\bot^bQ_\bot^c \right) \calR
\nonumber\\
+ & \epsilon\calR \left[
- \frac{2}{\mpl^2H^2} \left( V_aQ_\bot^a \right)^2
+ \frac{2}{\mpl^2 H^2} \left( \Delta^{-1}\partial_i\partial_j \left( V_a Q_\bot^a \right) \right)^2 \right.
\nonumber\\
& \quad\quad - 4\epsilon\calR (V_a Q_\bot^a)
+ \frac{2\epsilon}{H}\dot\calR(V_a Q_\bot^a)
- \frac{2\epsilon}{H} \left( \Delta^{-1} \partial^i \partial^j\dot\calR \right) \Delta^{-1} \partial_i \partial_j (V_a Q_\bot^a)
\nonumber\\
& \quad\quad - \frac{1}{2H}\dot\calR \left( V_aQ_\bot^a \right) + \frac{1}{2H}\calR \left( 3HV_aQ_\bot^a+V_a\dot{Q}_\bot^a \right)
\nonumber\\
& \quad\quad \left. + \frac12G_{ab} \left( \dot{Q}_\bot^a \dot{Q}_\bot^b - \partial^i Q_\bot^a \partial_i Q_\bot^b \right) - \frac12 M_{ab}^2 \left( Q_\bot^a Q_\bot^b - \frac{2\dot\phi^a_0}{H}\calR Q_\bot^b \right) \right]
\nonumber\\
+ & \left[ G_{ab}\partial^i Q_\bot^a \dot{Q}_\bot^b - \frac{1}{H} \left( \partial^i \calR V_a Q_\bot^a + \calR V_a \partial^i Q_\bot^a \right) \right] \Delta^{-1}\partial_i \left( \epsilon\dot\calR-\frac{V_aQ_\bot^a}{\mpl^2H} \right)
\nonumber\\
+ & \left. \frac{2}{H} \left( \epsilon \dot\calR\partial^i\calR - \delta_1 H \calR \partial^i \calR \right)  \Delta^{-1}\partial_i \left( V_aQ_\bot^a \right) \right\} \, .
\end{align}
Again, we can straightly compute the bispectrum of the curvature perturbation and the corrections from the interaction with the orthogonal modes $Q^a_\bot$ using the in-in formalism as we did in the previous section.

\subsection{Large-scale approximation}

Until now, we have considered the standard cosmological perturbation theory. Here, ``standard'' means that our perturbative expansion is based on the order of perturbations we are interested in. For example, with the typical order of perturbations being denoted by $\zeta$, for linear perturbation theory we are interested in $\calO(\zeta)$ and drop higher-order contributions, for second-order theory $\calO(\zeta^2)$, and so on. While perfectly legitimate, it is not the only and most convenient approach. Especially, if we are interested in non-linear perturbations, we have to go to the higher-order action or Einstein equations of our interest since the contributions of, say, second-order perturbations are otherwise not captured. Furthermore, as we have seen, finding the higher-order action is very tedious and usually requires further manipulations including integrations by parts and the background equation of motion.

In this section, we briefly discuss an alternative approach based on large-scale approximation. The benefit of this approach is twofold. During inflation, physical scales expand faster than the Hubble horizon. Thus all scales of our observational interest were once far outside horizon during inflation so that taking large-scale approximation for them is making very good sense to track their evolution on super-horizon scales. More importantly, the fully non-linear perturbation equations to be solved become surprisingly simple {\em irrespective} of the order of perturbations of our interest. This means we can follow the full non-linear evolution of perturbations, not necessarily order by order as we do in the standard manner.

\subsubsection{Non-linear equations}

First we decompose the energy-momentum tensor as measured by observers moving orthogonally (four-velocity is identical to the normal vector) to the slices set by the 3+1 decomposition of the ADM metric. Naturally, we can interpret the energy density $\calE$ as the pure temporal part, i.e. two indices are all contracted with the normal vector, the momentum density $\calJ^i$ as one index contracted with the normal vector while the other projected, and the spatial energy-momentum tensor $\calS_{ij}$ as the pure spatial part, i.e. two indices are all projected. In the ADM decomposition, we can write the unit timelike vector normal to the constant-time hypersurface $n^\mu$ has the components
\begin{align}
\label{eq:normalvector}
n^\mu & = \left( \frac{1}{\calN}, \frac{\beta^i}{\calN} \right) \, ,
\\
n_\mu & = \left( -\calN, 0 \right) \, .
\end{align}
With this, if we choose to write in the simplest forms, we have 
\begin{align}
\calE & = \calN^2T^{00} \, ,
\\
\calJ_i & = \calN T^0{}_i \, ,
\\
\calS_{ij} & = T_{ij} \, .
\end{align}

In the ADM decomposition scheme, we can also write the Einstein equation in terms of the components orthogonal and tangential to the constant-time hypersurface~\cite{Bardeen:1980kt}. The two constraint equations follow from those involving temporal components, i.e. both or one of the indices of the Einstein equation is contracted with the normal vector. The geometric parts of them are the well-known Gauss-Codazzi equations, and the corresponding matter parts are precisely the energy and momentum density $\calE$ and $\calJ_i$ found above. Further, these equations do not involve explicit time derivatives. Thus, they are equations of constraints which must be satisfied by the fundamental ADM variables, $\gamma_{ij}$ and $\dot\gamma_{ij}$, at all times. We can obtain the energy and the momentum constraints as
\begin{align}
R^{(3)} + \frac{2}{3}K^2 - \overline{K}^i{}_j\overline{K}^j{}_i & = \frac{2}{\mpl^2}\calE \, ,
\\
\overline{K}^j{}_{i;j} - \frac{2}{3}K_{;i} & = \frac{\calJ_i}{\mpl^2} \, ,
\end{align}
respectively, where
\begin{equation}
\overline{K}_{ij} \equiv K_{ij} - \frac{1}{3}\gamma_{ij}K
\end{equation}
is the traceless part of $K_{ij}$. The dynamical equations thus involve the spatial energy-momentum tensor $\calS_{ij}$. They can be formally written as the Lie derivatives of $\gamma_{ij}$ and $K_{ij}$, or more explicitly, the trace and traceless evolution equations can be found as
\begin{align}
K_{,0} - \beta^iK_{,i} & = -\calN^{;i}{}_{;i} + \calN \left( R^{(3)} + K^2 + \frac{1}{2\mpl^2}\calS - \frac{3}{2\mpl^2}\calE \right) \, ,
\\
\overline{K}^i{}_{j,0} - \beta^k\overline{K}^i{}_{j,k} + \beta^i{}_{,k}\overline{K}^k{}_j - \beta^k{}_{,j}\overline{K}^i{}_k & = -\calN^{;i}{}_{;j} + \frac{1}{3}\delta^i{}_j\calN^{;k}{}_{;k} + \calN \left( {\overline{R}^{(3)}}^i{}_j + K\overline{K}^i{}_j - \frac{1}{\mpl^2}\overline\calS^i{}_j \right) \, ,
\end{align}
where the trace and traceless parts of $\calS_{ij}$, denoted by $\calS$ and $\overline{\calS}_{ij}$, are given in the same way as $K_{ij}$, i.e. $\calS \equiv \gamma^{ij}\calS_{ij}$ and $\overline\calS_{ij} \equiv \calS_{ij} - \gamma_{ij}\calS/3$. The local energy and momentum conservation equations, given by $T^{0\mu}{}_{;\mu} = 0$ and $T^{\mu}{}_{i;\mu} = 0$ are
\begin{align}
\calE_{,0} - \beta^i\calE_{,i} & = \calN K \left( \calE + \frac{\calS}{3} \right) + \calN\overline{K}^i{}_j\overline{S}^j{}_i + \frac{1}{\calN} \left( N^2\calJ^i \right)_{;i} \, ,
\\
\calJ_{i,0} - \beta^j\calJ_{i,j} - \beta^j{}_{,i}\calJ_j & = \calN K\calJ_i - \left( \calE\delta^j{}_i + \calS^j{}_i \right)\calN_{;j} - \calN\calS^j{}_{i;j} \, .
\end{align}
With these non-linear equations with the energy-momentum tensor given by \eqref{eq:EMtensor}, we now can apply the large-scale approximation. That is, we assume a certain smoothing scale $1/k$ greater than which we can approximate well the actual observable universe. Then identifying
\begin{equation}
\varepsilon \equiv \frac{k}{aH}
\end{equation}
as the fictitious parameter associated with a spatial gradient $\partial_i$, we can expand the exact non-linear equations as a power series in $\varepsilon$. For more careful accounts on the conditions for this expansion, see e.g.~\cite{Lyth:2004gb}.

Before we proceed further, let us pause for a minute and consider the geometry set by the time slicing. A choice of a time coordinate determines a family of constant-time hypersurfaces $\Sigma$ in the perturbed space-time, which we refer to as time slicing. With each time slicing there are three important geometrical quantities: namely, the intrinsic scalar curvature of each $\Sigma$, the expansion rate $\theta$ and the shear $\sigma$ of the unit vector field normal to $\Sigma$. The ``curvature perturbation'' represents the amplitude of perturbation in the intrinsic curvature of $\Sigma$. The remaining two quantities $\theta$ and $\sigma$ are both connected with the behaviour of the vector normal to $\Sigma$. Here, we concentrate on the expansion rate $\theta$ being given by the divergence of the normal vector \eqref{eq:normalvector}. Then we can identify $\theta$ as the extrinsic curvature \eqref{eq:extrinsiccurvature}:
\begin{equation}
\theta = n^\mu{}_{;\mu} = -K \, .
\end{equation}
As $\theta$ denotes the expansion rate, we can interpret the extrinsic curvature $K$ as the ``local'' Hubble parameter $H = H(t,\mathbi{x})$:
\begin{equation}
\label{eq:localH}
K(t,\mathbi{x}) \equiv -3H(t,\mathbi{x}) \, .
\end{equation}
Then, by neglecting the second-order spatial gradient terms, the {\em fully non-linear} energy constraints, ADM trace and the scalar field equations of motion become~\cite{Salopek:1990jq}
\begin{align}
\label{eq:gradexp1}
& H^2 = \frac{1}{3\mpl^2} \left( \frac{G_{ab}}{2} \frac{\partial\phi^a}{\calN\partial{t}} \frac{\partial\phi^b}{\calN\partial{t}} + V \right) \, ,
\\
\label{eq:gradexp2}
& \frac{\partial{H}}{\calN\partial{t}} = -\frac{G_{ab}}{2\mpl^2} \frac{\partial\phi^a}{\calN\partial{t}} \frac{\partial\phi^b}{\calN\partial{t}} \, ,
\\
\label{eq:gradexp3}
& \frac{D}{\calN\partial{t}} \left( \frac{\partial\phi^a}{\calN\partial{t}} \right) + 3H\frac{\partial\phi^a}{\calN\partial{t}} + G^{ab}V_b = 0 \, .
\end{align}
Thus, in terms of the proper time $\calN dt$ the above equations are precisely the same as the {\em background} equations \eqref{eq:Friedmann}, \eqref{eq:Hdot} and \eqref{eq:BGphi}. That is, on sufficiently large scales, the information on the evolution of non-linear perturbations can be described by essentially background equations~\cite{Sasaki:1998ug}.

Another important point is that from \eqref{eq:gradexp1}, \eqref{eq:gradexp1} and \eqref{eq:gradexp3} the local Hubble parameter, which is essentially a geometric quantity, is related to the background-like field evolution. From the non-perturbative form of the spatial metric \eqref{eq:R}, using the identity
\begin{equation}
K = -\frac{d}{Ndt}\log\sqrt{\gamma} \, ,
\end{equation}
from \eqref{eq:localH} we notice
\begin{equation}
\label{eq:localH2}
H(t,\mathbi{x}) = \frac{d}{Ndt}\log\left(ae^\varphi\right) \, .
\end{equation}
Now, for each spatial point $\mathbi{x}$ we can define the integral with respect to the proper time:
\begin{equation}
N \equiv \int H \calN dt \, ,
\end{equation}
which is, with $H$ being the volume expansion rate of the three-hypersurfaces, the total expansion of the spatial volume, vix. the number of $e$-folds. From \eqref{eq:localH2}, given one and another moments $t_1$ and $t_2$, we can trivially obtain 
\begin{equation}
N = \int_1^2 H \calN dt = \log \left[ \frac{a(t_2)}{a(t_1)} \right] + \varphi(t_2) - \varphi(t_1) \equiv N_0 + \varphi(t_2) - \varphi(t_1) \, .
\end{equation}
Thus the total number of $e$-folds is given by the background number of $e$-folds $N_0$ given solely by the background scale factor, and the difference between the curvature perturbations evaluated at each moment. That is,
\begin{equation}
\delta{N} \equiv N - N_0 = \Delta\varphi \, .
\end{equation}
Notice that we have not specified the gauge conditions on the hypersurfaces at $t_1$ and $t_2$. Thus we are free to choose in such a way that the initial hypersurface at $t_1$ is flat, on which by definition $\varphi(t_1)=0$, and the final one at $t_2$ is comoving, $\varphi(t_2) = \calR(t_2)$. Then
\begin{equation}
\delta{N}(t_1,t_2) = \calR(t_2) \, ,
\end{equation}
i.e. the perturbation in the number of $e$-folds is identical to the non-linear final comoving curvature perturbation.

With multi-field inflation in our mind, let us take the initial time $t_1$ to be some time during inflation when all the modes of our interest are sufficiently outside the horizon so that the large-scale approximation is valid, and the final time $t_2$ to be some time after inflation when $\calR$ has reached its constant value. Then the dependence of $\delta{N}$ on $t_1$ appears through the phase space variable\footnote{As can be read from\eqref{eq:conjmom-phi} and \eqref{eq:BGmomphi-sol}, $\dot\phi^a$ is to be precise not the conjugate momentum of $\phi^a$. See~\cite{Remmen:2013eja} for more careful discussions on the phase space for inflationary dynamics.} $\phi^a(t_1)$ and $\dot\phi^a(t_1)$. But if we further make use of the slow-roll approximation since $t_1$ is some time during inflation, we can eliminate the dependence on $\dot\phi^a(t_1)$ so that $\delta{N}$ depends only on $\phi^a(t_1)$. Thus, we can relate the field fluctuations on the initial flat hypersurface $Q^a(t_1)$ and the final comoving curvature perturbation $\calR(t_2)$ as
\begin{equation}
\label{eq:deltaN}
\calR(t_2) = \delta{N} = N_a(t_1)Q^a(t_1) + \frac{1}{2}N_{ab}(t_1)Q^a(t_1)Q^b(t_1) + \cdots \, ,
\end{equation}
where $N_a \equiv \partial{N}/\partial\phi^a$ and so on and we have explicitly expanded up to second order. This is the so-called {\em $\delta{N}$ formalism}~\cite{multi-eom,Salopek:1990jq,Sasaki:1998ug,deltaNold}.

\subsubsection{Scale dependence in the $\delta{N}$ formalism}

\begin{figure}[h]
\begin{center}
 \includegraphics[width=10cm, trim = 5cm 20cm 5cm 1cm]{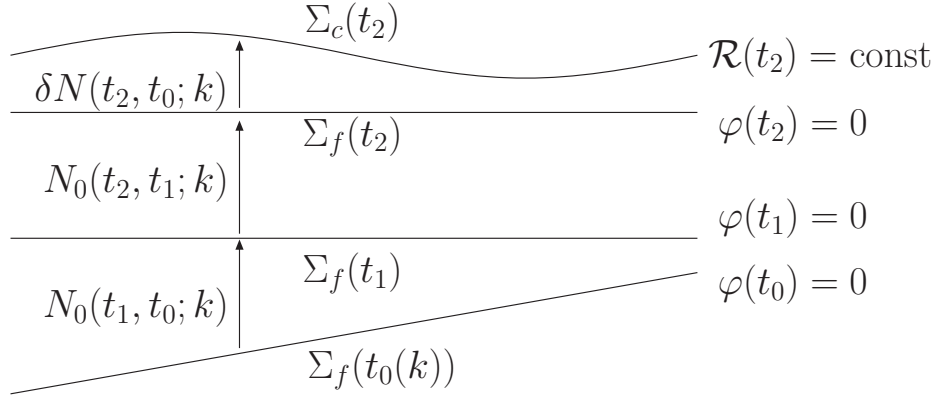}
\end{center}
\caption{A schematic figure showing different moments described in the main text: $t_0(k)$, $t_1$ and $t_2$. We evaluate the final comoving curvature perturbation $\calR(t_2)$ on the hypersurface $\Sigma(t_2)$ with comoving slicing after all isocurvature modes have decayed. We meanwhile choose flat slicing on which the curvature perturbation under this slicing condition, denoted by $\varphi$, of course vanishes. Since $t_1$ and $t_2$ are common to all modes of observational interest, the evolution of $\calR$ is identical between them. But $t_0(k)$ is different for different $k$-modes so the evolution of $\calR$ between $t_0$ and $t_1$ is distinctive for each mode.}
\label{fig:deltaN}
\end{figure}

In the previous section, we have seen that the $\delta{N}$ formalism is implemented in the configuration space with two fixed initial and final moments $t_1$ and $t_2$ respectively. These moments are common to all modes of observational interest, and hence the momentum dependence which is of crucial observational importance is not explicit. This however becomes manifest by taking into account the moment of horizon crossing for each mode as follows~\cite{deltaN_k-dep}. For a certain mode with $k$, the horizon crossing happens at $t_0 < t_1$, i.e. $k = (aH)_0$. Then, the number of $e$-folds elapsed between $t_0$ and $t_1$ is obviously $k$-dependent as
\begin{equation}
\Delta{N}_k \equiv \log \left( \frac{a_1}{a_0} \right) \approx \log \left[ \frac{(aH)_1}{k} \right] \, .
\end{equation}
Therefore, the field fluctuations at the initial moment $Q^a(t_1)$ in terms of which the $\delta{N}$ formalism is written as in \eqref{eq:deltaN} can be expanded as
\begin{equation}
\label{eq:Q_eq1}
Q^a(N_1 = N_0 + \Delta{N}_k) = Q^a(N_0) + \Delta{N}_k D_NQ^a(N_0) + \frac{1}{2} \left( \Delta{N}_k \right)^2 D_N^2Q^a(N_0) + \cdots \, ,
\end{equation}
where $N_0$ and $N_1$ are the numbers of $e$-folds corresponding to $t_0$ and $t_1$ respectively, and $D_N$ is a covariant derivative with respect to $N$. Thus, we can see that the non-trivial $k$-dependence is gained between the moment of horizon crossing $t_0(k)$ which is different mode by mode, and the initial moment for the $\delta{N}$ formalism $t_1$ which is common to all modes. In some sense, $t_1$ is an intermediate reference moment from which the evolution of each mode until $t_2$ is identical. Figure~\ref{fig:deltaN} shows $t_0(k)$, $t_1$ and $t_2$ and the evolution of the curvature perturbation.

For practical purpose, we have to find explicitly the derivatives $D_NQ^a$, $D_N^2Q^a$ and so on. From the equation of motion for $Q^a$ on very large scales with $N$ being the time variable~\cite{Elliston:2012ab},
\begin{align}
D_NQ^a & = w^a{}_bQ^b + \cdots \, ,
\\
w_{ab} & = u_{(a;b)} + \frac{\mathbb{R}_{c(ab)d}}{3} \frac{\dot\phi_0^c}{H}\frac{\dot\phi_0^d}{H} \, ,
\\
w_{a(bc)} & = u_{(a;bc)} + \frac{1}{3} \left[ \mathbb{R}_{(a|de|b;c)}\frac{\dot\phi^d_0}{H}\frac{\dot\phi^e_0}{H} - 4\mathbb{R}_{a(bc)d}\frac{\dot\phi^d_0}{H} \right] \, ,
\\
u_a & = -\frac{V_{a}}{3H^2} \, ,
\end{align}
where the indices between vertical bars are excluded from the symmetrization, then \eqref{eq:Q_eq1} becomes
\begin{equation}
\label{eq:Qevolution2}
Q^a(N_1) = Q^a + \Delta{N}_k \left( w^a{}_b Q^b + \frac{1}{2} w^a{}_{bc} Q^bQ^c + \cdots \right) + \frac{1}{2} \left( \Delta{N}_k \right)^2 \left[ \left( D_Nw^a{}_b \right)Q^b + w^a{}_bw^b{}_cQ^c \right] + \cdots \, .
\end{equation}
where all terms on the right hand side are evaluated at $N_0$, and
\begin{equation}
D_Nw^{ab} = w^{ab}{}_{;c}\frac{\dot\phi_0^c}{H} \, .
\end{equation}
Note that $\Delta{N}_k$ spans 5 - 10 for the current observational range of the CMB, but as we can see the coefficients of $(\Delta{N}_k)^2$ are of second order in slow-roll, so it is safely suppressed compared to $\Delta{N}_k$ terms as long as the expansion \eqref{eq:Q_eq1} or \eqref{eq:Qevolution2} remains valid.

\subsubsection{Correlation functions in the $\delta{N}$ formalism}

Now using the $\delta{N}$ formalism, we compute the correlation functions of the final comoving curvature perturbation $\calR(t_2)$. We first move to the Fourier space, where from \eqref{eq:deltaN} we can write the Fourier component of $\calR(t_2)$ as
\begin{equation}
\label{eq:deltaNformula}
\calR_k(t_2) = N_a(t_1)Q_k^a(t_1) + \frac{1}{2}N_{ab}(t_1) \left[ Q^a(t_1)\star Q^b(t_1) \right]_k + \cdots \, ,
\end{equation}
where star denotes a convolution. We first calculate the power spectrum $\calP_\calR$ and its momentum dependence, viz. the spectral index $n_\calR$. For the power spectrum it is sufficient to work to linear order in $Q^a$, in which case it is identical to $\delta\phi^a$. The two-point correlation function of $\calR(t_2)$ is, from \eqref{eq:deltaNformula}, related to that of $Q^a$ as
\begin{equation}
\left\langle \calR_\mathbi{k}(t_2)\calR_\mathbi{q}(t_2) \right\rangle = N_a(t_1)N_b(t_1) \left\langle Q_\mathbi{k}^a(t_1)Q_\mathbi{q}^b(t_1) \right\rangle \, .
\end{equation}
Using \eqref{eq:Qevolution2}, the two-point correlation function of $Q^a$ at $t_1$ can be written in terms of that at the moment of horizon crossing $t_0$ as
\begin{equation}
\label{eq:QQt1-QQt0}
\left\langle Q_\mathbi{k}^a(t_1)Q_\mathbi{q}^b(t_1) \right\rangle = \left\langle Q_\mathbi{k}^a(t_0)Q_\mathbi{q}^b(t_0) \right\rangle + 2\Delta{N}_k w^a{}_c \left\langle Q_\mathbi{k}^b(t_0)Q_\mathbi{q}^c(t_0) \right\rangle \, ,
\end{equation}
where {\em assuming} that $Q^a$ is light and follows slow-roll dynamics, $\left\langle Q_\mathbi{k}^aQ_\mathbi{q}^b \right\rangle$ is given by \eqref{eq:Q-2point}. Thus, from the definition of the power spectrum \eqref{eq:Pdef}, we can read the leading power spectrum of the curvature perturbation in the $\delta{N}$ formalism as
\begin{equation}
\calP_\calR = \left( \frac{H}{2\pi} \right)^2 N^aN_a \, .
\end{equation}
To compute the scale dependence to leading order in slow-roll, there are obvious two leading contributions: the scale dependence of $H^2(t_0)$ which gives rise to $-2\epsilon$, and $w^a{}_c$ which is multiplied by $\Delta{N}_k$. The other terms, e.g. the derivative of $\zeta^{ab}$, are further slow-roll suppressed so are sub-leading. Hence we can straightforwardly calculate the spectral index as
\begin{equation}
n_\calR-1 = \frac{D \log\calP_\calR}{d\log k} = -2\epsilon - 2\frac{N_aN_bw^{ab}}{N_cN^c} \, .
\end{equation}
An obvious new contribution is the field space curvature $\mathbb{R}_{abcd}$ contained in $w^{ab}$. This term must be $\calO(\epsilon)$ to be consistent with the observational constraint \eqref{eq:nR-obsbound}, barring an accidental cancellation with other terms.

The bispectrum of $\calR$ is defined in a similar manner to the power spectrum by
\begin{equation}
\label{eq:Bdef}
\left\langle \calR_{\mathbi{k}_1}(t_2)\calR_{\mathbi{k}_2}(t_2)\calR_{\mathbi{k}_3}(t_2) \right\rangle \equiv (2\pi)^3\delta^{(3)}(\mathbi{k}_1+\mathbi{k}_2+\mathbi{k}_3)B_\calR(k_1,k_2,k_3) \, .
\end{equation}
As we have seen, the $\delta{N}$ formalism concerns the evolution on super-horizon scales thus can describe the non-Gaussianity generated by non-linear evolution on very large scales. It is however blind to what happens on smaller scales, including the moment of horizon crossing. We thus usually assume when we compute the bispectrum using the $\delta{N}$ formalism that the intrinsic non-Gaussianity of the fields at the moment of horizon crossing is negligible, 
\begin{equation}
\left\langle Q^a_{\mathbi{k}_1}(t_0)Q^b_{\mathbi{k}_2}(t_0)Q^c_{\mathbi{k}_3}(t_0) \right\rangle = 0 \, .
\end{equation}
This is however not necessarily the case always, since there are models of inflation where the intrinsic non-Gaussianity at the horizon crossing is not negligible. Therefore to make a proper account of non-Gaussianity using the $\delta{N}$ formalism, we should make use of the form that gives the intrinsic non-Gaussianity as small as possible, by e.g. appropriate field redefinition~\cite{Domenech:2016zxn}.

From \eqref{eq:deltaNformula}, we can see that the three-point function consists of two terms, 
\begin{align}
\label{eq:RRR}
\left\langle \calR_{\mathbi{k}_1}(t_2)\calR_{\mathbi{k}_2}(t_2)\calR_{\mathbi{k}_3}(t_2) \right\rangle = & N_aN_bN_c \left\langle Q^a_{\mathbi{k}_1}Q^b_{\mathbi{k}_2}Q^c_{\mathbi{k}_3} \right\rangle + \frac{1}{2} \left\{ N_{ab}N_cN_d \left\langle \left[ Q^a\star Q^b \right]_{\mathbi{k}_1}Q^c_{\mathbi{k}_2}Q^d_{\mathbi{k}_3} \right\rangle + \text{2 perm} \right\} \, .
\end{align}
We can compute these terms very easily: for the first term, using \eqref{eq:Qevolution2} and \eqref{eq:Q-2point}, we can easily find
\begin{align}
\label{eq:B-1st}
& N_a(t_1)N_b(t_1)N_c(t_1) \left\langle Q^a_{\mathbi{k}_1}(t_1)Q^b_{\mathbi{k}_2}(t_1)Q^c_{\mathbi{k}_3}(t_1) \right\rangle
\nonumber\\
& = N_a(t_1)N_b(t_1)N_c(t_1) \left\{ \frac{1}{2}\Delta{N}_{k_1}w^a{}_{de} \left\langle \left[ Q^d(t_0)\star Q^e(t_0) \right]_{\mathbi{k}_1}Q^b_{\mathbi{k}_2}(t_0)Q^c_{\mathbi{k}_3}(t_0) \right\rangle + \text{2 perm} \right\}
\nonumber\\
& = (2\pi)^3 \delta^{(3)}(\mathbi{k}_1+\mathbi{k}_2+\mathbi{k}_3) N_a(t_1)N_b(t_1)N_c(t_1) \frac{H^4(t_0)}{4k_1^3k_2^3k_3^3} w^{abc} \left( k_1^3\Delta{N}_{k_1} + \text{2 perm} \right) \, .
\end{align}
Likewise, for the second term using \eqref{eq:QQt1-QQt0} trivially gives
\begin{align}
\label{eq:B-2nd}
& \frac{1}{2} N_{ab}(t_1)N_c(t_1)N_d(t_1) \left\langle \left[ Q^a(t_1)\star Q^b(t_1) \right]_{\mathbi{k}_1}Q^c_{\mathbi{k}_2}(t_1)Q^d_{\mathbi{k}_3}(t_1) \right\rangle
\nonumber\\
& = (2\pi)^3\delta^{(3)}(\mathbi{k}_1+\mathbi{k}_2+\mathbi{k}_3) N_{ab}(t_1)N_c(t_1)N_d(t_1) \frac{H^4(t_0)}{4k_1^3k_2^3} \left( \gamma^{ac}\gamma^{bd} + 2\Delta{N}_{k_1}w^{ac}\gamma^{bd} + 2\Delta{N}_{k_2}\gamma^{ac}w^{bd} \right) \, .
\end{align}
Thus putting both terms together, the leading bispectrum reads
\begin{equation}
\label{eq:B1}
B_\calR(k_1,k_2,k_3) = \frac{N^{ab}N_aN_b}{(N^cN_c)^2} \left[ P_\calR(k_1)P_\calR(k_2) + \text{2 perm} \right] \, ,
\end{equation}
where we have defined the dimensionfull power spectrum
\begin{equation}
P_\calR \equiv \frac{2\pi^2}{k^3}\calP_\calR \, ,
\end{equation}
so that the definition of the power spectrum \eqref{eq:Pdef} reads similar to \eqref{eq:Bdef}:
\begin{equation}
\left\langle \calR(\mathbi{k})\calR(\mathbi{q}) \right\rangle \equiv (2\pi)^3 \delta^{(3)}(\mathbi{k}+\mathbi{q}) P_\calR(k) \, .
\end{equation}

A convenient way of parametrizing non-Gaussianity is to introduce a set of so-called non-linear parameters. It is defined by the local expansion of the curvature perturbation around its linear, Gaussian component~\cite{Komatsu:2001rj}:
\begin{equation}
\label{eq:fNL}
\calR = \calR_g + \frac{3}{5}\fnl\calR_g^2 + \cdots \, .
\end{equation}
Using the $\delta{N}$ formalism, we are interested in the form of the non-linearity generated during the super-horizon evolution of the curvature perturbation. On super-horizon scales, perturbations at different locations cannot communicate with each other, and in this sense the non-linearity produced during this stage is local. Conversely, in the Fourier space, the momenta that constitute the momentum triangle of the Fourier transformation of the three-point correlation function, i.e. the bispectrum, are very different. Typically, one of three momenta is very small in magnitude, and the remaining two are pointing almost the opposite directions with nearly equal magnitude. Because of the form of the triangle these momenta constitute, it is frequently referred to as the ``squeezed limit''. From the expansion \eqref{eq:fNL}, we can easily compute the bispectrum as
\begin{equation}
\label{eq:B2}
B_\calR(k_1,k_2,k_3) = \frac{6}{5}\fnl \left[ P_\calR(k_1)P_\calR(k_2) + \text{2 perm} \right] \, ,
\end{equation}
where we have assumed for simplicity that $\fnl$ is a constant. Comparing \eqref{eq:B1} and \eqref{eq:B2}, we can read the leading, scale-independent $\fnl$ as~\cite{Lyth:2005fi}
\begin{equation}
\frac{6}{5} \fnl = \frac{N^{ab}N_aN_b}{(N^cN_c)^2} \, .
\end{equation}
Furthermore, as we did for the power spectrum, we can also compute the scale dependence of $\fnl$ by incorporating $\Delta{N}_k$ term explicitly. In the equilateral configuration in which we are sensibly calculate $n_{\fnl}$, the running of $\fnl$~\cite{fNLrunning}, we can trivially find
\begin{equation}
\label{eq:nfNL}
n_{\fnl} \equiv \frac{D\log\fnl}{d\log{k}} = -\frac{N_aN_bN_cw^{abc}}{N^{de}N_dN_e} + 4w^{ab} \left( \frac{N_aN_b}{N^dN_d} - \frac{N_{ac}N_bN^c}{N^{de}N_dN_e} \right) \, .
\end{equation}
The first term in \eqref{eq:nfNL} comes from \eqref{eq:B-1st}, i.e. the non-linearity generated during the evolution of the field fluctuations between $t_0$ and $t_1$, while the other terms in \eqref{eq:nfNL} arises from the non-linear couplings.

\section{Conclusions}
\label{sec:conc}

Inflation offers a simple and consistent framework in which both the homogeneity and isotropy of the current observable universe as well as the tiny initial perturbations as can be observed from the temperature anisotropies in the CMB. Indeed, with the recent observations on the CMB we have entered the era of data-driven cosmology and the concordance cosmological model, $\Lambda$CDM, provide a very good fit to the power spectrum of the CMB anisotropies. The constraints on the properties of the primordial perturbations, \eqref{eq:P-obsbound} - \eqref{eq:fNL-obsbound}, tell us that the primordial adiabatic perturbation has nearly scale-invariant power spectrum and follows almost perfect Gaussian statistics, with the contribution of the tensor perturbation, i.e. the primordial gravitational waves, occupying less than a few percent of power on large scales relevant for the CMB observations. These are mostly consistent with the predictions of simple single field inflation models.

While single field inflation provides the best fit to the data without free parameters, then what should be the merit to consider multi-field inflation? As mentioned at the beginning of this article, multi-field inflation can open a rich possibilities beyond the predictions in single field models that could be constrained and even detected by observations. Some of them include the isocurvature perturbations, correlated or anti-correlated with the curvature perturbation, detectable level of non-Gaussianity that may have non-trivial scale dependence, and possibly residual signatures that survive elusive post-inflationary dynamics such as reheating. Beyond these exciting observational possibilities there are a number of theoretical motivations to study multi-field inflation. The early universe is a huge particle accelerator, and the trail of yet unknown parent physics, whose low-energy effective theory is the standard model of particle physics, should be left on cosmic scales with a multiple number of inflaton fields. In this article we have considered a few basic bricks to understand the dynamics of multi-field inflation.

With advanced observation programs in operation and in plan on the CMB as well as large-scale structure, we are entering the precision era for cosmological observations with unprecedented quality and quantity of data in the coming decade. By that time we should a good understanding about the inflationary dynamics with strong observational supports and/or constraints, along with further lessons about the nature of the physics underlying inflation. At the same time we may well witness the interesting possibilities of multi-field inflation, with further observational and theoretical opportunities widely open.

\subsection*{Acknowledgements}

I thank Ana Achucarro, Christian Byrnes, Ki-Young Choi, Xian Gao, Jai-chan Hwang, Donghui Jeong, Gonzalo Palma, Subodh Patil, Misao Sasaki, David Seery, Min-Seok Seo, Gary Shiu, Ewan Stewart, Spyros Sypsas and Takahiro Tanaka for sharing their insights with me on the subjects discussed in this article. I am also grateful to Misao Sasaki and Takahiro Terada for helpful comments on earlier drafts.
I acknowledge the support from the Korea Ministry of Education, Science and Technology, Gyeongsangbuk-Do and Pohang City for Independent Junior Research Groups at the Asia Pacific Center for Theoretical Physics. I am also supported in part by a Starting Grant through the Basic Science Research Program of the National Research Foundation of Korea (2013R1A1A1006701) and by a TJ Park Science Fellowship of POSCO TJ Park Foundation.

\newpage

\appendix

\renewcommand{\theequation}{\Alph{section}.\arabic{equation}}

\section{Gauge transformations}
\label{app:gauge}
\setcounter{equation}{0}

In this appendix section, we address the issue of gauge transformation which was not discussed in the main text. For that, we first consider the issue of ``background'' and ``perturbation''. In the background universe $U$, there is no ambiguity in choosing the time coordinate on the homogeneous and isotricpic spatial hypersurfaces in such a way that time is constant: $t=t_1$ corresponds to the moment when the homogeneous scalar field has a specific value of $\phi(t=t_1)$, and so on. However, in a perturbed universe $\widehat{U}$, our choice of time is arbitrary in the sense that we can choose arbitrary coordinate system where the deviation from homogeneity and isotropy is small. In different coordinate systems, the notion of perturbations is different too. For example, we can choose spatial hypersurfaces on which the density perturbation vanishes. Thus, just saying that the density perturbation is such and such is not enough. We have to also specify the coordinate system in describing the density perturbation.

Let us consider in a more detail. How can we define the perturbation in a scalar quantity $\widehat\phi$ at a point $p$ in the perturbed universe $\widehat{U}$? To define the perturbation, we need to specify the corresponding background value $\phi_0$: the difference between $\widehat\phi$ and $\phi_0$ is the perturbation $\delta\phi(p)$. But what is the corresponding background $\phi_0$? For this, we have to specify a coordinate system, or {\em mapping} in such a way that each point in the perturbed universe $\widehat{U}$ is associated with the corresponding point $x^\mu$ in the background universe $U$. Once this mapping is specified, the perturbation 
\begin{equation}\label{scalar_pert}
\delta\phi(p) = \widehat\phi(p) - \phi_0(x^\mu)
\end{equation}
is meaningful.

\begin{figure}[h]
 \centering
  \includegraphics[angle = -90, width=10cm]{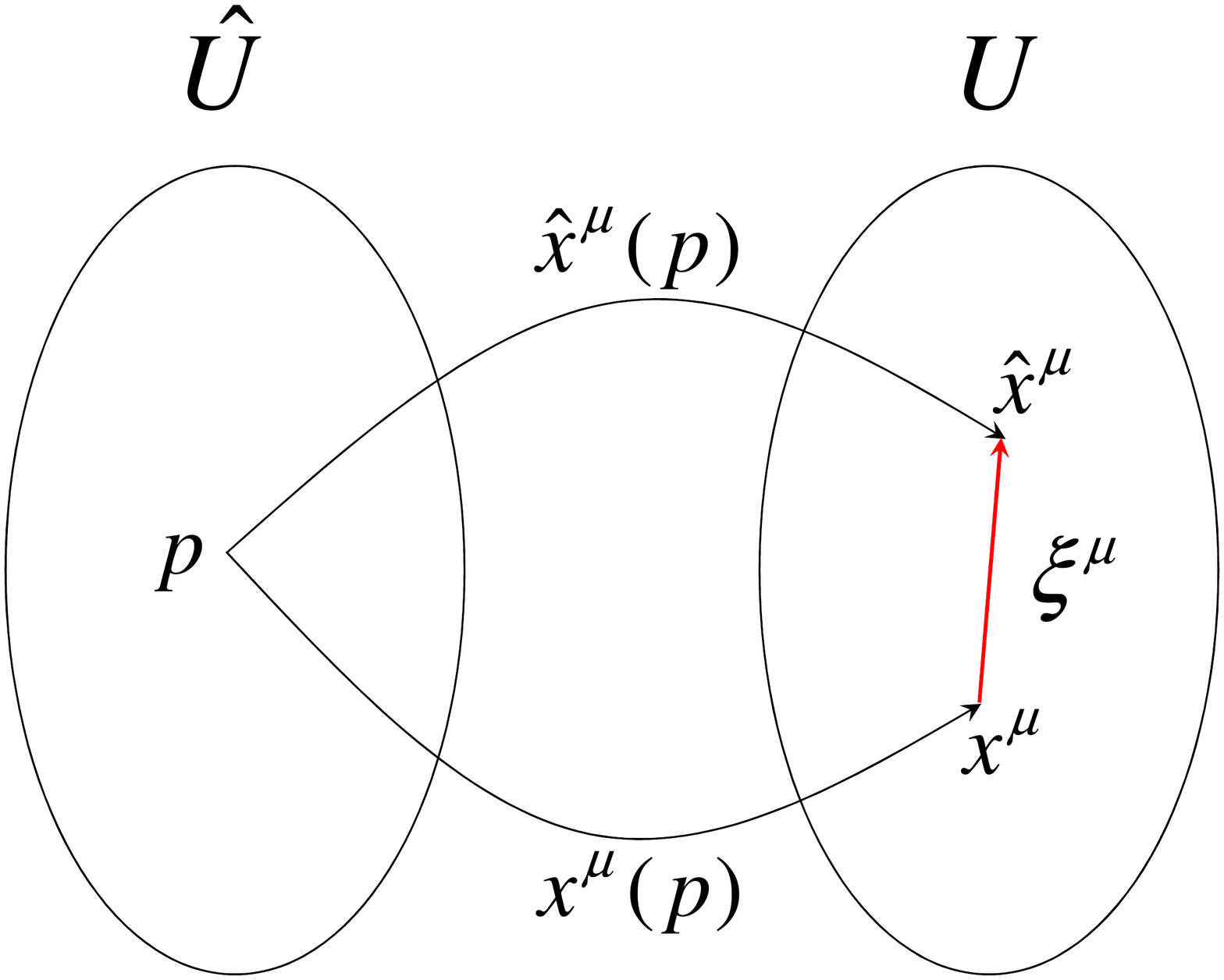}
  \caption{A schematic image of gauge transformation.}
 \label{fig:gauge_transformation}
\end{figure}

So to specify perturbations we only need to specify the coordinate system, or the mapping between $\widehat{U}$ and $U$. The problem is, as stated before, there is no natural choice of this mapping and one is as good (or bad) as the others. Thus we need to know how one mapping is related to another. It is very important to note that any change induced by a change in the mapping is {\em not} physical: it is simply a transformation because we have changed the coordinate, or ``gauge'', to describe the same thing. In this sense, this non-physical change is called {\em gauge transformation}. Suppose two coordinate systems, $x^\mu$ and $\hat{x}^\mu$, which map $p$ in  $\widehat{U}$ to the corresponding {\em different} points in $U$, are related by
\begin{equation}\label{coordinate_transformation}
\widehat{x}^\mu(p) = x^\mu(p) + \xi^\mu(x^\nu(p)) \, .
\end{equation}
If this transformation is infinitesimal, we have
\begin{align}
\widehat{\delta\phi}(p) & = \widehat\phi(p) - \phi_0(\hat{x}^\mu(p))
\nonumber\\
& = \delta\phi(p) - \left[ \phi_0(\hat{x}^\mu(p)) - \phi_0(x^\mu(p)) \right]
\nonumber\\
& = \delta\phi(p) - \xi^\nu \frac{\partial\phi_0}{\partial{x}^\nu}(x^\mu(p)) \, .
\end{align}
Since the background universe $U$ is spatially homogeneous and isotropic, $\phi_0 = \phi_0(x^0)$ so we simply have
\begin{equation}\label{scalar_transformation}
\widehat{\delta\phi}(p) = \delta\phi(p) - \dot\phi_0(t(p))\xi^0(x^\mu(p)) \, ,
\end{equation}
where we have taken $x^0 = t$. It is very important to note that we are comparing two different mappings, $x^\mu(p)$ and $\hat{x}^\mu(p)$, from the {\em same} point $p$ in $\widehat{U}$. It is schematically shown in Figure~\ref{fig:gauge_transformation}.

Note that we can extract the gauge transformations of the metric perturbations by requiring that $ds^2$ be invariant under the gauge transformation\footnote{Generic gauge transformation law for an arbitrary tensor quantity $Q$ under the coordinate transformation $x^\mu \to \hat{x}^\mu = x^\mu+\xi^\mu$ can be written in terms of the Lie derivatives as
\begin{equation*}
\widehat{\delta{Q}} -\delta{Q} = -\calL_\xi Q \, ,
\end{equation*}
where $\calL_\xi$ is the Lie derivative in the direction of the vector $\xi$. By component, a Lie derivative on a tensor $T^{\alpha_1\cdots}{}_{\beta_1\cdots}$ is given by
\begin{equation*}
\calL_\xi \left( T^{\alpha_1\cdots}{}_{\beta_1\cdots} \right) = \xi^\gamma \partial_\gamma T^{\alpha_1\cdots}{}_{\beta_1\cdots} - \left( \partial_\gamma\xi^{\alpha_1} \right) T^{\gamma\alpha_2\cdots}{}_{\beta_1\cdots} + \left( \partial_{\beta_1}\xi^\gamma \right) T^{\alpha_1\cdots}{}_{\gamma\beta_2\cdots} + \cdots \, .
\end{equation*}
For example, the perturbations in the metric tensor transform as 
\begin{equation*}
\widehat{\delta{g}_{\mu}} - \delta{g}_{\mu\nu} = -g_{\mu\nu,\rho}\xi^\rho - g_{\rho\nu}\xi^\rho{}_{,\nu} - g_{\mu\rho}\xi^\rho{}_{,\mu} \, .
\end{equation*}}: with the coordinate transformations
\begin{align}
\label{transformation_t}
t & \to \widehat{t} = t + \xi^0(t,\mathbi{x}) \, ,
\\
\label{transformation_x}
x^i & \to \widehat{x^i} = x^i + \xi^i(t,\mathbi{x}) \, ,
\end{align}
we can easily see that in the linear order
\begin{align}
a\left(\hat{t}\right) \equiv \widehat{a} & = \left( 1 + H\xi^0 \right) a(t) \, ,
\\
\widehat{dt} & = \left( 1 + \dot{\xi^0} \right)dt + \xi^0_{,i}dx^i \, ,
\\
\widehat{dx^i} & = dx^i + \dot\xi^idt + \xi^i{}_{,j}dx^j \, .
\end{align}
From the fact that the line element in space-time is the same irrespective of the coordinate transformation, we can write using the perturbed metric \eqref{eq:metric-pert}
\begin{align}
\widehat{ds}^2 = & -\left( 1 + 2\widehat{A} \right) \widehat{dt}^2 + 2\widehat{a}\widehat{\mathcal{B}}_i\widehat{dt}\widehat{dx^i} + \widehat{a}^2 \left[ \left( 1 + 2\widehat{\varphi} \right) \delta_{ij} + 2\widehat{\mathcal{E}}_{ij} \right] \widehat{dx^i}\widehat{dx^j}
\nonumber\\
= & -\left[ 1 + 2 \left( \widehat{A} + \dot{\xi^0} \right) \right] dt^2 + 2a \left( \widehat{\mathcal{B}}_i - \frac{\xi^0{}_{,i}}{a} + a\dot\xi_i \right)dtdx^i
\nonumber\\
& + a^2 \left\{ \left[ 1 + 2 \left( \widehat{\varphi} + H\xi^0 \right) \right] \delta_{ij} + 2 \left( \widehat{\mathcal{E}}_{ij} + \frac{\xi_{i,j}+\xi_{j,i}}{2} \right) \right\} dx^idx^j \, ,
\end{align}
where $\xi_i = \delta_{ij}\xi^j$.
Equating this expression with \eqref{eq:metric-pert}, we can find that under the coordinate transformation given by \eqref{transformation_t} and \eqref{transformation_x}, the new metric perturbations are given by
\begin{align}
\widehat{A} & = A - \dot\xi^0 \, ,
\\
\widehat{\calB}_i & = \calB_i + \frac{\xi^0_{,i}}{a} - a\dot\xi_i \, ,
\\
\widehat\varphi & = \varphi - H\xi^0 \, ,
\\
\widehat\calE_{ij} & = \calE_{ij} - \frac{\xi_{i,j}+\xi_{j,i}}{2} \, .
\end{align}
Further, we can decompose the spatial gauge transformation vector $\xi^i$ into the scalar and transverse vector components as we did for the metric perturbation,
\begin{equation}
\xi^i = \delta^{ij}\xi_{,j} + \xi^{\text{(tr)}i} \, ,
\end{equation}
where $\xi^{\text{(tr)}i}{}_{,i} = 0$. Then, we can find trivially that the scalar, vector and tensor components of $\calB_i$ and $\calE_{ij}$ given by \eqref{eq:0ipert} and \eqref{eq:spatialmetricpert} transform as
\begin{align}
\label{gauge_transformation:g0i}
\widehat{B} & = B + \frac{\xi^0}{a} - a\dot\xi \, ,
\\
\widehat{S}_i & = S_i - a\dot\xi^\text{(tr)}_i \, ,
\\
\label{gauge_transformation:gij}
\widehat{H}_T & = H_T - \xi \, ,
\\
\widehat{F}_i & = F_i - \xi^\text{(tr)}_i \, ,
\\
\widehat{h}^{TT}_{ij} & = h^{TT}_{ij} \, .
\end{align}
Notice that the tensor perturbation $h^{TT}_{ij}$ as well as the combination $S_i - a\dot{F}_i$ remain the same under the gauge transformation, i.e. it is {\em gauge-invariant}. Thus, when we consider the vector and tensor perturbations, we need not worry about the gauge ambiguity because the variables we are dealing with are from the beginning gauge invariant. Gauge ambiguity only matters for scalar perturbations, and we will explicitly discuss this issue in the following section. Also, it is fruitful to notice that for the scalar components of the Einstein equation, $B$ and $H_T$ only appear in the specific combination $aB-a^2\dot{H}_T$~\cite{Noh:2004bc}. From (\ref{gauge_transformation:g0i}) and (\ref{gauge_transformation:gij}), we can see that
\begin{equation}
\label{shear_transform}
\widehat{a}\widehat{B} - \widehat{a}^2\dot{\widehat{H}}_T = a \left( B + \frac{\xi^0}{a} - a\dot\xi \right) - a^2 \frac{d}{dt} \left( H_T - \xi \right) = aB- a^2\dot{H}_T + \xi^0 \, ,
\end{equation}
so that although the transformations of $B$ and $H_T$ include the spatial component of the gauge transformation $\xi$, in practice only $\xi^0$, the time translation matters.

\end{document}